\documentclass[preprint]{elsarticle}

\usepackage{lineno,hyperref}
\modulolinenumbers[5]

\journal{Computer Science Review}

\usepackage[english]{babel}
\usepackage[utf8x]{inputenc}
\usepackage[T1]{fontenc}
\usepackage{graphicx}
\usepackage{url}
\usepackage{subfigure}
\usepackage{multirow}
\usepackage[colorinlistoftodos]{todonotes}
\usepackage{soul}
\usepackage{pifont}
\usepackage{balance}
\usepackage{url}
\usepackage{siunitx}

\newcommand{\cmark}{\ding{51}}%
\newcommand{\xmark}{\ding{55}}%

\newcommand{\aclBased}{Encryption-based}
\newcommand{\policyBased}{Allocation-based}
\newcommand{\lkhBased}{LKH-based}

\begin{document}
\begin{frontmatter}

\title{Content Privacy Enforcement Models in 
Decentralized Online Social Networks: 
State of Play, Solutions, Limitations, and Future Directions}
%\thanks{Grants or other notes}





\author[1]{Andrea De Salve\corref{cor1}%
}
\ead{andrea.desalve@cnr.it}
\author[2]{Paolo Mori}
\ead{paolo.mori@iit.cnr.it }
\author[3]{Laura Ricci}
\ead{laura.ricci@unipi.it}
\author[4,2]{Roberto Di Pietro}
\ead{rdipietro@hbku.edu.qa}

\cortext[cor1]{Corresponding author}
\address[1]{Institute of Applied Sciences and Intelligent Systems - CNR, DHITECH Scarl Campus Universitario, Lecce, Italy}
\address[2]{Istituto di Informatica e Telematica - CNR, Via Giuseppe Moruzzi, 1, Pisa, Italy}
\address[3]{Universit\`a di Pisa - Dipartimento di Informatica, Largo Bruno Pontecorvo, 3, Pisa, Italy}
\address[4]{Hamad Bin Khalifa University, College of Science and Engineering, Doha, Qatar}

%\author{\IEEEauthorblockN{Andrea De Salve\IEEEauthorrefmark{1}, Paolo Mori\IEEEauthorrefmark{2}, Laura Ricci\IEEEauthorrefmark{3} and Roberto Di Pietro\IEEEauthorrefmark{4}\IEEEauthorrefmark{2}}\\
%\IEEEauthorblockA{\IEEEauthorrefmark{1}Institute of Applied Sciences and Intelligent Systems - CNR, DHITECH Scarl Campus Universitario, Lecce, Italy}\\
%\IEEEauthorblockA{\IEEEauthorrefmark{2}Istituto di Informatica e Telematica - CNR, Via Giuseppe Moruzzi, 1, Pisa, Italy}\\
%\IEEEauthorblockA{\IEEEauthorrefmark{3}Universit\`a di Pisa - Dipartimento di Informatica, Largo Bruno Pontecorvo, 3, Pisa, Italy}\\
%\IEEEauthorblockA{\IEEEauthorrefmark{4}Hamad Bin Khalifa University, College of Science and Engineering, ICT Division, Doha, Qatar}\\ e-mail: andrea.desalve@isasi.cnr.it - paolo.mori@iit.cnr.it - 
%laura.ricci@unipi.it - rdipietro@hbku.edu.qa}

\begin{abstract}
In recent years, Decentralized Online Social Networks (DOSNs) have been attracting the attention of many users because they reduce the risk of censorship, surveillance, and information leakage from the service provider. 
In contrast to the most popular Online Social Networks, which are based on centralized architectures (e.g., Facebook, Twitter, or Instagram), DOSNs are not based on a single service provider acting as a central authority.
Indeed, the contents that are published on DOSNs are stored on the devices made available by their users, which  cooperate to execute the tasks needed to provide the service. 
%RDP la frase che segue non e' chiara
To continuously guarantee their availability, the contents published by a user could be stored on the devices of other users, simply because they are online when required.
%To continuously guarantee their availability, sometimes user contents could be stored on the devices of other users  that should access them according to the privacy preferences expressed by the publishers,
%potentially untrusted devices  simply because they are online when required.
Consequently, such contents must be properly protected by the DOSN infrastructure, in order to ensure that they can be really accessed only by users who have the permission of the publishers. 
As a consequence, DOSNs require efficient solutions for protecting the privacy of the contents published by each user with respect to the other users of the social network. 
In this paper, we investigate and compare the principal content privacy enforcement models adopted by current DOSNs evaluating their suitability to support different types of privacy policies based on user groups. 
Such evaluation is carried out by implementing several models and comparing % simulations and by measuring the 
their performance %they achieve 
for the typical operations performed on groups, i.e., content publish, user join and leave. Further, we also highlight the limitations of current approaches and show future research directions. This contribution, other than being interesting on its own, provides a blueprint for researchers and practitioners interested in implementing DOSNs, and also highlights a few open research directions.
\end{abstract}

\begin{keyword}
Data Privacy, Access Control, Authorization, Online Social Networks, Distributed Systems
\end{keyword}
\end{frontmatter}
%\linenumbers

\section{Introduction}
Decentralized Online Social Networks (DOSNs) \cite{yeung2009decentralization} have been proposed as alternative solutions to the currently widespread centralized Online Social Networks (OSNs) because they give back to the users the control on their own data. 
As a matter of fact, the  most popular OSNs are based on centralized architectures, where service providers (e.g., Facebook, Twitter, or Instagram) have the full control over the data published by their users, thus increasing the risk of censorship, surveillance, and information leakage \cite{jain2021online}. 
DOSNs, instead, are typically based on a P2P architecture (such as a network of trusted servers, or a Distributed Hash Table), where there is no central service provider which
has full control on user data. 
For instance, Diaspora \cite{bielenberg2012growth}, one of the most popular DOSNs which attracted about 400K users\footnote{\url{https://diasp.eu/stats.html}}, is based on a network of independent and federated servers that are provided and managed by the users themselves.
%\todo[inline,color=pink]{Citare Mastodon? Fare riferimento a cose blockchain-based come Steemit, anche solo per mostrare le differenze? Inoltre Diaspora e' federated, non proprio P2P}
%\todo[inline]{ok.Aggiungo dei riferimenti nelle sezioni seguenti. Qui ho menzionato solo diaspora perche' il messaggio che volevo mandare era: esistono e vengono utilizzate.}
Hence, DOSNs give back to users the control over the contents they publish by storing these contents on users' devices, and by running an instance of the service on these devices as well.
For the above reasons, DOSNs have to adopt efficient and effective solutions for guaranteeing contents availability on the one hand, and contents privacy on the other hand \cite{siddula2018empirical}. 
To guarantee their availability, the contents of each user are typically kept available on the DOSN by replicating them on the devices of a number of other users that are online, and by migrating them on the devices of other online ones as soon as the former go offline. 
This strategy raises relevant privacy and security issues, because the users hosting the contents of other users on their devices 
%for ensuring availability,
could not be actually authorized to access those contents according to the privacy preferences of the publishers. Consequently, DOSNs require the definition of proper security strategies and mechanisms to preserve contents privacy.

Most of the popular DOSNs, in an effort to help users smoothly regulate content sharing in adherence to their privacy preferences, allow to organize users in groups.
In this way, each user can choose to share the contents she/he publishes only with the users belonging to a given group.
%\st{ Groups are widely adopted to describe in different scenarios to represet different aspects of the users.}
Groups are widely adopted in social networks for several purposes \cite{gyarmati2009characterizing}.
For instance, some DOSNs allow their users to define some kind of %\st{private} 
groups typically aimed to link together a number of users who have the same interests (such as sports, school, work, hobbies), enabling content sharing among them. The creator is the group administrator, and the other users either have to ask for being admitted, or they are invited by the administrator.
Once members, users can publish contents (such as posts, images, or videos) that will be reserved only to the other members of the group. 
%\todo[inline]{P: i gruppi pubblici non vanno bene per questo discorso, perche' tipicamente i contenuti li possono leggere tutti. In facebook ci sono i closed e i secret. Per ora ho scritto kind of private. I gruppi secret non ci sono piu', quindi va cambiata la frase} 
Another kind of groups are the ones that are defined by users to organize their contacts into a private address book, typically according to circles \cite{gay2017relationship} or the type of relationships with them (e.g., schoolmates, colleagues, family, acquaintances, etc.).  
These groups can be seen only by the user who defined them \cite{cutillo2009safebook,shakimov2011vis}, who can restrict the access to the contents she/he published to the members of one or more of her/his groups.  
%\todo[inline]{Aggiunti riferimenti e tabella che riporta i vari gruppi forniti da alcune DOSNs}
%\todo[inline, color=pink]{Buono, una criticita' potrebbe essere che i riferimenti sono molto pochi, 3 mi pare e tra questi 3 uno e' un riferimento ad un nostro lavoro}
%\todo[inline]{Ho cercato di sistmare ma, relativamente a questi tipi di gruppi, ci sono dolo due DOSNs che li implementano \cite{cutillo2009safebook,shakimov2011vis}.}
The privacy enforcement models adopted by existing DOSNs  
%(e.g., LifeSocial.KOM \cite{graffi2011lifesocial}, and PeerSoN \cite{buchegger2009peerson}) 
to guarantee the privacy of the contents are very different from each other, each one having its own pro and cons. For instance, several approaches \cite{buchegger2009peerson,graffi2011lifesocial,nilizadeh2012cachet,koll2017socialgate,aiello2012lotusnet,sharma2012supernova} 
%\todo[inline,color=yellow]{sopra, come esempio di approccio che usa i gruppi viene utilizzato il 38. PErche' qui non c'e'? Non utilizza la crittografia?}
%\todo[inline]{ho spostato il riferimento perche' 38 non e' un approccio. }
\textcolor{black}{allow their users to define groups and }
exploit encryption mechanisms for making contents visible only to the members of the group they are reserved to.
%In particular, most DOSNs combine  Asymmetric and Symmetric cryptography. %\st{Moreover, to achieve fine-grained access control, each content should be encrypted with its specific content key before being replicated on other users' devices. 
Consequently, such approaches are typically computationally demanding both for publishing and accessing contents and, especially, for changing the privacy preferences.

\begin{table}[!t]
\footnotesize
\centering
\caption{Main characteristics of the surveys existing in the current literature\label{tab:surveySummary}}
\begin{tabular}{ccccp{7.5cm}}
\hline
\multicolumn{1}{|c|}{\textbf{Survey}} 
& \multicolumn{1}{c|}{\rotatebox{90}{\textbf{Privacy enforcement model }}} &
%\multicolumn{1}{c|}{\rotatebox{90}{\textbf{Architecture}}} &
\multicolumn{1}{c|}{\rotatebox{90}{\textbf{Theoretical model}}}  & \multicolumn{1}{c|}{\rotatebox{90}{\textbf{Experimental evaluation}}}  & \multicolumn{1}{c|}{\textbf{Details}} \\ \hline
\cite{datta2010decentralized} &  &  &  & study the architecture to implement a general purpose DOSN\\
\cite{zuo2016survey} &  &  &   & classify the types of social information\\
\cite{koll2017good}  &  &  &   & challenges preventing the adoption of DOSNs\\
\cite{paul2014survey}& \cmark &  &   & analysis of design choices\\
\cite{chowdhury2015taxonomy} & \cmark&  &    & analysis of design choices\\
\cite{masinde2020peer}  & \cmark &  &  & analyze technical requirements and the overlay networks\\
\cite{guidi2018managing} & \cmark &  &   & study of data storage management in DOSN \\
\cite{troncoso2017systematizing}  & \cmark&   &   & analysis of the advantages and disadvantages of the decentralization techniques\\
\cite{schwittmann2013privacy} & \cmark &   &   & discussing the benefit and the performance penalties of decentralized approaches \\
\cite{bahri2018decentralized}  & \cmark  &  & & evaluation of the privacy issues \\
\cite{kumaraguru2007survey} & \cmark& &  & study the properties of privacy policy languages \\
\cite{taheri2015security}  & \cmark &  &  & study of the data privacy, data integrity, and secure social search solutions\\
\cite{de2018survey} & \cmark & \cmark &   & study of the overhead introduced by privacy policy mechanisms \\
%Aggiungere riferimento ad articoli che non sono survey 
{\bf this paper} & \cmark & \cmark & \cmark  & evaluation of the group privacy management mechanisms  \\
\hline
\end{tabular}
\end{table}

\subsection{Motivations}
%Analisi overhead
%Architettura
%Sicurezza
%Scenario
%Scability
In the last few years, several papers have been devoted to review the available literature on decentralized privacy preserving mechanisms for DOSNs, analysing several aspects of such approaches. In the following, we briefly mention the most important ones, highlight the main differences between them and with respect to the work proposed in this manuscript. 
%\todo[inline,color=yellow]{la frase precedente potrebbe essere piu' chiara. Forse "work" non e' la parola giusta}
%\todo[inline]{possiamo scambiare works con researchers o research}

Table \ref{tab:surveySummary} summarizes in a convenient format the distinguishing features taken into account by the previously surveys on DOSNs, namely: the privacy enforcement models to protect users' contents (\textit{Privacy enforcement model}), the definition of an analytical cost-model describing the overhead introduced by the privacy preserving mechanisms
\textcolor{black}{implementing the enforcement models}
(\textit{Theoretical model}), and the experimental evaluation of the  privacy preserving mechanisms (\textit{Experimental evaluation}). 
%\todo[inline, color=pink]{il dubbio mi rimane: gli approcci in tabella non si riferiscono ai gruppi}
%\todo[inline]{In generale, sono tutti survey relativi alle DOSNs, tra questi, molti studiano aspetti di privacy.}
In particular, the works proposed in \cite{datta2010decentralized,zuo2016survey,koll2017good} were among the first ones to provide an overview of current privacy preserving mechanisms to ensure the independence from the centralized service provider in DOSNs. However, they do not take into account the privacy enforcement models used by DOSNs for contents sharing, but mainly focus on investigating the architecture used to allow independence from the centralized service provider \cite{datta2010decentralized}, the types of social information that have been used to improve the design of current DOSNs \cite{zuo2016survey}, and the major technical aspects that prevent widespread adoption of current DOSNs \cite{koll2017good}.
%proposing a general architecture for DOSNs. Similarly, the work proposed in \cite{zuo2016survey} investigates, from an architectural point of view, the types of social information that have been used to improve the design of current p2p systems and the challenges arising from the use of such information.
%the benefits provided by socially aware . i.e., decentralized systems that integrate social knowledge in their design. In particular, they focused on characterizing \\
%Authors of \cite{koll2017good} further investigate the major technical aspects that prevent widespread adoption of current DOSNs, by focusing mainly on the architecture used to allow independence from the centralized service provider.

Authors of \cite{paul2014survey} study the impact caused by design decisions % that have to be made for defining 
in DOSN architectures. The work is mainly intended to investigate the general architectural models of DOSNs while the privacy enforcement models of the different solutions are only slightly described.\\
The work proposed in \cite{chowdhury2015taxonomy} identifies a set of criteria that are used to construct a taxonomy for the classification of DOSNs. Architecture,  types of service,  social application development,  availability and  scalability are among the most important criteria discussed by the authors. As per the security criteria of DOSNs, they  mainly focused on the authentication, the confidentiality of messages, and the data integrity.\\
Authors of \cite{troncoso2017systematizing} provide literature review of about 165 papers related to DOSNs by focusing on how decentralization is achieved and the advantages/disadvantages of decentralization. They analyse the decentralized designs of current DOSNs and cluster them based on their infrastructure, network topology, authority relations, and privacy properties. In particular, the main privacy property considered in the paper is the confidentiality from both third parties and peers.
The work proposed in \cite{guidi2018managing} investigates how current DOSNs manage contents generated by users, discussing the data format and describing the type of privacy mechanisms used to protect them.\\ 
Recently, authors of \cite{masinde2020peer} provide a comprehensive survey on DOSNs, discussing their technical requirements and the mechanisms used to define the overlay network.  
However, the problem of protecting the privacy of contents in DOSNs has been partially addressed in \cite{masinde2020peer} because they only focused on listing security features provided by DOSNs.\\
A discussion of the privacy and architectural features provided by current DOSNs is presented in \cite{schwittmann2013privacy}, where authors mainly focused on investigating their suitability for mobile devices.\\
%\todo[inline,color=yellow]{frase non chiara. Intendi l'overhead dovuto alle diverse scelte architetturali?}
%\todo[inline]{Aggiornato}
Instead, authors of \cite{bahri2018decentralized} compare and discuss available literature on privacy preserving mechanisms for DOSNs based on three main areas:  data storage and replication,  data access control management, and  fake accounts and fake content management. The work does not cover in more details the characteristics of the privacy enforcement models and, as authors claim, there is still need for more research efforts, especially w.r.t. dynamic group membership management.\\
The work proposed in \cite{taheri2015security} provides fine-grained classification of various state-of-the-art approaches related to data privacy, data integrity, and secure social search solutions for DOSNs.\\
Another relevant work for user's privacy is proposed in \cite{kumaraguru2007survey}, where the authors survey the features of the most important privacy policy languages available in current literature for representing privacy preferences of users.\\
Finally, in \cite{de2018survey} the authors analyse the most popular DOSNs and, for each of them, they describe the specific features and architectural choices, analyse the privacy enforcement models, and formulate an analytical cost model for evaluating the complexity of executing the main operations on groups (adding/removing a member, publishing a content) in terms of number of group members.

\textcolor{black}{Despite most of the popular DOSNs adopt groups-based solutions for preserving contents privacy, Table \ref{tab:surveySummary} reveals that none of the available works provides a complete evaluation of the  main current privacy enforcement models w.r.t. %\st{the different types of groups.}
}
\textcolor{black}{the different privacy requirements characterizing  distinct types of groups defined in DOSNs.} \textcolor{black}{In fact, only the authors of \cite{de2018survey} present an analytical cost model evaluating the complexity of executing the main operations on groups in several privacy enforcement models adopted in DONSs.}
\textcolor{black}{However, they do not provide an experimental assessment of such cost-models which, instead, is actually the best way to find}
\textcolor{black}{which privacy enforcement models are suitable for efficiently implementing the privacy requirements characterizing each group type.}

%\textcolor{blue}{\st{Instead, this paper experimentally evaluates}} \textcolor{red}{\st{how the privacy requirements  defined by the different group types affect the cost of executing the group operations in each privacy enforcement model.}}
%\todo[inline,color=yellow]{commento 1: commento stupido: nella tabella il fatto che ci siano le X per dire che un paper NON tratta un aspetto, non evidenzia graficamente il fatto che il nostro e' l'unico a fare la valutazione sperimentale. Io lascerei vuote le caselle dove ora c'e' la X per evidenziare che il nostro e' l'unico a tre spunte.}
%\todo[inline,color=yellow]{osservazione: qui dicevamo group type per la prima volta, allora ho voluto specificare che i group type dipendono dai privacy requirements (che erano: backward o forward secrecy sulle actions)}
%\todo[inline]{Ok}
%\todo[inline]{Riportare in contributions}
%However, despite the widespread adoption of groups in DOSNs, the previous studies does not reveal how the different group types affect the cost of privacy management mechanisms and which mechanisms are suitable to be used for each group type.
%RDP relativamente a quanto sopra, va detto quali sono gli shortcomings di \cite{de2018survey} rispetto agli obj di questo lavoro --- RISOLTO
Hence, this work is motivated by the fact that the design of DOSNs privacy enforcement models deserves to be further analyzed 
%\st{both} 
in terms of 
%\st{performance and} 
suitability to accommodate the peculiar privacy requirements of groups defined in the DOSNs, 
\textcolor{black}{taking also into account performance aspects.}

\textcolor{black}{
%In the following, we identify the main differences between this paper and the existing ones, highlighting the striking contributions of the former. 
In particular, the main differences between this paper and the existing ones, %which determine the striking contribution of the former, 
are highlighted in the following. 
%This manuscript differs from the previous works in several ways. 
First, we  do not compare DOSNs among themselves. Instead, we first extract a set of features idiosyncratic for each DOSN, and later propose 
%this paper at first defines 
a taxonomy of the user groups based on the cited
\textcolor{black}{privacy requirements}---note that user groups are typically adopted in DOSNs (and in OSNs in general) to define privacy preferences. 
%The proposed taxonomy is leveraged to identify ---in order to classify them in a number of different types [by properly combining the features defined in the taxonomy] and to identify the subset of the ones that are of interest for our study. 
Moreover, based on the above introduced taxonomy, we identify three main classes of privacy enforcement models representing the solutions adopted by most of the currently available DOSNs and, for each of the cited classes, we analyse how the privacy requirements for each of the group types %previously defined 
could  be implemented. 
Finally, we perform an extensive set of experiments based on realistic simulations to evaluate the real applicability of such solutions.}
\textcolor{black}{In particular, we experimentally evaluate how the privacy requirements  defined by the different group types affect the cost of executing the main operations on groups in each privacy enforcement model.}
Open research directions are also highlighted.

\subsection{Contributions}
%\st{We believe that the design of current DOSNs may be further improved in terms of performance by adopting enhanced privacy enforcement models which take into account the}
%We believe that the design of current privacy enforcement models for DOSNs may be further analyzed in terms of performance, by evaluating their suitability to take into account the
%\textcolor{blue}{peculiarity} \st{complexity} of the Online Social Network (OSN) scenario. 
%\todo[inline,color=yellow]{1) da questa frase sembra quasi che in questo paper noi vogliamo proporre nuove soluzioni che migliorano quelle esistenti. 2) qui potremmo ricordare che una delle peculiarita' e' che ci sono i gruppi}
%\todo[inline]{Aggiornato. Sono indeciso tra 'to take into account' o 'to support'.}
%\todo[inline,color=yellow]{pero' questa frase deve essere spostata nelle motivation.}

The main novel contribution brought by this paper is an extensive analysis of the solutions adopted to preserve contents privacy in existing DOSNs following an alternative approach with respect to the existing surveys. This way, we are able to shed light on some of the less known aspects of privacy in DOSNs, as well as to provide a solid experimental support for the performance of the the most common functionalities provided by the different models. Some novel research directions are also discussed.
%taking  into account those solutions that define groups of users in order to grant them access rights on contents.
%and we classify these groups in a number of distinct types according to their privacy related features.
%Then, we classify privacy enforcement models adopted by the existing DOSNs for protecting user contents into three main categories. %based, respectively, on an enhanced cryptography strategy which reduces the encryption overhead, and on a proper  allocation strategy of the contents on the user peers which allows to avoid content encryption. %The first privacy enforcement model used to guarantee the privacy of the contents \cite{buchegger2009peerson,graffi2011lifesocial,nilizadeh2012cachet,koll2017socialgate,aiello2012lotusnet,sharma2012supernova} exploits encryption mechanisms for making contents visible only to the members of the group they are reserved to.
%The second privacy enforcement model is still cryptography based, and it improves the performance of the DOSN platform by adopting a hierarchical data structure for a smart management of encryption keys \cite{de2017logical,wong2000secure,gunther2011key,kwak2007decentralized,de2016logical,sherman2003key}.
%The third privacy enforcement model, instead, avoids content encryption by adopting proper content allocation and replication strategies. In particular, contents remain unencrypted, but they are stored on the devices of the content owner or of trusted (or authorized) users only \cite{narendula2011my3,conti2014trusted,guidi2015didusonet,shakimov2011vis,de2015privacy,de2016privacy,de2017privacy}. 
In detail, we provide %, the proposed approach brings 
the following contributions: 
\begin{enumerate}
\item We study the features of the user groups provided by current DOSNs, 
%\st{We study the features of the privacy management mechanisms based on user groups provided by current DOSNs,} 
and we identify 4 different types of groups, each having different requirements related to content privacy;
\item We classify the privacy enforcement solutions adopted by the existing DOSNs for protecting user contents into three main models. The first privacy enforcement model %used to guarantee the privacy of the contents 
\cite{buchegger2009peerson,graffi2011lifesocial,nilizadeh2012cachet,koll2017socialgate,aiello2012lotusnet,sharma2012supernova} exploits encryption mechanisms for making contents visible only to the members of the group they are reserved to.
The second privacy enforcement model is still cryptography based, and it improves the performance of the DOSN platform by adopting a hierarchical data structure for a smart management of encryption keys \cite{de2017logical,wong2000secure,gunther2011key,kwak2007decentralized,de2016logical,sherman2003key}.
The third privacy enforcement model, instead, 
avoids content encryption by adopting proper content allocation and replication strategies. 
In particular, contents remain unencrypted, but they are stored on the devices of the content owner or of trusted (or authorized) users only \cite{narendula2011my3,conti2014trusted,guidi2015didusonet,shakimov2011vis,de2015privacy,de2016privacy,de2017privacy}.

\item We study the capability of each of the 3 previously listed privacy enforcement models to implement the 4 types of groups available in DOSNs;

\item We execute a set of experiments based on realistic models simulations to give an estimation of the overhead introduced by the 3 privacy enforcement models in performing the typical DOSN operations 
%required to ensure the privacy preferences of the users 
(publication of a content in a group, user joining or leaving a group)
\textcolor{black}{on each of the group types we defined};

\item We compare the \textcolor{black}{costs of the 3 privacy enforcement models obtained from our experiments in order to find which is the most suitable one for implementing each of the group types we defined. We believe that this is the best way of making a realistic comparison among the 3 privacy enforcement models. } 
We highlight the lessons learned from our analysis, providing important information for increasing effectiveness and efficiency of privacy enforcement models;

\item We 
%\st{compare the results of the 3 privacy enforcement models by discussing in more details their implications, providing}
provide interesting future research directions and discuss open challenges.
%\todo[inline,color=yellow]{forse in questo ultimo punto i risultati sperimentali non vanno sottolineati???}
%\todo[inline]{possiamo fare il merge tra 5) e 6)?}
%\todo[inline,color=yellow]{io lascerei le future directions senza dire che derivano dai risultati sperimentali. Quindi ho barrato un pezzo}

\end{enumerate}

\begin{figure*}[tb] 
\centering 
\includegraphics[width=0.90\textwidth]{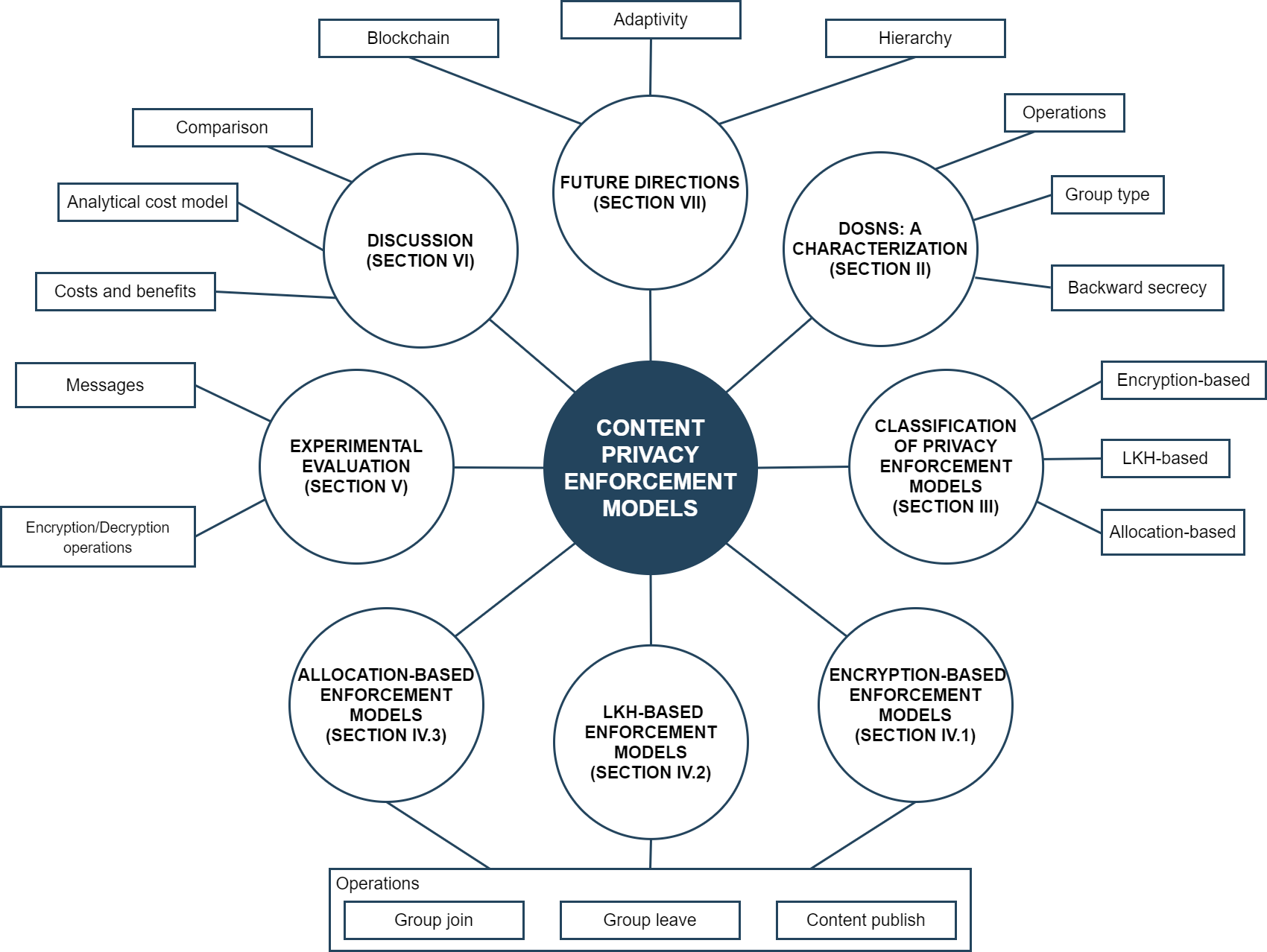}
\caption{Graphical description of the organization of our paper.\label{fig:organization}}
\end{figure*}

It is worth noting that, in order to reduce the analysis to a manageable dimension, while preserving the generality of results, rigor, and formalism, the following approach was selected. Once the relevant policies of interest to manage privacy in DOSNs are identified, we focused on the representative mechanisms instrumental to implement  the related privacy  enforcement models.
As such, we were called to choose among the many privacy enforcement mechanisms available. We decided to go for the most used ones, because of their representativeness and possible impact of our findings. Note that the choice, while reducing the domain to a manageable dimension, does not hamper either the  generality or applicability of the achieved results, given the fact that all of the selected mechanisms fully represent  the expressiveness  of the underlying policy.

%In particular, we investigate the type of privacy policies provide by DOSNs to their users and we give an estimation of the overhead introduced by the related enforcement model in order to ensure the privacy preferences of the users. The evaluation of the enforcement models is achieved by developing a set of P2P simulations which aim to measure the overhead introduced by security mechanisms of each model, for different operations. Finally, we compare the enforcement models by discussing in more details their implications and benefits.

\subsection{Outline}
The contributions previously defined, as well as the general structure of the manuscript, are graphically represented in Figure \ref{fig:organization}. In particular, the remainder of this paper is organized as follows.
In Section \ref{sec:scenario} we characterize the properties of current DOSNs by providing a taxonomy for groups (based on operations, group type, and backward secrecy property), that will be used to structure the evaluation of the privacy enforcement models. In Section \ref{sec:relatedWork} we introduce the reader to the typical privacy solution adopted by DOSNs (Encryption-based, LKH-based, and Allocation-based). In Section \ref{sec:privacyControl} we describe how the selected DOSNs' solutions \textcolor{black}{implement the privacy policies enforcement during the execution of the  typical operations (group join, group leave, and content publish) and how they implement the modification of such policies}.
%\todo[inline,color=yellow]{HO modificato perche' nelle sottosezioni della sezione 4 vengono descritte le implicazioni del'approccio adottato nelle operazioni di publish, join e leave, e dalla frase precedetente secondo me non si capiva.}
In Section \ref{sec:evaluation} we illustrate the evaluation of the selected approaches \textcolor{black}{taking into account the messages
%\todo[inline,color=yellow]{siamo sicuri che sia number of messages? e non size? Forse qui all'inizio del paper è meglio essere vaghi e scrivere semplicemente "messages" qui e nella figura 1??}
%the number of keys,
%\todo[inline,color=yellow]{sarebbe keys encrypted e decrypted? a questo livello lo metterei assieme alla voce successiva "number of enc/dec operations"}
and the encryption/decryption operations required to execute the previously listed DOSN typical operations},
%\todo[inline,color=yellow]{manca il tempo per effettuare le operazioni}
%\todo[inline,color=yellow]{se si modifica questo testo, va modificato di conseguenza la figura 1}
%\todo[inline]{modificato. }
while in Section \ref{sec:discussion} we compare the performance of the different approaches and discuss their advantages and limitations. In Section \ref{sec:futureDirection} we draw future research directions, i.e., based on the recent advent of blockchain technology, the adaptivity, and the hierarchy property. Finally, in Section \ref{sec:conclusion} we report some conclusions. %and discuss future works.

\section{DOSN{\scriptsize s}: A  Characterization} % The current DOSNs scenario
\label{sec:scenario}
%\todo[inline]{L'intento di questa sezione e' quello di identificare le caratteristiche dei gruppi utilizzati dagli attuali servizi di OSNs e valutare come vengono supportati nelle DOSNs}
%\todo[inline,color=yellow]{L'ho riletta diverse volte, e penso che parlare delle OSN non sia la soluzione migliore, visto anche che la sezione si intitola DOSN caracterization. Quindi lascerei solo le cose relative alle DOSN}
In order to allow their users to protect the contents they publish, current DOSNs enable them to
%\st{define privacy control on such contents in order to} 
specify their privacy preferences to determine who can access each of these contents. 
%As for instance, the DOSN services offer a set of default privacy options that allow user to publish a content as private (i.e., intended for a specific user) or public (i.e., visible to all contacts).\todo{QUI SEMBRA CHE TUTTE LE DOSN ADOOTINO L'APPROCCIO PRIVATE O PUBLIC, LA RISCRIVEREI COSI':}  
%\st{As for instance, some DOSNs adopt a basic privacy preferences model which allows their users to publish a content as private (i.e., intended for a specific user) or public (i.e., visible to all the contacts of the publisher). Besides this basic setting,}
%\todo[inline,color=yellow]{la frase prima non serve ai nostri scopi}
The typical privacy controls provided by the most part of existing DOSNs are based on the group communication model, where a user shares his/her contents with a previously defined group of users. Despite its simplicity, this privacy model involves communication between a user, the content producer, and a possibly large set of contacts, i.e., the members of the group. 
In addition, each DOSN
%\todo[inline,color=yellow]{qui c'e' scritto OSN invece che DOSN}
service implements its own type of groups, by providing variants of this basic setting.
In this section, we investigate the characteristics of the groups provided in existing DOSNs and, based on the identified characteristics, we defined a taxonomy for groups that will be used to structure
the evaluation of the privacy enforcement models in DOSNs.

\begin{figure*}[t] 
\centering 
\includegraphics[width=0.90\textwidth]{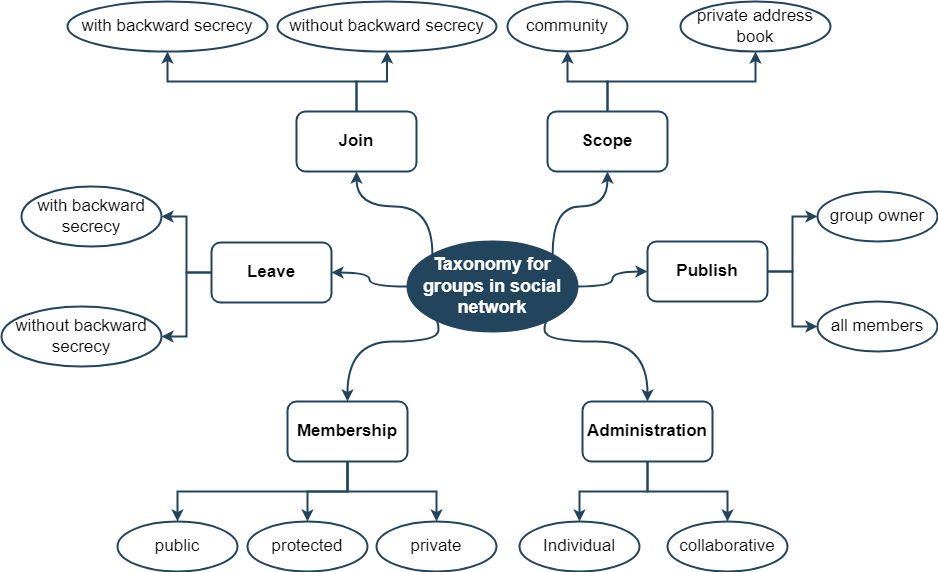}

\caption{Taxonomy for groups in Social Networks.\label{fig:groupTaxonomy}}
\end{figure*}
Since we are interested in analyzing privacy enforcement models which restrict access to contents, we do not consider public groups where anyone can see all contents without restrictions. 
Indeed, groups exposing their contents publicly to all users do not need to regulate access to such contents. 
\textcolor{black}{Instead, for non public groups,} the main common characteristics we identified are: the scope, the visibility of the membership information, the number of administrators of the groups, the enforcement of the backward secrecy property for the join and leave operations, and which users are allowed to publish contents on the group (group owner or all members). %\st{the type of operation used to join, to leave or to publish a content on the group.} %\todo[inline,color=yellow]{"type of operation" non si capisce, melgio andare in dettaglio e dire backward secrecy for join e leave insieme e publish da solo}
Figure \ref{fig:groupTaxonomy} graphically represents these characteristics as well as the options within each of them.

As concern the scope, DOSNs enable users to create groups in order to either organize their own contacts in a private address book 
%\todo[inline,color=yellow]{nella figura 2 quasi quasi preferirei mettere "organize contacts" e "communities" piuttosto che "relationships" e "interests", perche' non e' detto che quei gruppi effettivamente riflettano le relazioni di chi li crea.  }
%\todo[inline]{Ok cambiare il nome ma la definizione secondo me non dovrebbe cambiare. Se un gruppo non riflette le relazioni di chi lo crea e' un gruppo basato con scope 'communities' per definizione.}
%\todo[inline, color=pink]{Sono d'accordo col cambiare il nome}
(such as, Safebook \cite{cutillo2009safebook} or Vis-a-Vis \cite{shakimov2011vis})
%{\it lists} in Facebook
%\todo[inline,color=yellow]{all'inzio di questa frase abbiamo detto esplicitamente  DOSN, quindi prendere Facebook come esempio non mi pare corretto}
%and Flickr %\todo[inline,color=yellow]{facebook ha anche quel tipo di gruppi?}
%\todo[inline]{Si, hai la possibilita' di assegnare una label (es, friends (di default), follower, family,...) alle relazione che puoi utilizzare per la condivisione dei contenuti}
%, or {\it aspects} in Diaspora) 
or to create a community for connecting users among them (such as, groups in LifeSocial.KOM \cite{graffi2011lifesocial}, ProofBook \cite{biedermann2014proofbook}, or LibreSocial \cite{graffi2020libresocial}). 
%\todo[inline]{La definizione che ho usato e' la seguente: Nel caso di tipi di gruppi basati su interest non e' necessario che i membri abbiano una certa relazione pre-esistente con il proprietario del gruppo (es, i Gruppi di facebook, linkedIn, e le liste in Twitter. Invece, i gruppi basati su relazione richiedono l'esistenza di una relazione tra proprietario e membri del gruppo (es. le liste di Flickr o di Facebook. Ti torna? Possiamo comunque cambiare il nome da tipo a scopo del gruppo se sei d'accordo. Dubbio: tra questi ho considerato anche WhatsApp}
%\todo[inline,color=yellow]{Poi, qui direi anche che ne primo caso il creatore sceglie i membri a loro insaputa, senza loro consenso, e senza che loro lo sappiano mai, mentre nel secondo caso i membri vengono  invitati dal creatore e devono dare consenso oppure si propongono essi stessi al creatore, ( in alcuni casi' il creatre li deve accettare, in altri l'accettazione e' automatica). Vi torna? Sbaglio? Infine, questa divisione per tipo mi fare un po forzata}
%because they are used to define different groups of contacts 
\textcolor{black}{In the first case, groups are privately defined by the group creator based on the relationship he/she has with the group members
(such as, family members, classmates, or colleagues), and the members are not aware of the groups they belong to, } 
\textcolor{black}{they cannot see the other members of the group, and they cannot decide to join/leave them or to publish a content in the group, because these groups are for the exclusive use of the group creator.  
Hence, each DOSN user sees and uses only the set of groups he/she created, which is different from the set of groups defined by the other users.}

\textcolor{black}{Instead, community groups are} 
\textcolor{black}{meant to build communities where users interact by }
\textcolor{black}{sharing contents with other users, typically having the same preferences (such as, the supporters of a football team or a political party). }
\textcolor{black}{In this case, DOSN users are able to see the groups created by the other users, and they explicitly ask (or they are asked by) the administrators to join such groups, thus being always aware of belonging to a community group.}\\
At the time of creation, the group owner indicates whether the group is administrated also by other members. 
%\todo[inline,color=yellow]{credo che questo valga solo per i gruppi del secondo tipo}
%\todo[inline]{Esatto. Tuttavia abbiamo un caso a parte, avendo considerato anche WhatsApp. Alcuni utenti del gruppo whatsapp che hai creato a partire dai tuoi contatti in rubrica possono essere designati come amministratori del gruppo}
If the group is configured to be managed collaboratively by multiple users, then the group owner must specify the identity of the other administrators. 

The group owner can also select the initial members of the group and must configure the visibility of the membership information. 
In particular, the group owner can decide to reveal the identities of the group members to everyone (i.e., public), to group members only (i.e., protected) or to no one else than the group owner (i.e., private). 
%\todo[color=pink, inline]{anyone else in che senso?}
%\todo[inline]{nessun altro. modificato}
It should be noted that in the case of private membership information, the members of a group are not aware of the group itself, they simply have access to the content shared in that group, \textcolor{black}{seeing them as contents published by the group owner}.
We identify three operations   involving groups typical of DOSNs: publication of contents in a group, join of a new user to a group, and leave of a member from a group.

\paragraph{Publication of a content} the publication of  contents on a group determines the privacy level of such contents. Indeed, by publishing a  content on a group, the content producer gives access rights to all (and only) the members currently belonging to these groups. Depending on the configuration of the group, it can accept for publication contents created by the group owner or by any of the group members. 
%\st{In such a way, the publisher of a content can chose where to publish the content: on a group she/he has created or on a group to which she/he belongs to.}
%\todo[inline,color=yellow]{per essere pignoli,  chi da i diritti dovrebbe essere chi pubblica il contenuto, e li da scegliendo il gruppo daassociare a quel contenuto. In alcuni casi il gruppo lo ha definito lui, in altri casi pubblica in gruppi a cui appartiene.}
%\todo[inline,color=yellow]{qui bisogna fare attenzione a come presentiamo questa cosa dei gruppi, perche' se fosse un gruppo pubblico di facebook, in realta' il contenuto lo potrebbero vedere tutti. Solo i gruppi privati restringono gli accessi ai contenuti in facebook}
\paragraph{Group Join} Groups defined in DOSNs are dynamic because new members can be added to the group. The join operation involves at least the group owner and the users who are being added to the group. It can be either initiated by the group owner or requested by a user who does not yet belong to the group.
If the join operation is successfully executed, a new user is added to a group in order to allow her/him to access the new contents that will be posted to this group. Consequently, every time a new user is added to a group, the membership information needs to be updated, in order to ensure that the new member can access such contents.
Instead, the contents published in the group prior to the join of the new user can be made either accessible (join without backward secrecy) or not accessible (join with backward secrecy) to the joining member. 

\begin{table*}[!tb]
\footnotesize
\centering
\caption{Mapping of the types of groups available in DOSNs\label{tab:mappingDOSNs}}
\begin{tabular}{lcclcllllll}
\hline
\multicolumn{1}{c}{\multirow{2}{*}{\rotatebox[origin=c]{45}{\textbf{OSN }}}}  & \multicolumn{1}{c}{\multirow{2}{*}{\begin{tabular}[c]{@{}l@{}} \rotatebox[origin=c]{0}{\textbf{Join}} \\ Back-\\ward \\  Secrecy\end{tabular}}} & \multicolumn{1}{c}{\multirow{2}{*}{\begin{tabular}[c]{@{}l@{}} \rotatebox[origin=c]{0}{\textbf{Leave}}\\ Back-\\ward \\ Secrecy\end{tabular}}}& \multicolumn{1}{c}{\multirow{2}{*}{\rotatebox[origin=c]{0}{\textbf{Group Scope}}}} & \multicolumn{2}{c}{\textbf{\textbf{Publish}}} & \multicolumn{2}{c}{\textbf{\textbf{Admin}}} & \multicolumn{3}{c}{\textbf{\textbf{Membership}}} \\ \cline{5-11} 
\multicolumn{1}{c}{} & \multicolumn{1}{c}{} & \multicolumn{1}{c}{} & \multicolumn{1}{c}{} & \multicolumn{1}{l}{\rotatebox[origin=c]{90}{ all members }} & \rotatebox[origin=c]{90}{ group owner} & \rotatebox[origin=c]{90}{individual} & \rotatebox[origin=c]{90}{  collaborative } & \rotatebox[origin=c]{90}{public} & \rotatebox[origin=c]{90}{protected} & \rotatebox[origin=c]{90}{private} \\ \hline
\multirow{2}{*}{Safebook \cite{cutillo2009safebook}}  & \xmark & \xmark & community & \cmark & \xmark & \cmark & \xmark & \cmark & \xmark & \xmark \\ & \xmark & \xmark & priv addr book & \xmark & \cmark & \cmark & \xmark & \xmark & \xmark & \cmark \\
PeerSoN \cite{buchegger2009peerson}& \xmark & \cmark & community & \cmark & \xmark & \cmark & \xmark & \cmark & \xmark & \xmark \\
LibreSocial \cite{graffi2020libresocial} & \xmark & \cmark &community & \cmark & \xmark & \cmark & \xmark & \cmark & \xmark & \xmark \\
\multirow{2}{*}{Vis-a-Vis \cite{shakimov2011vis}} & \xmark & \xmark & community & \cmark & \xmark & \cmark & \xmark & \cmark & \xmark & \xmark \\
& \xmark & \xmark & priv addr book & \xmark & \cmark & \cmark & \xmark & \xmark & \xmark & \cmark \\ 
ProofBook \cite{biedermann2014proofbook}&  \xmark & \xmark & community & \cmark & \xmark & \cmark & \xmark & \xmark & \cmark & \xmark \\
Contrail \cite{stuedi2014contrail} & \cmark & \xmark & community & \xmark & \cmark & \cmark & \xmark & \xmark & \xmark & \xmark \\
\hline
\end{tabular}
\end{table*}
\paragraph{Group Leave} An existing member of a group can be removed at any time by the group owner in order to deny her/him the access to the new contents that will be posted to the group (forward secrecy), or can voluntarily decide to leave the group. 
%Consequently, when an existing member is removed from a group,
As a consequence of the group leave, 
the group membership information needs to be updated in order to ensure that the removed member 
%of the group 
cannot access the new contents that will be posted on this group after the execution of the remove operation.
Instead, the contents published in the group before the execution of the remove operation can be either still accessible (leave without backward secrecy) or not (leave with backward secrecy) to the removed member through the DOSN infrastructure. 

Unlike the join operation, which requires asynchronous interaction between the group owner and the joining user, the leave operation is initiated by either the group owner or the leaving user, and it typically does not need any other action from the leaving user.
%\todo[inline,color=yellow]{chi sarebbe the other party? Il group owner deve sempre intervenire per cambiare le chiavi.}
%\todo[inline]{updated}

%In the case of the join and the leave operation, the backward secrecy property specifies whether the joining/removed user can or cannot access the previous contents published on the group.
%As for instance, Facebook allows its users to create public or close\todo[inline,color=yellow]{ora i close si chiamano private in facebook. Ho cercato nella documentazione di facebook la pagina dove dice chi puo' fare le join per mettere il link qui ma non l'ho trovata}
%group: the former can be joined (or left) by any user of the OSN while the latter requires the invitation from the group owner (or from an internal member).

Table \ref{tab:mappingDOSNs} summarizes the features of the group types supported and implemented by a subset of the most popular existing DOSNs,  where the taxonomy is used to classify each group.
We can observe that existing DOSNs support a very heterogeneous set of groups having different characteristics.
%\todo[inline,color=yellow]{se si cambia relationships/interest nella figura si deve cambiare pure qui}
For instance, the majority of the DOSNs allow their users to create groups based on community that do not support the backward secrecy on the join operation, i.e., they allow a joining user to access all the contents previously published in the group. 
Instead, groups aimed to organize contacts are provided by few DOSNs (i.e., Safebook \cite{cutillo2009safebook} and Vis-a-Vis \cite{shakimov2011vis}) and they have private membership information because they are used by owners to regulate the access to the contents they share. For what concerns the group leave operation, most of the existing DOSNs allow the users removed from a group to still access the contents published prior to the execution of the leave operation (leave without backward secrecy). In a few number of DOSNs (such as PeerSoN \cite{buchegger2009peerson} and LibreSocial \cite{graffi2020libresocial}), instead, the group leave operation revokes the access rights to these contents to the removed members (leave with backward secrecy). In this case,  although after the execution of the leave operation the DOSN prevents the removed members from accessing such contents anymore, this cannot be really considered as an effective countermeasure for protecting the privacy of such contents. As a matter of fact, the removed users could have created a local copy of such contents on their local nodes when they had the right to do that, and such copies will be always available to them. Hence, we could state that a group leave fully guaranteeing backward secrecy is not possible.

\begin{table*}[!tb]
\footnotesize
\centering
\caption{Mapping of the types of groups available in centralized OSNs\label{tab:mapping}}
\begin{tabular}{lcclcllllll}
\hline
\multicolumn{1}{c}{\multirow{2}{*}{\rotatebox[origin=c]{43}{\textbf{OSN }}}} & \multicolumn{1}{c}{\multirow{2}{*}{\begin{tabular}[c]{@{}l@{}} \rotatebox[origin=c]{0}{\textbf{Join}} \\ Back-\\ward \\ Secrecy\end{tabular}}} & \multicolumn{1}{c}{\multirow{2}{*}{\begin{tabular}[c]{@{}l@{}} \rotatebox[origin=c]{0}{\textbf{Leave}}\\ Back-\\ward \\ Secrecy\end{tabular}}} & \multicolumn{1}{c}{\multirow{2}{*}{\rotatebox[origin=c]{0}{\textbf{Group Scope}}}} & \multicolumn{2}{c}{\textbf{\textbf{Publish}}} & \multicolumn{2}{c}{\textbf{\textbf{Admin}}} & \multicolumn{3}{c}{\textbf{\textbf{Membership}}} \\ \cline{5-11} 
\multicolumn{1}{c}{} & \multicolumn{1}{c}{} & \multicolumn{1}{c}{} & \multicolumn{1}{c}{} & \multicolumn{1}{l}{\rotatebox[origin=c]{90}{ all members }} & \rotatebox[origin=c]{90}{ group owner } & \rotatebox[origin=c]{90}{ individual } & \rotatebox[origin=c]{90}{ collaborative } & \rotatebox[origin=c]{90}{public} & \rotatebox[origin=c]{90}{protected} & \rotatebox[origin=c]{90}{private} \\ \hline
Anobii & \xmark & \cmark & community & \cmark & \xmark & \cmark & \cmark & \xmark & \cmark & \xmark \\
Chess & \xmark & \xmark & community & \cmark & \xmark &\cmark & \cmark & \cmark & \xmark &\xmark\\
LinkedIn & \xmark & \cmark & community & \cmark & \xmark & \cmark & \cmark & \xmark & \cmark& \xmark  \\
Goodreads & \xmark & \cmark & community & \cmark & \cmark & \cmark & \cmark & \xmark & \cmark  & \xmark \\
Fotki & \xmark & \cmark & priv addr book & \xmark & \cmark & \cmark & \xmark & \xmark& \xmark & \cmark \\
\multirow{2}{*}{Flickr }& \xmark & \cmark & priv addr book & \xmark & \cmark & \cmark & \xmark & \xmark & \xmark & \cmark \\
 & \xmark & \cmark & community & \cmark & \cmark & \cmark & \cmark & \cmark & \xmark & \xmark \\
Fitocracy & \xmark & \cmark & community & \cmark & \cmark & \cmark & \xmark & \xmark & \cmark & \xmark \\
\multirow{2}{*}{Facebook} & \xmark & \cmark & community & \cmark & \xmark & \cmark & \cmark & \xmark& \xmark & \cmark \\
 &\xmark & \cmark & priv addr book & \xmark & \cmark & \cmark & \xmark & \xmark & \xmark & \cmark \\
Ning & \xmark & \cmark & community & \cmark & \cmark & \cmark & \xmark & \xmark & \cmark & \xmark \\
Printerest & \xmark & \cmark & community & \xmark & \cmark & \cmark & \xmark & \xmark & \cmark & \xmark \\
Twitter & \xmark & \cmark & community & \xmark & \cmark & \cmark & \xmark & \xmark & \xmark & \cmark \\ \hline
%WhatsApp & messaging & with backward secrecy & with backward secrecy & relationship & \cmark& \xmark & \cmark & \cmark & \xmark & \cmark& \xmark \\ 
\end{tabular}
\end{table*}

\textcolor{black}{Furthermore, although this paper is focused on DOSNs, we noted that the taxonomy for groups we defined actually have general characteristics, which are common to groups implemented by existing centralized OSNs as well. 
%\st{Indeed, the groups provided by DOSNs can be also supported by centralized OSNs, but the way in which they are implemented are different, depending on the mechanisms provided by the different architectures.}
%\todo[inline,color=yellow]{la frase precedente non e' per niente chiara, e io la leverei perche' ancora qui non si parla di implementazione ma solo delle caratterstiche}
%\todo[inline]{ok}
%\textcolor{red}{Although we focus on DOSNs, we noted that  groups have general characteristics, which are common to both centralized OSNs and DOSNs, }
For instance, Table \ref{tab:mapping} summarizes the group types commonly supported and implemented by a subset of the most popular centralized OSNs. 
We observe that the type of group supported by the most part of current OSNs, such as Facebook, Flickr, Twitter, and LinkedIn allows a joining user to access all the contents previously published in the group, while former members cannot access anymore the contents published in a group if they leave (or are removed from) it.
%Instead, groups supported by Chess are different because a joined user can access the previous contents shared in the group before his arrival while the removed user will still be able to access old contents in the group.
}

\textcolor{black}{Based on the combination of the previous common characteristics shown in Figure \ref{fig:groupTaxonomy}, 4 different types of groups can be defined} %Although all these characteristics concur to define different group types, 
and, in the following, we focus only on the most common ones. Table \ref{tab:groupType} summarizes the types of groups considered in our analysis, by specifying the type of operations (join, leave, and publish), the contents visible to the joining/removed users, the type of administration and membership.
In particular, let $t$ the time when a join or leave operation terminates successfully, we define the set of contents published on the group before time $t$ as old contents. Instead, the contents published on the group after time $t$ are referred as new contents. 
\textcolor{black}{Groups of type G1 ensure that both the removed and the joined users cannot be able to access old contents published in the group. Since, this type of groups does not found use in current DOSNs (and OSNs), it is not considered in our analysis.}
Groups of type G2 ensure that the new member cannot access old contents published in the group. Instead, the leave operation on groups of type G2 allows the removed user to still access the contents previously published in the group. 
Groups of type G3 ensure that a member removed from the group cannot access old contents published in the group. Instead, the join operation on groups of type G3 allows the new user to access both future and old contents published in the group. 
Finally, groups of type G4 allow a new member to access the contents already published on the group before she/he joined, and a removed member can still access the contents published in the group before the removal operation.
\textcolor{black}{As concern the publication of the contents, the group administration, and the membership of the group, we
consider the case where contents can be published by all members of the groups, only the group owner is responsible for the administration of the group, and the memberships information are public. Indeed, as shown in Table \ref{tab:mappingDOSNs} and \ref{tab:mapping}, such configuration have been found wide spread usage in most DOSNs (more than 50\%), and OSNs in general.}

\begin{table*}[tb]
\centering
\footnotesize
\caption{Characteristics of the types of groups considered in our analysis\label{tab:groupType}}
\begin{tabular}{lcccccc|c|c|c}
\hline
\multicolumn{1}{c}{\multirow{2}{*}{\textbf{\begin{tabular}[c]{@{}c@{}}Group\\ Type\end{tabular}}}} & \multicolumn{1}{c}{\multirow{2}{*}{\textbf{\begin{tabular}[c]{@{}c@{}} Join\\ Backward \\ Secrecy\end{tabular}}}} & \multicolumn{2}{c}{\textbf{\begin{tabular}[c]{@{}c@{}}Contents \\visible to\\ the joining \\ user\end{tabular}}} & \multicolumn{1}{c}{\multirow{2}{*}{\textbf{\begin{tabular}[c]{@{}c@{}} Leave\\ Backward \\ Secrecy\end{tabular}}}} & \multicolumn{2}{c|}{\textbf{\begin{tabular}[c]{@{}c@{}}Contents \\visible to\\ the removed \\ user\end{tabular}}} & \multirow{2}{*}{\rotatebox[origin=l]{90}{\textbf{Publis.}}} & \multirow{2}{*}{\textbf{\rotatebox[origin=l]{90}{Admin}}} & \multirow{2}{*}{\textbf{\rotatebox[origin=l]{90}{Memb.}}} \\ \cline{3-4} \cline{6-7} 
\multicolumn{1}{c}{} & \multicolumn{1}{c}{} & \multicolumn{1}{c}{\textbf{Old}} & \multicolumn{1}{c}{\textbf{New}} & \multicolumn{1}{c}{} & \multicolumn{1}{c}{\textbf{Old}} & \multicolumn{1}{c|}{\textbf{New}} & &  &  \\ 
&&&&&&&& \\ \hline
G1 & \cmark & \xmark & \cmark & \cmark & \xmark & \xmark & \multirow{4}{*}{\rotatebox[origin=c]{90}{members}} & \multirow{4}{*}{\rotatebox[origin=c]{90}{individual}} & \multirow{4}{*}{\rotatebox[origin=c]{90}{public}}\\ \cline{1-7}
G2 & \cmark & \xmark & \cmark & \xmark & \cmark & \xmark & & &\\ \cline{1-7}
G3 & \xmark & \cmark & \cmark & \cmark & \xmark & \xmark & & &\\ \cline{1-7}
G4 & \xmark & \cmark & \cmark & \xmark & \cmark & \xmark & & &\\ 
    \hline
\end{tabular}
\end{table*}

\begin{table}[tb]
\small
\centering
\caption{Summary of the enforcement models proposed in the current literature}
\label{tab:enforcModelRelatedWork}
\begin{tabular}{lll}
\hline
\textbf{\aclBased~} & \textbf{\lkhBased~} & \textbf{\policyBased~} \\ \hline
PeerSoN \cite{buchegger2009peerson} & LKH+TGDH\cite{kwak2007decentralized} & Vis-a-Vis \cite{shakimov2011vis}\\
LifeSocial.KOM \cite{graffi2011lifesocial} & DOSN LKH \cite{de2017logical,de2016logical} & My3 \cite{narendula2011my3}\\
DECENT \cite{jahid2012decent} & LKH+OFT \cite{gunther2011key}& DiDuSoNet \cite{guidi2015didusonet} \\
SocialGate \cite{koll2017socialgate} & Shi R.H. et al. \cite{shi2016novel} & Privacy Policy \cite{de2016privacy,de2017privacy} \\
LotusNet \cite{aiello2012lotusnet} & ELK \cite{penrig2000elk} & Trust \cite{conti2014trusted} \\
Contrail \cite{stuedi2014contrail}& OFT \cite{sherman2003key} & Zeng S. et al. \cite{zeng2019decentralized}  \\
ProofBook \cite{biedermann2014proofbook} & LKH \cite{wong2000secure} & eXO\cite{loupasakis2011exo} \\
Safebook \cite{cutillo2009safebook}& DGKD \cite{adusumilli2005dgkd} & Solid \cite{sambra2016solid} \\
SuperNova \cite{sharma2012supernova} &  & Bortoli S. et al. \cite{bortoli2011decentralised}\\
LibreSocial \cite{graffi2020libresocial} & & \\
BCOSN \cite{jiang2019bcosn} & & \\
WebP2P \cite{disterhoft2015protected} & & \\
Megaphone \cite{perfitt2010megaphone} & & \\
SEDOSN \cite{fang2015sedosn} & & \\
ReClaim \cite{zeilemaker2014reclaim} & & \\
PSON \cite{klukovich2016posn} & & \\
\hline
\end{tabular}
\end{table}
\section{Classification of the content privacy enforcement models}
\label{sec:relatedWork}
Although several DOSNs exist in current literature, the most part of the solutions they adopt to protect content privacy can be classified into three distinct types of privacy enforcement models: \aclBased, \policyBased, and \lkhBased.
%\todo[inline,color=yellow]{da questa frase sembra che esistano tante soluzioni e noi ci concentriamo su un numero ristretto. Potremmo almeno dire che la maggior parte delle soluzioni esistenti ricade nelle 3 categorie che noi definiamo?}
%\todo[inline]{Aggiornato}
\textcolor{black}{Without loss of generality, we can assume that the selected privacy enforcement models allow users to protect the contents they publish by defining %privacy preferences that specifies 
the group of users who 
%\todo[inline,color=yellow]{forse qui dovremmo gia' dire che le preferenze di privacy si basano sui gruppi, anche nelle sottosezioni dopo si parla di quelli}
%\todo[inline]{Qui l'assunzione e' piu' generale. Altrimenti diamo per scontato che i modelli supportano i gruppi.}
%\todo[inline,color=yellow]{secondo me e' una ipotesi che dobbiamo fare gia' qui, perche' nel seguito del paper parliamo solo dei modelli che supportano gruppi. Inoltre nella prossima sezione (2.1) diamo gia' per scontato che si utilizzino i gruppi per la privacy.}
can access these contents.} 
\textcolor{black}{As a matter of fact, Section \ref{sec:scenario} shows that a very common solution to express privacy preferences in DOSNs (and OSNs) is the definition of groups of users. An alternative solution to specify the users who can access a content is the one that expresses privacy preferences through attribute based access control policies.}
%\todo[inline,color=yellow]{Vi piace la frase in viola che cerca di prepare il campo alle policy abac?}
Each privacy enforcement model adopts a specific technique to enforce such privacy preferences.
Table \ref{tab:enforcModelRelatedWork} summarizes the privacy enforcement models adopted by a number of existing DOSN platforms and by the approaches proposed in the scientific literature, and in the following, we describe in more detail such models.

\subsection{\aclBased~\ approaches}
The typical solution for the enforcement of content privacy adopted by a large number of DOSN is to have each group $G$ paired with its symmetric encryption key $K_G$, which is created by the group owner and  distributed to the members of $G$ \textcolor{black}{when they join the group} \cite{harney1997rfc2093}. 
%\st{The symmetric group key is distributed to the members of $G$ by encrypting it with the public key of each of such members.}
%\todo[inline,color=yellow]{le tecniche per distribuire la chiave vengono dette piu' in dettaglio alla fine del paragrafo}
In this way, each member of G is able to publish a content addressed to the other members of $G$ by encrypting it with the symmetric group key $K_G$ before being stored on the peers of the DOSN. 
Consequently, the members of group $G$ use $K_G$ for accessing the contents published in the group. 
%\todo[inline,color=yellow]{nel paragrafo successivo c'e' scritto che invece il contenuto non si cripta direttamente con la chiave di gruppo, ma si cripta con una chiave di contenuto che, a sua volta, viene criptata con la chiave di gruppo}
%\todo[inline,color=yellow]{siamo sicuri che sia solo il group owner a pubblicare content nel gruppo? Oppure tutti i membri possono pubblicare? Dipende dal modello di gruppo. Pensiamoci bene.}
%\todo[inline]{Aggiornato. Mi sono mantenuto sulle generali: chiunque possieda la chiave può pubblicare o leggere}
For instance, PSON \cite{klukovich2016posn}, Safebook \cite{cutillo2009safebook}, LotusNet \cite{aiello2012lotusnet}, Contrail \cite{stuedi2014contrail}, ProofBook \cite{biedermann2014proofbook}, SuperNova \cite{sharma2012supernova}, and Megaphone \cite{perfitt2010megaphone}
exploit the symmetric group key to encrypt all the contents shared in a group, i.e., $C_0,C_1,.....C_n$. 

Instead, a similar approach is exploited by PeerSoN \cite{buchegger2009peerson,bodriagov2013encryption}, LifeSocial.KOM \cite{graffi2011lifesocial}, Cachet \cite{nilizadeh2012cachet}, DECENT \cite{jahid2012decent}, SocialGate \cite{koll2017socialgate}, LotusNet \cite{aiello2012lotusnet}, SuperNova \cite{sharma2012supernova}, LibreSocial \cite{graffi2020libresocial}, and BCOSN \cite{jiang2019bcosn}, which create a new specific symmetric key, $K_{C_i}$ for each content to be protected, i.e., $C_i$ with $i\in \{0,1,...,n\}$. 
In this case, each content is encrypted with the corresponding symmetric content key $K_{C_i}$ which is, in turn, encrypted with the symmetric group key $K_G$.
%\todo[inline,color=yellow]{la chiave specifica del contenuto viene poi criptata con quella del gruppo. MA a che serve? Forse per mettere lo stesso contenuto in piu' gruppi senza criptarlo 2 volte?}
%\todo[inline]{Le motivazioni possono essere diverse, alcune non molto chiare: una e' sicuramente quella che hai detto, in altri casi viene introdotta per dare più controllo sul contenuto all'autore.}
%\todo[inline,color=yellow]{prima mettiamo le DOSN che fanno come abbiamo spiegato all'inizio della sottosezione, una chiave simmetrica per ogni gruppo, poi mettiamo come caso particolare le DOSN che  creano una chiave simmetrica per ogni content}
%\todo[inline]{Aggiornato}

The most part of current DOSNs exploit asymmetric encryption (such as, PeerSoN \cite{buchegger2009peerson}, Safebook \cite{cutillo2009safebook}, LifeSocial.KOM \cite{graffi2011lifesocial}, LibreSocial \cite{graffi2020libresocial}) or Attribute Based Encryption (ABE) (such as, Cachet \cite{nilizadeh2012cachet}, DECENT \cite{jahid2012decent}, SocialGate \cite{koll2017socialgate}, SEDOSN \cite{fang2015sedosn}, and  BCOSN \cite{jiang2019bcosn}) to securely distribute the symmetric group key $K_G$ to the members of group $G$.\\
An encryption-based approach 
%\todo[inline,color=yellow]{io non direi similar, anche se e' sempre cryptography based}
%\todo[inline]{Aggiornato}
is also used in \cite{disterhoft2015protected,zeilemaker2014reclaim}, where asymmetric cryptography is now used to protect the confidentiality of contents.
%\todo[inline,color=yellow]{nel senso che anche i contenuti sono criptati con crittografia asimmetrica?  Lo specificherei esplicitamente}
%\st{by using the destination's public key.}
%Every time a new user is added to the group, the symmetric group key needs to be changed in order to ensure that the new member joining the group cannot access the contents previously posted in the same group (i.e., enforcing backward secrecy). Similarly, every time an existing member is removed from a group, the corresponding symmetric group key must be changed as well, in order to prevent the leaving member from accessing the new contents that will be posted to the same group (i.e., enforcing forward secrecy).  In the case of addition of a user to a group, the new symmetric group key $K_G'$ can be securely distributed to old members by encrypting it with the previous group key $K_G$ while for the joining user the new key $K_G'$ must be encrypted with his public key. As a result, the group owner create one message for the old members and one for the joining user. Instead, in the case of removal of a user from a group, the old symmetric group key $K_G$ is known by the removed user and it cannot be used to encrypt the new symmetric group key $K_G'$. As a result, the group owner creates a message which contains the key $K_G'$ encrypted with all the public keys of the users currently belonging to the group. 

\subsection{\lkhBased~}
The \lkhBased~ enforcement model enhances the encryption based one exploiting a
hierarchical structure for managing symmetric group symmetric keys, such as  Logical Key Hierarchical Tree \cite{wong2000secure} (LKH). 
\textcolor{black}{This structure is used to reduce the cost of redistributing the symmetric group key to the group members when it is updated, as shown in Section \ref{sec:evaluation}}.
For instance, the approach proposed in \cite{de2017logical,de2016logical} exploits the LKH Tree proposed in \cite{wong2000secure}. When a group $G$ is created, the group owner creates the related key-tree $KT (d, h, G)$, 
\textcolor{black}{where $d$ is the maximum number of children for a node.}%, and $h$ is the \st{maximum} height of the tree}.  
Each node of $KT$ is paired to a symmetric key and each group member is paired to a leaf of  $KT$. In particular, a leaf node is paired with a \textit{symmetric user key}, the root of the tree is paired with a \textit{symmetric group key} while the intermediate nodes are paired with \textit{symmetric intermediate keys}. As for the other cases, each user of the DOSN is paired to an individual asymmetric key-pair.
The main characteristics of the group are stored and kept available by the DOSNs storage system while the $KT$ is stored on the local device of the group owner.
The idea behind this approach is 
\textcolor{black}{that the group owner manages the $KT$, updating it every time a new user joins the group or a member leaves it}. %\st{to securely distribute to each group member the symmetric keys on the path from the root of $KT$ to the leaf node corresponding to that member. 
%In particular, the symmetric user key is distributed to each group member by asymmetrically encrypting it using his/her individual asymmetric public key while the symmetric group key and the symmetric intermediate keys are encrypted with the symmetric user key.}
%cosi' come per l'approccio encryption based, i dettagli della distribuzione delle chiavi verranno spiegati nella prossima sezione, quindi li ho rimossi da qui
The join or removal of a user also requires the group owner to redistribute both the symmetric group key and the symmetric intermediate keys of $KT$ located on the path from the root to the leaf representing the removed/joined user, as described in more details in Section \ref{sec:privacyControl}.
%\todo[inline,color=yellow]{nell'encryption based non serve questa struttura per "spedire" le chiavi?}
%\todo[inline]{non serve}
%contains also a \textit{message list} which is used by the group owner for exchanging notification messages with the group members). The group descriptor is stored and kept available by the DOSNs storage system while the $KT$ is stored on the local device of the group owner. 
%The secure distribution of the new group keys to the group members is implemented by exploiting both the group descriptor and the individual public key of the users.

%\todo[inline,color=yellow]{la frase precedente resta un po' sospesa. E comunque, la descrizione che e' stata fatta di questo approccio non e' chiara e non e' dello stesso grado di dettaglio della descrizione dell'approccio precedente. Forse va un po' estesa.}
%\todo[inline]{Aggiornato. Potrebbe bastare? non sono andatto in dettaglio sulla join e la leave perche' forse conviene specificarle nelle sezioni successive.}
Authors of \cite{kwak2007decentralized} proposes a decentralized group key management algorithm which combines the LKH and the TGDH scheme. In particular, a group of $n$ users is divided into $s$ subgroups, each with $n/s$ members.
Each subgroup is managed by an individual LKH scheme while TGDH is employed for inter-subgroup key management. Every node of the TGDH is paired with a symmetric secret key and a blinded key.
Finally, One-way Function Tree (OFT) schemes \cite{sherman2003key} are also bases on the LKH approach but OFT assigns keys on the tree based on the key of the parent \cite{gunther2011key}.

A similar approach is exploited by Shi R.H. et al. \cite{shi2016novel}, where a large group is
divided into several subgroups, each managed by a subgroup key manager. A group key generator center is in charge of managing all the key manager, generating, distributing and updating
group keys for secure communications by all group members.\\
Authors of \cite{penrig2000elk} proposed ELK, a secure group communication protocol which combines LKH and OFT mechanisms.\\
The DGKD protocol proposed by \cite{adusumilli2005dgkd} adopts a tree structure 
where the leaf key of a node is paired the public key of the corresponding group member while all the intermediate nodes are paired to symmetric secret keys. The protocol introduces also the concept of sponsors and co-distributors. The former are group members initiating the key generation and rekeying process. The latter, are group member that receive the new keys from the sponsor and help distribute the new keys to group members.\\
The key management approaches for secure group communication have been also proposed and studied in different contexts and scenarios (such as, centralized and distributed architectures). A complete review of such approaches is presented in \cite{rafaeli2003survey}.

\subsection{\policyBased~}
In the \policyBased~ enforcement model, the privacy of the contents is enforced by implementing a privacy preserving content storage strategy.
The idea behind this model is to avoid encryption of contents still preserving the producers' privacy preferences by 
%\st{by defining a privacy-preserving strategy for the}
% and to exploit the privacy policies paired with the contents in order to define a privacy-preserving 
properly choosing the peers of the DOSN (replica peers) where allocating contents. %\st{In particular, the \policyBased~ enforcement model stores a copy of a content $c$ on a set of replica peers, which can be selected according to different approaches. }

%\st{The general approach used by most of the DOSNs in Table ref{tab:enforcModelRelatedWork} is to}
\textcolor{black}{Some approaches, such as the one proposed in \cite{de2015privacy,de2016privacy} and extended in \cite{de2017privacy}},
exploit the privacy preferences defined by the producer of a content $c$ in order to select a set of replica peers where a copy of $c$ can be stored. 
In particular, such replica peers are chosen among the ones that belong to users who are allowed to access $c$ according to the privacy preferences defined for $c$.
\textcolor{black}{In this way, if one user 
%\st{$r_i$} 
tries to directly access the unencrypted contents stored in his/her device bypassing the DOSN infrastructure, she/he
%\st{$r_i$} 
cannot acquire more information than the ones she/he is allowed to access according to contents' privacy preferences.} 
%\textcolor{red}{In this way, if one of the users $r_i$ 
%who has been chosen to store hosting a copy of  $c$ exploits the fact $c$ is stored on his peer to directly  access it, bypassing the DOSN infrastructure, she/he actually accesses a content that she/he was allowed to access through the DOSN infrastructure according to the related privacy preferences.} 
%\todo[inline,color=yellow]{La modifica proposta mi fa capire che la frase originale non ha dato il messaggio che avrebbe dovuto dare. LA frase originale voleva dire che il fatto di memorizzare i dai sui peer degli utenti che sono autorizzati a vederli fa si che se anche un utente cercasse di barare e, invece di accedere ai dati tramite il programma della DOSN, vi accedesse andando direttamente nella directory dove sono memorizzati in chiaro, tale utente non potrebbe acquisire piu' informazioni di quelle che ha effettivamente  il diritto di vedere secondo la policy perche' sul suo peer viene memorizzata solo roba che lui ha il diritto di vedere secondo la policy}
%\todo[inline]{Ads: ho provato a sistemare}
%\st{A similar approach is proposed in cite{de2015privacy,de2016privacy} and extended in cite{de2017privacy}, where the allocation strategy takes into account both the permissions granted to a user and the temporal pattern of user's activities (such as, the average session length). }

Instead, a number of DOSNs such as Solid \cite{sambra2016solid}, eXO \cite{loupasakis2011exo}, Vis-a-Vis \cite{shakimov2011vis}, My3 \cite{narendula2011my3}, DiDuSoNet \cite{guidi2015didusonet}, and \cite{tran2016decentralized,bortoli2011decentralised} %\st{avoid the encryption of contents}
select the replica peers by simply asking their users to select a set of trusted peers in the DOSN. 
In this case  contents are replicated on the peers belonging to users explicitly declared as trusted by \textcolor{black}{the contents producers.}\\
%\todo[inline, color=pink]{Vedo un problema: gli approcci allocation based sono basati sul fatto che i contenuti siano allocati su utenti fidati, come facevamo noi in DiDuSonet. Ma se gli utenti sono quelli di un gruppo, in cui ci si incontra per interessi comuni, nessuno conosce gli altri. Diverso e' il caso in cui io organizzo i miei amici in gruppi}
The system proposed in \cite{zeng2019decentralized} exploits blockchain to verify the correctness of the data and users contents are stored in clear on both the local database of the contents' owners and the random selected peers.
Finally, authors of \cite{conti2014trusted} exploits sociological trust model derived from real OSNs (known as the Dunbar model) to automatically select the set of devices trusted by users. The Dunbar trust model relies on the fact that users establish relationships with different levels of intensity (strong or weak ties) and the interactions occurred on social relationships can be used to measure the degree of trust between individuals \cite{sutcliffe2015modelling}. 
%\st{Another important implication of this model is that any user can sustain about 150 stable relationships with other users.}
%\todo[inline,color=yellow]{il 150 non so se e' rilevante per i nostri scopi. Se lo fosse, andrebbe spiegato esplicitamente il perche'.}

In the following, we focus on \policyBased\ enforcement models where the replica peers are selected according to the privacy preferences defined by users. Indeed, such allocation mechanism
%has good security features and it
does not introduce a risk for the privacy of the contents because the users asked to store a copy of a given content are among the ones authorized to access it. 

Finally, replica peers are in charge of enforcing the security preferences expressed by the producers of the contents they store. As a matter of fact, when a user $v$ requests access to the content $c$, the replica peer  storing $c$ evaluates the privacy preferences of the content in order to decide whether to allow $v$ to access $c$.

\section{Privacy Preferences Enforcement Models}
\label{sec:privacyControl}
\textcolor{black}{In this section, we briefly describe how the privacy enforcement models defined in Section \ref{sec:relatedWork} can be exploited by current DOSN systems in order to support the definition of different types of groups commonly used in DOSNs and OSNs, namely, the groups G2, G3, and G4 discussed in Section \ref{sec:scenario}.} \textcolor{black}{In particular, we select a representative approach for each enforcement model of Section \ref{sec:relatedWork} and, for each of them, we discuss how the following three operations are implemented: content publish, group join with/without backward secrecy, and group leave with forward secrecy and with/without backward secrecy.} 
%\st{Then, the implementations of these operations are used to provide the types of group defined in Sec. \ref{sec:scenario} (namely G1, G2, G3 and G4).}

Since users can disconnect from the system at any time, all the approaches we selected exploit a Distributed Hash Table (or DHT \cite{balakrishnan2003looking}) to enable asynchronous communication between users, even if one of the parties is disconnected from the DOSN.

%\textcolor{red}{Each group is paired to a shared data structure: the Group Message List  (i.e., a message list kept available by the DOSN storage system), that is exploited by the group administrator to notify group members with messages containing the nodes' keys that have been updated.}
%\todo[inline,color=yellow]{SAREBBE SOLO QUESTO DA SPOSTARE NEL IV}

\subsection{\aclBased~enforcement model}
\label{sec:aclBased}
\textcolor{black}{The reference approach we take into account for the \aclBased~enforcement model is the one that have each group $G$ paired with its symmetric group key, $K_G$, which is created by the group owner and securely distributed to the group members \cite{buchegger2009peerson,graffi2011lifesocial,nilizadeh2012cachet,jahid2012decent,koll2017socialgate,aiello2012lotusnet,sharma2012supernova,graffi2020libresocial,jiang2019bcosn}.}

\textcolor{black}{In addition, each group is also paired to a shared data structure: the Group Message List  (i.e., a message list kept available by the DHT), exploited by the group administrator to notify all group members with messages containing symmetric group key that have been updated.\\ Instead, direct messages between the group administrator and a group member are exchanged by using a Private  Mailbox service \cite{kangasharju2003secure,mezo2011distributed} implemented by the DHT. In particular, each user $u$ is paired to a Private  Mailbox data structure which supports the append operation and allows the other users to append new (encrypted) messages (such as notifications or private messages).}
%\todo[inline]{Il pezzo sopra e' piu' o meno simile anche per l'lkh ma ci sono delle differenze per come vengono utilizzati}

%\todo[inline,color=yellow]{SAREBBE SOLO QUESTO DA SPOSTARE NEL IV}

\subsubsection{Content publishing} 
%\todo[inline,color=yellow]{qui abbiamo il problema della nuova tassonomia, ovvero dobbiamo dire qualcosa riguardante la differenza tra i gruppi in cui pubblica solo l'owner e quelli in cui pubblicano tutti i membri}
In the reference scenario considered, all the members of a group $G$ can publish a content $c$ in the group.
In particular, the content producer creates a new symmetric content key, $K_c$ to protect $c$. In fact, the content $c$ is encrypted with the symmetric content key $K_c$, 
while the current symmetric group key, $K_G$, is used to encrypt and securely communicate $K_c$.
%\todo[inline,color=yellow]{qui diciamo che la chiave di contenuto criptata viene comunicata. In realta' potrebbe essere piu' preciso dire che viene memorizzata pure quella sui peer assieme al contenuto. No?}
Finally, the encrypted content is published on the DOSNs, 
\textcolor{black}{i.e., stored along with  the encrypted symmetric content key on a number of peers of the DHT.}

\subsubsection{Group Join}
%\begin{description} 

\paragraph{Without backward secrecy} 
%\todo[inline,color=yellow]{qui e nelle altre "intestazioni di paragrafo" anche se levo Gx, Gy non cambio il significato, perche' tanto il focus e' sulla backward secrecy}
%\todo[inline]{esatto. Tolto}
%\todo[inline,color=yellow]{qui si potrebbe dire che secondo le tabelle 2 e 3 questo e' il piu' diffuso}
%\todo[inline]{aggiunto}
The join without backward secrecy is used to implement groups of type G3 and G4. As shown in Table \ref{tab:mappingDOSNs} and \ref{tab:mapping}, the group join operation provided by almost all the considered DOSNs and OSNs does not provide backward secrecy. 
The \aclBased~ approach where the join operation does not support backward secrecy is quite easy to implement. 
In particular, when a new user is added to a group, the group owner does not change the current symmetric group key.
The group owner simply encrypts the 
symmetric group key to the joining user by
encrypting it with the public key of the joining user \textcolor{black}{ and stores the encrypted message on the Private Mailbox of the joining user.}

In order to join the group, the joining user retrieves the message containing the encrypted group key \textcolor{black}{from the Private MailBox} and decrypts 
%\st{the symmetric group key} 
\textcolor{black}{it} by using her/his private key. 
%\todo[inline]{Da normalizzare: his/her - she/he}
Since the symmetric group key does not change \textcolor{black}{when a new user is added to the group}, both the old and the future contents that were/will be published in the group are/will be accessible to the new member, as well as to the already existing members of the group.

\paragraph{With backward secrecy}
%\st{In the \aclBased~ enforcement model,} 
The join with backward secrecy is used in groups of type G2. %\st{In particular,}
\textcolor{black}{In the \aclBased~ enforcement model,}
the join operation with backward secrecy must change the group key every time a new user is added to the group.  
%If the \aclBased~ enforcement model provides the user join operation with backward secrecy, the group key is changed every time a new user is added to the group. 
As a result, when a new user is added to a group, the group owner creates a new symmetric group key that will be used to encrypt the contents that will be shared for this group from this moment on. 
\textcolor{black}{The group owner encrypts the new group key with the public key of the joining user and stores the encrypted message to the Private MailBox of the joining user.}
In order to join the group, the joining user retrieves the new symmetric group key and decrypts it by using his/her private asymmetric key. Since the already existing content published in the group are encrypted with the old group key, the new member cannot access them. \textcolor{black}{In order to communicate the new symmetric group key to the already existing members of the group, the group owner encrypts it with the old group key and stores the encrypted message in the corresponding Group Message List.}

\subsubsection{Group Leave}
In the \aclBased\ approach, when a user leaves the group, the group owner updates the symmetric group key, and sends it to the remaining group members in encrypted form, i.e, the group owner asymmetrically encrypts the new symmetric group key with each individual public key of the members remaining in the group. \textcolor{black}{Finally, each encrypted copy of the new symmetric group key is stored on the Private MailBox of each member remaining in the group.}
The new group key will be used to protect new contents that will be published in the group. 
Consequently, future content published in the group can be accessed only by the members of the group who received the new symmetric group key ({\it forward secrecy}). 

\paragraph{Without backward secrecy%(G2, G4)
} 
%\todo[inline,color=yellow]{anche qui si puo' levare Gx,Gy perche' il focus e' sulla backward secrecy}
%By default, 
%\todo[inline,color=yellow]{perche' "by default"? le tabelle 2 e 3 dicono che nelle DOSN e' il piu' adottato mentre nelle OSN e' il meno adottato}
In most of the existing DOSNs, the group leave operation does not guarantee the backward secrecy property because it is used to implement groups of type G2 and G4.
In the \aclBased~ approach, the leave operation without backward secrecy can be implemented by simply changing the current group key when a user leaves the group,
%\st{In particular, the group owner updates the symmetric group key and encrypts the new symmetric group key with each individual public key of the members still belonging to the group. The new group key will be used to protect new contents that will be published in the group.}
\textcolor{black}{leaving the already published contents unchanged.
As a result, those users who are members of a group when a content is published will be enabled to access such content forever.} 
Hence, the removed member can still access the contents published in the group when she/he was a group member because such contents remain encrypted by exploiting the old symmetric group key. \textcolor{black}{Instead, the removed member will not be able to access the content published after the leave operation because they will be encrypted with a new group key she/he does not know}. 

\paragraph{With backward secrecy%\st{(G1, G3)}:
} 
\textcolor{black}{The \aclBased~ approach can also provide the leave operation with backward secrecy, which is used to implement groups of type G3. 
%\st{In this case, when a user leaves the group, the group owner updates the symmetric group key, and sends it to the remaining group members in encrypted form, i.e, the group user encrypts the new symmetric group key with each individual public key of the members remaining in the group.Future content published in the group can be accessed only the members of the group who received the new symmetric group key.}  
In this case, besides changing the current group key to guarantee the forward secrecy property,}
%Since contents already published in the group remain encrypted with the old group key, they can be accessed by the removed user. 
%\todo[inline,color=yellow]{dato che questa parte appena descritta e' in comune con l'approccio precedente e serve per ottenere la forward secrecy, si potrebbe mettere subito all'inizio del 4.1.3, e poi solo dopo fare la distinzione tra without e with the backward secrecy?}
%\todo[inline]{ok, fatto. Quindi le parti da rimuovere sono due. Lo facciamo per tutte?}
%For this reason, 
\textcolor{black}{the group owner needs to update all the contents already published in the group by re-encrypting them with a new symmetric content key which, in turn, is encrypted with the latest group key. %\st{ 
%with properly updated group keys.
%As a matter of fact, in principle, each distinct content key could be encrypted with a distinct version of the group key, in order to grant distinct access rights to distinct contents, and the update of such keys should not alter the access rights of the users remaining in the group.}
Hence, in case of groups with a large number of contents and/or where the members change frequently, this solution is very expensive from the computational point of view, because it requires to re-encrypt all the contents every time a member leaves the group.}
%\todo[inline,color=yellow]{il paragrafo precedente era scritto per descrivere sia G1 che G3, mentre ora si parla solo del G3. Quindi si potrebbe "semplificare"}
%Moreover, this solution also supposes that the group owner is able to cause the cancellation of all the out of date copies of such contents.} 
%\todo[inline]{Nell'ultima frase non mi torna il discorso della cancellazione.}
%\todo[inline,color=yellow]{intendevo rimuovere dalla rete p2p tutte le copie criptate con le chiavi vecchie. Ovvero, che l'utente rimosso non riesca a trovare nella (oppure gli venga proprio proposto dalla) sottostante rete p2p una copia del contentuo ancora criptata con le vecchie chiavi, che quindi lui puo' accedere. Evidentemente non e' scritto chiaro.}
\textcolor{black}{We also recall that, besides being computationally expensive, this solution does not really protect content privacy, as explained in Section \ref{sec:scenario}}.

\subsection{\lkhBased~enforcement model}
\label{sec:lkhBased}
\textcolor{black}{The reference approach we take into account for the \lkhBased~ enforcement model used for our evaluation is based on \cite{de2017logical,de2016logical}
%,wong2000secure} 
and it exploits the Logical Key Hierarchical Tree data structure.} 

\textcolor{black}{As for the \aclBased~ enforcement model, private messages between the group administrator and a group member are exchanged by using the Private Mailbox service.
Similarly, each group is also paired to a shared data structure: the Group Message List  (i.e., a message list kept available by the DHT), exploited by the group administrator to notify all group members with messages containing the parts of the Logical Key Hierarchical Tree data structure that have been updated.}

\subsubsection{Content publish} 
\textcolor{black}{The implementation of the content publish operation supporting the LKH-based enforcement model is quite the same as the implementation supporting the Encryption-based one.}
A group member creates a content $c$ to publish in the group, which is paired with an LKH data structure.
In order to share the content in the group, 
%\st{the publisher can directly encrypts $c$ by using the symmetric group key paired to the root of the LKH. Alternatively, }
the content producer creates a new symmetric content key, $K_c$, to protect $c$, encrypts $c$ with $K_c$, and uses the current symmetric group key paired to the root of the LKH to encrypt $K_c$. Finally, the encrypted content \textcolor{black}{and content key} are published on a number of peers of the DHT.
%\todo[inline]{Anche qui vogliamo cambiare con DHT?}

\begin{figure}[t] 
\centering 
\subfigure[Group join with backward secrecy.\label{fig:JoinWithBS}]{
        \includegraphics[width=0.65\textwidth]{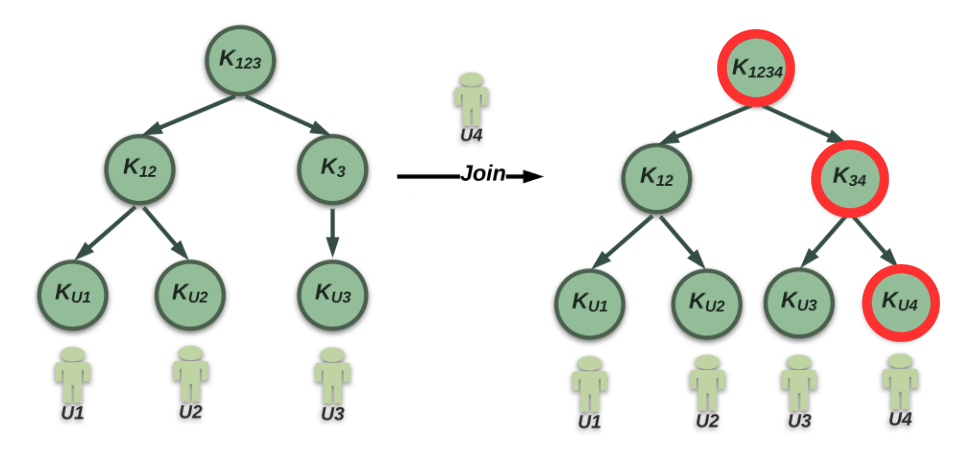}}
\subfigure[Group join without backward secrecy\label{fig:JoinWithoutBS}]{
        \includegraphics[width=0.65\textwidth]{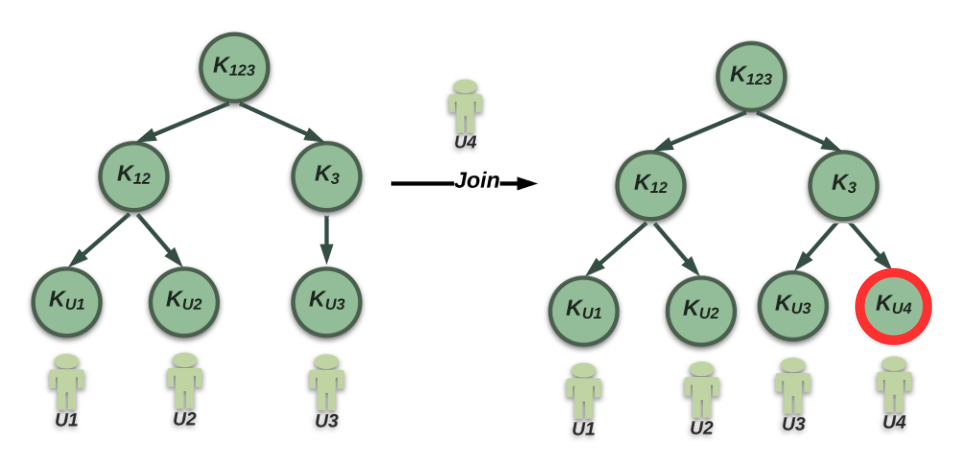}}
\caption[Graphical description of the \lkhBased~ join operation with and without backward secrecy]{Graphical description of the \lkhBased~ join operation with and without backward secrecy.\label{fig:treeJoin}}
\end{figure}
\subsubsection{Group Join}
\paragraph{Without backward secrecy} The \lkhBased~ enforcement model can be easily adopted in order to implement the join operation without guaranteeing backward secrecy, used in groups of type G3 and G4. In fact, since the joining user should have access to all the contents already published in the group, adding a new user to the group simply means that the current symmetric group key is shared with the new user. 
In particular, as shown by Figure \ref{fig:JoinWithoutBS}, to add the user $U4$ to the group, the group owner $o$ simply creates a new leaf node $K_{U4}$ of the key tree $KT(d,h,G)$ for $U4$, while the symmetric keys of the existing nodes remain the same. 
The group owner securely communicates the nodes of KT from the root $K_{123}$ 
%\todo[inline,color=yellow]{nella figura b la root e' K123}
to the leaf $K_{U4}$ to the joined user by using \textcolor{black}{the Private Mailbox of $U_4$.}
\textcolor{black}{In particular, the group owner exploits the asymmetric public key of $U4$ to asymmetrically encrypt the symmetric leaf key $K_{U4}$, and uses the symmetric leaf key $K_{U4}$ to symmetrically encrypt the rest of the keys paired to the nodes on the path from the father of $U_4$ to the root of the $KT$.\\}
The group owner does not send anything to existing members of the group because the nodes of $KT$ concerning them have not been changed.

\paragraph{With backward secrecy} In order to guarantee backward secrecy in groups of type G2, each time the group owner $o$ adds a new user $U4$ to the group $G$, $o$ must change the symmetric group key paired with $G$ and must communicate the new key to $U4$ and to the previously exiting group members.
In this way, $o$ prevents $U4$ from accessing the contents previously published in the group.
To this aim, $o$ updates the key tree $KT$ by changing the symmetric group key paired with the root of $KT$, $K_{123}$, and by adding the leaf node $K_{U4}$ corresponding to the new user. Figure \ref{fig:JoinWithBS} shows how the key tree $KT$ is reorganized when the user $U4$ joins the group, and the nodes of $KT$ highlighted in red color are the ones that are created or updated because of  the join operation.
In order to ensure backward secrecy, $o$ also updates the symmetric keys of the nodes on the path from the new root $K_{1234}$ to the joining node $K_{U4}$. 
\textcolor{black}{Then, the group owner communicates these keys to $U4$ by exploiting the Private Mailbox of $U_4$.
%provided by the DOSN infrastructure.
In particular, to securely transfer such keys, the group owner encrypts the symmetric leaf key $K_{U4}$ by using the asymmetric public key of $U4$, and uses the symmetric leaf key $K_{U4}$ to symmetrically encrypt the rest of  the  keys paired to the nodes on the path from the father of $U_4$to the root of the $KT$.}
%\st{Moreover, $o$ securely communicates the updated nodes to both $U4$ and the existing members of the group, separately. In particular, $o$ encrypts the symmetric key $K_{U4}$ paired with the leaf representing $U4$ with the asymmetric public key of $U4$.}
\textcolor{black}{Instead, to communicate the updated keys (including the group key) to the existing members of the group, the group owner encrypts each of them with its previous version, and creates a message including all such encrypted updated keys which is sent to group members by storing it on the Group Message List.
}

Once she/he received such data from $o$, $U4$ exploits her/his asymmetric private key to decrypt the symmetric leaf key $K_{U4}$. 
Then, the key $K_{U4}$ is used to decrypt the new symmetric intermediate key $K_{34}$ %which, in turn, is used to decrypt
and the new symmetric group key, $K_{1234}$, which is paired with the root.
This key will be used by $U4$ and by the other users for the publication of the new contents of the group and for accessing them. 
The new symmetric group key, $K_{1234}$, does not reveal anything about the contents published before $U4$ joined the group, because they have been encrypted with the previous version of such key ($K_{123}$). 

\textcolor{black}{Finally, the existing members of the group who have some nodes on the path from their leave to the root of the $KT$ updated, exploit the old versions of the keys of such nodes to decrypt the new version of such keys. This is also valid for the new symmetric group key $K_{1234}$ of the $KT$, which can be decrypted by existing members  using the key $K_{123}$.}

\subsubsection{Group Leave}
\begin{figure}[t] 
\centering 
\includegraphics[width=0.80\textwidth]{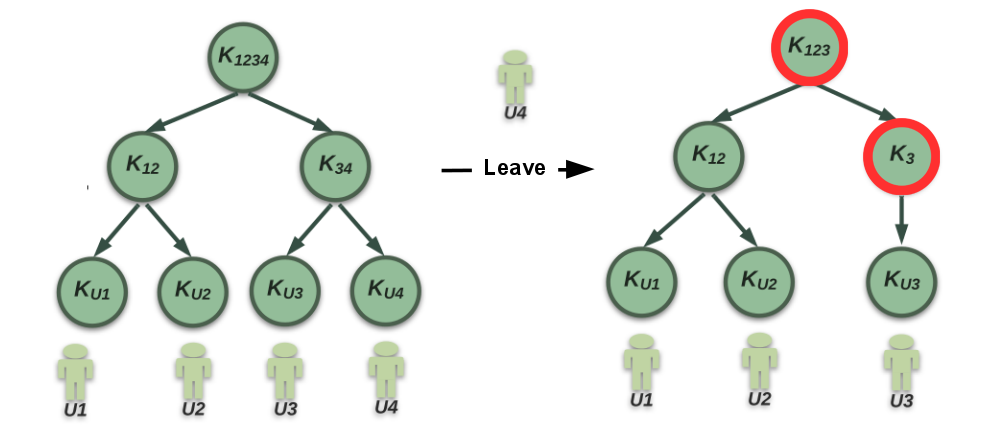}
\caption[Graphical description of the \lkhBased~ leave operation with and without backward secrecy]{Graphical description of the \lkhBased~ leave operation with and without backward secrecy.\label{fig:treeLeave}}
\end{figure}

\paragraph{Without backward secrecy} The leave operation without backward secrecy used in groups of type G2 and G4 can be easily implemented by the \lkhBased~ enforcement model. Indeed, when a member leaves the group, the group owner removes the corresponding leaf  from the key tree (see Figure \ref{fig:treeLeave}) and updates the symmetric keys of the nodes along the path from the root to the father of the removed leaf (including the symmetric group key).
In particular, supposing that the user $U4$ leaves the group, the group owner removes the leaf $K_{U4}$ of the user from the key tree and updates the symmetric keys of the nodes along the path from the root $K_{1234}$ (i.e., the symmetric group key) to the father of the removed leaf, $K_{34}$. Figure \ref{fig:treeLeave} shows the key tree resulting from the leave operation while the updated nodes are highlighted in red color. 
The group owner creates a  notification message
\textcolor{black}{and stores it in the Group Message List}
in order to distribute such new node \textcolor{black}{keys} to the members left in the group. 
The notification message consists of such updated 
%\st{nodes of the key tree} 
\textcolor{black}{keys},
%\st{(the nodes highlighted with red color in Figure ref{fig:treeLeave})}, 
each of which is encrypted with the keys on their children nodes.
\textcolor{black}{This encryption procedure starts from the father of the node corresponding to the leaving user, and proceeds towards the root of the $KT$ in order to guarantee that the key of each node is encrypted with the latest versions of the keys of its children nodes.}
As a result, future contents of the group will be encrypted with the new symmetric group key, which is unknown to the removed user.
The removed user can still access the old contents published in the group by exploiting the old symmetric group key.
%As a result, future contents of the group will be encrypted with the new symmetric group key, which is unknown to the removed user.

\paragraph{With backward secrecy} 
%\st{As explained for the \aclBased\ enforcement model, it is not really possible to define an effective solution to protect from the users removed from a group the contents published in such group before the execution of the group leave operation, because removed users could have made a local copy of such contents when they had the rights to read them and, consequently, they will always have this copy available. Since, to the best to our knowledge, none of the currently available DOSNs adopting the \lkhBased~enforcement model implements the group leave with backward secrecy,  we skip its description here.}
%\todo[inline,color=yellow]{non c'e' nessuno degli approcci esistenti che fa questo (lhk con leave con backward secrecy), giusto? Avevamo detto che era una buona scusa per non descriverlo nemmeno}

The \lkhBased~ enforcement model can be easily adapted to implement group G2, providing a leave operation with backward secrecy.In order to guarantee the forward secrecy property, when a member $U4$ of a group $G$ leaves the group (see Figure \ref{fig:treeLeave}), the symmetric group key must be changed following the same procedure described for the group leave without backward secrecy. 
Moreover, in order to ensure that the removed member cannot access the old contents published in the group, the group owner must re-encrypt all the contents by exploiting the new symmetric group key created in the previous step. For this reason, the group owner retrieves all the $p$ contents already published in the group, decrypts them by exploiting the old symmetric group key $K_{1234}$, obtaining the plain contents. 
%\todo[inline,color=yellow]{la vecchia e' K1234}
Then, the group owner re-encrypts each content by using a new symmetric content key which is, in turn, encrypted with the current group key $K_{123}$. 
%\todo[inline,color=yellow]{la nuova e' K123}
Finally, the group owner store the $p$ contents on the storage service of the DOSNs. As a result, the removed user cannot access the current version of the contents published in the group because they are encrypted with a new content key and a new group key. 
%\st{This assumes that all the contents encrypted using the old content keys and group key are removed from the DOSN's peers.} 
%\textcolor{red}{However, the removed user can access an outdated version of the group's contents cached on his/her peer by using the old symmetric content key and group key.}
\textcolor{black}{Again, we recall that this solution does not really protect content privacy, as explained in Section \ref{sec:scenario}}.
%\todo[inline,color=yellow]{la frase in rosso l'ho rimessa al posto della frase cancellata, ma se vi sembra una ripetizione si leva}

\begin{table*}[t]
\small
\centering
\caption{Strategies for the maintenance of data structure in the \lkhBased~ enforcement model\label{tab:maintenanceLKH}}
\begin{tabular}{cclll}
\hline
 \textbf{\begin{tabular}[c]{@{}c@{}}Rekeying \\ strategy\end{tabular}} & \textbf{\begin{tabular}[c]{@{}c@{}}Rekeying \\ time\end{tabular}} & \multicolumn{1}{c}{\textbf{Issues}} & \multicolumn{1}{c}{\textbf{Advantages}} & \multicolumn{1}{c}{\textbf{Applications}} \\ \hline
individual& immediately &\begin{tabular}[c]{@{}l@{}}- scalability\\ - out-of-sync\end{tabular} & \begin{tabular}[c]{@{}l@{}}+ security\\+ easy to implement\end{tabular} & \begin{tabular}[c]{@{}l@{}}OSN,\\healtcare,\\ messaging\end{tabular} \\ \\
bach & periodically & \begin{tabular}[c]{@{}l@{}}- rekey interval\\- security\end{tabular} & \begin{tabular}[c]{@{}l@{}}+ low cost\\+ customizable\end{tabular} & \begin{tabular}[c]{@{}l@{}}teleconferencing,\\ pay-per-view,\\ pay-tv,\\ streaming\end{tabular} \\ \hline
\end{tabular}
\end{table*}

\subsubsection{Lessons learned}
The analysis of the \lkhBased~ enforcement model allowed us to learn several lessons that are worth mentioning in this section.
\paragraph{\textbf{Maintenance of data structure}} The data structure used in the \lkhBased~ enforcement model changes over time because of users joining and leaving the group.
For this reason, the efficiency of the model depends on whether the underlying data structure remains balanced. 
Several works \cite{ng2007dynamic,briscoe1999marks,pham2007efficiency,li2001batch} have focused on proposing new algorithms to reduce the rekeying cost and they can be classified into two groups, depending on the proposed rekeying strategy.
Table \ref{tab:maintenanceLKH} summarizes the characteristics of the strategies used for the maintenance of the data structure.
In individual rekeying strategy \cite{briscoe1999marks,pham2007efficiency,chang1999key} the operation for updating the data structure (known as rekeying) must be performed whenever a member joins or leaves the group. In particular, the key used to encrypt the data is changed immediately and the sender has to reveal some secrets to each receiver in order to enable the reconstruction the data. For this reason, the  suffers from scalability issue in the case of dynamic groups, because several new keys may be generated, distributed, but not used by the group members. In addition, the out-of-sync problem
is also present in individual rekeying strategy because a group member could receive a content encrypted with a group key that it has not received yet.
However, this strategy is very easy to implement and ensures that the users joining/leaving the group can/cannot instantly access to the content published in the group after their addition/removal. Consequently, the individual rekeying strategy is suitable to be used in several scenarios that are relevant for the security and privacy, e.g., online social network, healthcare, and messaging applications. 

The batch rekeying strategy \cite{li2001batch,ng2007dynamic,bruhadeshwar2009balancing,vijayakumar2016efficient} is another approach used to reduce the overhead of the join and leave operations and it consists in executing the rekeying periodically, only when a certain number of requests have been collected. For the join operation, the rekeying of the data structure can be executed either when a number of requests are ready to be processed or a new content is published on the group.
However, the leave operation exposes a critical security issue because the removed users can still get access to the contents published on the group 
until the rekeying of the data structure has been performed \cite{li2001batch}. Another important issue for the batch rekeying strategy, is to determine a proper rekey interval, i.e., the length of the time interval to wait for group access/removal requests \cite{li2010batch}. In addition, the batch rekeying strategy introduces a lower overhead compared to the individual rekeying strategy and it can be dynamically adapted to the behavior of group members. For these reasons, the batch rekeying strategy is widely used in the field of teleconferencing, pay-per-view, and streaming applications.

\paragraph{\textbf{Optimal tree structure}}  
An important task in the key tree model is to find the optimal configuration parameters of the data structure for a certain pattern of user behaviors, minimizing the cost of individual or batch updates. The approaches used to solve this problem are summarized in Table \ref{tab:optimalDataStructure} and they aim to model the optimal parameters of the key tree for a specific rekeying strategy and problem setting.\\
For instance, the works proposed in \cite{guo2015optimal,wong2000secure} investigate the optimal tree structure with minimal cost for individual rekeying strategy in the case of a single deletion.

Instead, the authors of \cite{wu2008optimal,wu2013optimal,chen2008optimizing,wu2009optimal,chan2011optimal,li2009approximately,li2001batch} focused on the optimal tree structure for the batch rekeying strategy. In particular, authors of \cite{wu2008optimal,wu2013optimal} study the degree bound for the problem of deleting two users while authors of \cite{chen2008optimizing,wu2009optimal} investigate the optimal tree structure under the assumption that $k$ members arrive in the initial setup period and only member deletions are allowed after that period. Authors of \cite{li2001batch} prove that 4 is the best key degree under the assumption that the batch consists of 
$k$ joins and $j$ leaves.
Instead, the work proposed in \cite{li2009approximately} study the optimal data structure when the size of a batch is not large and each user has a fixed probability $p$ of being replaced during the batch period. Finally, authors of \cite{chan2011optimal} extended the work of \cite{li2009approximately} by considering also the case of loyal users, i.e., users who have zero probability to leave the group.

\begin{table*}[tb]
\small
\centering
\caption{Characteristics of the optimal data structure for the \lkhBased~ enforcement model\label{tab:optimalDataStructure}}
\begin{tabular}{lcp{3.3cm}p{4.7cm}}
\hline
\bf{Work}& \bf{Rekeying} &\bf{Problem}& \bf{Optimal data structure} \\ \hline
\cite{wong2000secure} & individual & single user deletion & the optimal node degree of the key tree is 4 \\
\cite{guo2015optimal} & individual & single user deletion & the degree of all nodes is at most 3, the number of leaves in the branches differs by at most 1\\
\cite{wu2008optimal,wu2013optimal} & batch & 2-deletion problem & root degree is between 5 and 7, the number of leaves in the branches differs by at most 1 and each subtree is a 2-3 tree \\
\cite{chen2008optimizing} & batch& k-insertion followed by k-deletion & internal node has degree at most 5, the children of nodes which have degree 3 are all leaves \\
\cite{wu2009optimal} & batch & k-insertion followed by k-deletion & internal node has degree at most 7, children of nodes with degree not equal to 2 or 3 are all leaves \\
\cite{chan2011optimal} & batch & normal users have probability $p$ to leave the group, loyal users have probability zero & when $p \geq 0.43$ the optimal data structure is a star when $p<0.43$ the optimal data structure is a tree \\
\cite{li2009approximately} & batch & users have probability $p$ to be replaced by another user & a ternary tree is a 2-approximation of the optima data structure \\
\cite{li2001batch} & batch & k-insertion and \mbox{j-deletion} & when the size of a batch is not large, 4 is the best key tree degree, otherwise, key star outperforms key tree \\
\hline
\end{tabular}
\end{table*}

\subsection{\policyBased~ enforcement model}
\label{sec:policyBased}
\textcolor{black}{The \policyBased~enforcement model considered for our evaluation is based on the approaches described in \cite{de2015privacy,de2016privacy,de2017privacy}.
These approaches protect the privacy of a content $c$ by taking into account the privacy preferences defined by the content owner in order to choose the replica peers for storing the copies of $c$.
Indeed, with respect to other models, such approach guarantees that contents will be always stored on the replica peers belonging to users who are authorized to access them. For this reason, content encryption is not necessary.
Moreover, in \cite{de2015privacy,de2016privacy,de2017privacy}, the privacy preferences of the users are specified in XML through the XACML standard \cite{xacml}, %\cite{xacml}, 
and they are stored 
%and made available to other users through a
in a policy repository, called Policy Administration Point (PAP) according to the XACML reference architecture. Whenever a privacy policy is created or updated by a user, the group owner has to notify it to the policy repository by sending the updated data.}
%\todo[inline,color=yellow]{aggiungere una frase sul PAP non appena abbiamo deciso come e'. Dato che e' una questione implementativa, la}
%\todo[inline,color=yellow]{Altra cosa. In questo punto della sezione potremmo dire esplicitamente che in questa sezione parliamo di come si implementerebbero i gruppi con le policy per essere paragonabile alle sezioni precedenti, no? Ho scritto la frase qui sotto. Vi torna?}

%\textcolor{red}{\st{In order to compare the \policyBased~ enforcement model with the previously described ones, in the following we suppose to enforce a identity-based XACML policy which mimics a group and we discuss how the operations content publish, group join and group leave can be implemented.}}

In order to compare the \policyBased\ enforcement model with the previously described ones, in the following we suppose to have an XACML privacy policy for each group, consisting of a set of identity-based privacy rules, one for each content published  in the group, which lists  the identities of the members who are authorized to access such content. %\todo[inline,color=yellow]{la prossima frase e' stata aggiunta a seguito della call del 19 febbraio}
\textcolor{black}{Hence, in this scenario the list of peers authorized to access the content $c$ is explicitly available, thus making not necessary to evaluate the policy on a (possibly very large) set of candidates to find the peers for storing the copies of $c$.}

In the following, we discuss how the operations content publish, group join and group leave can be implemented.

\subsubsection{Content publish} 
In the \policyBased~ enforcement model, a group member creates a content $c$ and a
privacy rule for $c$
%\st{a privacy policy for $c$} 
which lists the identities of current group members who are authorized to access the content $c$. Then, the group member %\st{configures the privacy policy of $c$ by granting access to the current members of the group} 
\textcolor{black}{asks the PAP to add such rule}
%\st{adds such rule}
%\todo[inline,color=yellow]{mi pare piu' naturale dire che il mebro chiede al PAP di aggiungere la regola piuttosto che il membro lo faccia da solo e poi delega il PAP}
to the privacy policy of the group. 
%\st{pairs $c$ with the privacy policy ID and delegates the management of such privacy policy to the PAP.}
Once the content will be published, the privacy policy of the group will be evaluated each time a user will request to access it, by using the authorization component of the privacy-preserving framework.
Such component checks whether the requesting user actually holds the required access right and returns authorization decision.
%The privacy policy of the content is evaluated by using the authorization component of the privacy-preserving framework in order to decide the peers that can host a replica of such content. 
Moreover, to guarantee the availability of contents, the privacy policy of the group is evaluated by using the authorization component of the privacy-preserving framework in  order to define the set of peers that can host a replica of such contents. 
As matter of fact, each content $c$ is replicated by the allocation mechanism of the DOSN on a number of peers of authorized users available in the system. 
%Moreover, once the content will be published, the privacy policy of the content will be evaluated each time a user will request to access it in order to check whether she/he actually holds the required access right.

\subsubsection{Group Join}

\paragraph{With backward secrecy} The \policyBased~ enforcement model is well suited to implement the join operation with backward secrecy provided by groups of type  G2. % as it requires only to update the list of group members with the identity of the new user. 
As a matter of fact, when a new member joins a group, the group owner updates the list of the current group members with the identity of the new user, and each new content that will be published on the group after the join of the user will be paired to a privacy rule which grants access to the identities of both the new user and the members that were already in the group.
%\st{creates a new privacy policy that grants access to both the joining user and the members that were already in the group.} 
%\st{The new contents that will be published on the group after the new user joined it will be paired to such new privacy policy.}
Instead, to guarantee backward secrecy, the privacy rules of the group privacy policy related to the contents already published in the group remain the same, i.e., they only grant access to the old group members while they deny access to the  joining user. As a result, the joining users cannot access the contents published in the group before she/he joined it.

\paragraph{Without backward secrecy} The \policyBased~ enforcement model can be easily adapted to implement join operation without backward secrecy provided by groups of type G3 and G4. 
As in the previous case, when a new user joins the group, the group owner updates the list of current group members by adding the identity of the new user. %\st{creates a new privacy policy that grants access to both the joining user and the members that were already in the group. }
%updates the list of group members by adding the identity of the joining user. 
As a result, the joining user will be able to access future contents that will be published on the group because a new rule will be added to the group privacy policy for such contents which grants access to both the joining user and the members that were already in the group.  
In addition, in order to not have the backward secrecy, the group owner must ensure that the new user can access also the old contents published in the group. For this reason, the group owner must update the privacy rules paired to the contents already published in the group as well, by adding the ID of the new user among the IDs of the authorized ones. %This can be done by exploiting the built-in functionalities that the privacy policy language provided for this purpose. A first approach is to change the privacy policy of each content of the group by adding the identity of the new user. However, the time required for such modifications can be quite high because the number of privacy policies is typically equal to the number of contents published in the group. 

\subsubsection{Group Leave} 
\paragraph{Without backward secrecy} %(G2, G4):} 
The \policyBased~ enforcement model is well suited to implement the leave operation without backward secrecy provided by groups of type G2 and G4.
\textcolor{black}{As a matter of fact, when a member $u$ leaves the group, the group owner 
%\st{creates a new policy where he} 
removes $u$ from the list of current group members 
%\st{the authorized users,} 
so as to guarantee that $u$ will not be allowed to access the contents that will be published in the future (\textit{forward secrecy}).} 
The rules of the group privacy policy related to the existing contents of the group remain the same, so that the removed user can still access the corresponding contents.

\paragraph{With backward secrecy:} %(G1, G3):} 
The \policyBased~ enforcement model can be adapted in order to implement a leave operation with backward secrecy, as provided by group G3. 
As in the previous case, to ensure the forward secrecy property, when a user leaves the group, the group owner %\st{creates a new policy for the contents that  will be published in the future which does not allow the leaving user to access them.}
removes the intended user from the list of the current group members so as to guarantee that she/he will not be considered as authorized member for the publication of future contents. 
%However, the removed user can still access old contents published in group. 
In order to ensure also the backward secrecy property, the group owner changes the rules of the group privacy policy related to the contents already published in the group to deny the access to the leaving user as well. 
%\todo[inline,color=yellow]{qui si potrebbe aggiungere il discorso che dicevo prima, ovvero che se la ioin e' senza backward secrecy, allora devo sempre (ovvero sia per la join che per la leave) cambiare la polity a tutti i contenuti del gruppo, per cui potrei avere una policy sola per gruppo e quell'ID viene puntato da tutti i contenuti del gruppo}
%\todo[inline]{Questo e' il caso della leave CON backward secrecy. Per la leave senza backward secrecy non e' necessario cambiare la politica. E' sicuramente una possibile implementazione ma non sono sicuro sia fattibile e performante perchè non e' stata valutata.}
%As for the case of the join without backward secrecy, we exploit the built-in functionalities of the privacy policy language which allows to combine the decisions produced by different privacy policies into a single authorization decision. For this reason, the group owner creates a new privacy policy for the removed user, that denies access to all the $p$ contents published in the group. As a result, when the removed users try to access old contents of the group, the most recent privacy policy applicable to the user request (i.e. the one previously created by the group owner) is evaluated and it will deny access to the user. The leave operation involves only the group owner, who removes the user from the list of group members and creates the new privacy policy. 

\subsubsection{Lessons learned}
The analysis of the \policyBased~ enforcement model provided us with new insights and highlighted challenges that are summarized below.

\begin{figure*}[t] 
\centering 
\includegraphics[width=0.98\textwidth]{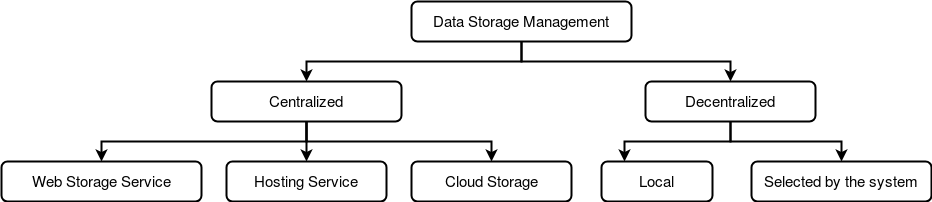}
\caption{Summary of the data storage management approach used in the \policyBased~ enforcement model.\label{fig:summaryDataManagement}}
\end{figure*}
\paragraph{\textbf{Data storage management}} The \policyBased~ enforcement model takes advantage of a data storage model to ensure that authorized users can access unencrypted contents they are interested in and several proposals of data storage managements for DOSNs exist. Figure \ref{fig:summaryDataManagement} classifies the data storage management approaches used in the \policyBased~ enforcement model. In general, existing approaches assume that the underlying storage service responsible for data management is either implemented in a decentralized way by the DOSNs \cite{narendula2011my3,guidi2015didusonet,conti2014trusted,de2015privacy,de2016privacy,de2017privacy,loupasakis2011exo} or provided by a centralized entity \cite{shakimov2011vis,zeng2019decentralized,bortoli2011decentralised,sambra2016solid}.\\ 
In the case of a centralized data storage management, the DOSN assumes that users have cloud-computing utility that is used for running virtual machine instance implementing the OSNs services. For instance, Vis-a-Vis \cite{shakimov2011vis} provides a virtual machine instance running in a paid cloud-computing utility such as Amazon Elastic Compute Cloud (EC2) or Rackspace Cloud Servers while Zeng S. et al. \cite{zeng2019decentralized}, Bortoli S. et al. \cite{bortoli2011decentralised}, and Solid \cite{sambra2016solid} require an accessible storage repository, which can be deployed on personal servers by the users themselves, or on any cloud storage provider (e.g., Dropbox).
The users must trust the centralized provider of the storage service because it can access any user data. In addition, the cloud-computing utility should support a Trusted Platform Module \cite{shen2010cloud} for verifying the software stack running under the instances.\\
Instead, in the case of a decentralized data storage management, the contents generated by users can be stored either on their local devices \cite{loupasakis2011exo} or on the other users' devices automatically selected by the DOSN, based on different factors (such as, trust relationships between users \cite{narendula2011my3,guidi2015didusonet,conti2014trusted} or privacy policies defined by users \cite{de2015privacy,de2016privacy,de2017privacy}).

\paragraph{\textbf{Data availability}}
Another important challenge for the \policyBased~ enforcement model is to keep the contents available as much as possible while securing them from unauthorized access by other users.
\begin{figure}[t] 
\centering 
\includegraphics[width=0.70\textwidth]{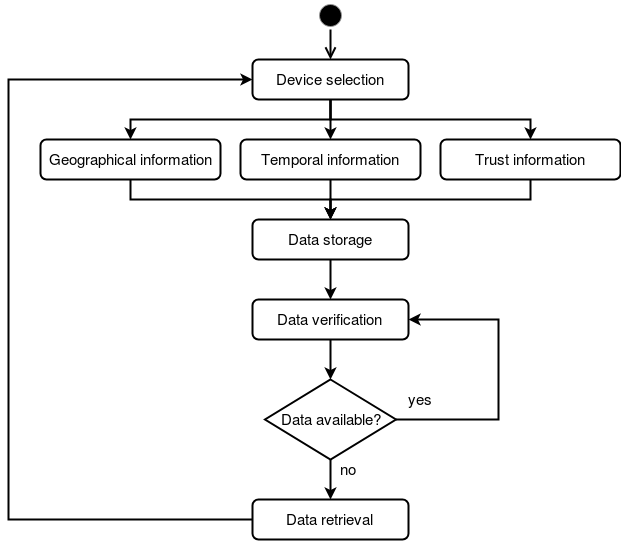}
\caption{Summary of the workflow used by the \policyBased~ enforcement model to guarantee the availability of the contents.\label{fig:summaryDataAvailability}}
\end{figure} 
In the case of a centralized data storage management, such as Vis-a-Vis \cite{shakimov2011vis}, Zeng S. et al. \cite{zeng2019decentralized}, Bortoli S. et al. \cite{bortoli2011decentralised}, and Solid \cite{sambra2016solid}, the users are responsible for configuring the quality of service offered by cloud provider by specifying upload and download bandwidths or cloud downtime. Consequently, the data availability and the reliability of the data storage service is guaranteed by the cloud provider, which ensures also the compliance with agreed terms of the service.
In contrast to the previous approach, the decentralized data storage management stores the contents on some users' devices which can disconnect from the system whenever they want. For this reason, the classical approach used to increase the availability of the contents is to replicate a copy of them on different users' devices. An overview of the general workflow for guaranteeing data availability in P2P system was described in \cite{oualha2010secure}. However, in order to discuss all the challenges of the data availability for the \policyBased~ enforcement model, we further refine the general workflow for data availability as shown in Figure \ref{fig:summaryDataAvailability}. 
The first phase of the workflow is to properly selects the devices on which to store users' contents according to different types of information which could improve the availability and the security of the data. For instance,
authors of \cite{guidi2015didusonet,conti2014trusted} select the devices based on the trust relationships and the online time period of the user while authors of \cite{narendula2011my3} exploits also geographical locations. Instead, the approach proposed in \cite{de2015privacy,de2016privacy,de2017privacy}
selects the devices according to trust information derived by the privacy policies defined by users. Once the device is selected, the data storage phase is responsible for storing a copy of the content on the selected device while the data verification phase periodically checks if the device (and the contents stored on it) remains available in the system. When content is no longer available a data repair phase is executed, which retrieves a copy of the content and repeat the entire workflow.

\section{Experimental evaluation}
\label{sec:evaluation}
In order to evaluate the three privacy preferences enforcement models (\aclBased~, \policyBased~, and \lkhBased), we developed a set of simulations by using the P2P Peersim simulator\footnote{Available at: \url{http://peersim.sourceforge.net/}}. \textcolor{black}{For each privacy enforcement model, the simulation implements the corresponding reference approach described in Section \ref{sec:privacyControl} and creates several groups of types G2, G3, and G4 in order to evaluate its performance.} As observed in Section \ref{sec:scenario}, the groups of type G1 are not considered in our analysis because this type does not found use in current DOSNs and OSNs. For each type of group, we evaluate separately the operation of content publication, group join, and group leave,  varying the number $n$ of members in the group. In particular, we consider groups having 
%number of members, 
$n$ equal to 10, 50, 100, \num{1000}, and \num{10000} \textcolor{black}{and we suppose that they are obtained by a sequence of join operations}. 
For the content publication operation, the number $p$ of contents to be published in each group is set to 10, 50, and 100. For the group join and leave operations, we consider the addition or removal of a single user from several groups having different number of contents and members.\\ 
For each operation and privacy enforcement model, we measure the execution cost for the group owner, for the joining/leaving users, and for the other members
in terms of time spent to execute the operation and number of bytes sent and received. We consider contents of size $c$ fixed to 100KB, since it is the maximum image size you can upload on Facebook\footnote{\url{https://www.facebook.com/help/266520536764594?helpref=uf_permalink}}.
%The operations provided by each group are implemented by exploiting the \aclBased~, \policyBased~, and \lkhBased~ enforcement models described, respectively, in Sec. \ref{sec:aclBased}, \ref{sec:policyBased}, and  \ref{sec:lkhBased}.

\begin{table}[tb]
\centering
\small
\caption[The parameters of the simulation]{Configuration of the simulation parameters\label{tab:simPar}}
\begin{tabular}{cll}
    \hline
    \textbf{Parameter} & \textbf{Values} & \textbf{Description} \\ \hline
    $GT$ &  \{G2,G3,G4\} & type of the groups \\
    $n$ & \{\num{10}, \num{50}, \num{100}, \num{1000}, \num{10000}\}&  size of the groups \\ 
    $p$ & \{10, 50, 100\} & number of contents  \\ \hline
\end{tabular}
\end{table}

\textcolor{black}{In the  \policyBased~ enforcement model, the Balana\footnote{\url{https://github.com/wso2/balana}} open source framework is used as reference implementation of the XACML standard specification \cite{xacml} and for the evaluation of the privacy policies. Moreover, we assume that each peer hosts its own PAP, which stores the privacy policies of all the groups related to the contents stored by such peer.} 

%\textcolor{black}{\st{The  \lkhBased~ based    enforcement  model creates a key-tree $KT (d, h, G)$ for each group $G$ where the maximum height $h$ of the tree is 8 and the maximum number $d$ of children for a node is equal to 4. Consequently, using the previous configuration, the \lkhBased~ based enforcement model is able to manage groups consisting of at most $d^h$ (65536) members.}}
%\todo[inline,color=yellow]{MAGGIO 2021: la frase precedente non vale piu' perche' mettiamo il paragrafo successivo.}

\textcolor{black}{The \lkhBased\ enforcement model creates a key-tree ($KT$) for each group $G$. In the following, when we measure the costs of this enforcement model,
%In the following, we discuss the fact that 
we do not account for the overhead that would be needed  to keep such key-tree balanced. In particular,
%we would not consider the cost for re-balancing the tree. 
such an approach is justified by the following reasons: \textit{(i)} tree re-balancing operations typically occur in batch, when a given number of evictions is reached \cite{di2002efficient}; \textit{(ii)} the rate of leave is not triggered by m2m communications, but mainly by humans, as such the total number of evictions is not expected to be relevant over relatively short period of times \cite{de2019analysis}; \textit{(iii)} when the rate of join and leave is roughly balanced, the tree data structure does not need frequent re-balancing, since join basically contribute to keep it balanced \cite{ng2006scalable}; and, finally, \textit{(iv)} the balancing operations could occur during periods of low utilization of the data structure. For the above cited reasons the introduced overhead can be considered manageable, and hence we will not consider it, while  we will focus on the overhead related to the management of the core operations for the different scheme under analysis.}
\begin{table}
\small
\centering
\caption[Cost of the symmettric encryption algorithms]{Cost of the symmetric encryption algorithms provided by the Crypto++ Library \label{tab:encryptionCostSy}}
\begin{tabular}{ccccc}
\hline
\textbf{Algorithm} & \textbf{\begin{tabular}[c]{@{}c@{}}Throughput \\ (MB/s)\end{tabular}} & \textbf{Cycles/Byte} & \textbf{\begin{tabular}[c]{@{}c@{}}Setup Key \\ ($\mu$s)\end{tabular}}& \textbf{\begin{tabular}[c]{@{}c@{}}Setup Key\\(cycles)\end{tabular}}  \\ \hline
\begin{tabular}[c]{@{}c@{}}AES/CTR\\(256-bit key)\end{tabular}&\num{2496}&\num{0.8}&\num{0.278}&611 \\ 
\begin{tabular}[c]{@{}c@{}}AES/CBC\\(256-bit key)\end{tabular}&447&4.7&0.216&475 \\ \hline
\end{tabular}
\end{table}

The implementation of the symmetric and asymmetric scheme used by the \lkhBased~ and \aclBased~ enforcement models we exploit the Crypto++ library\footnote{\url{https://www.cryptopp.com/}}: a well-known open-source cryptography library written in C++ which implements many ciphers with consistently good performance on all of them. Table \ref{tab:encryptionCostSy} summarizes the performance of the symmetric cryptographic algorithms considered by the enforcement models investigated in this manuscript. For each cryptographic algorithm we reported the following performance measures: the number of MB encrypted/decrypted per second (Throughput), the number of cycles-per-byte required by encryption/decryption (which depends upon the CPU frequency), and the number of microseconds ($\mu$s) and cycles required for key setup. The cryptographic algorithm based on symmetric schema does not increase the size of the encrypted data compared to the input data and it can be used with a CTR configuration block mode or with a CBC configuration block mode \cite{verma2011peformance}. 
Instead, Table \ref{tab:encryptionCostAs} summarizes the performance of the asymmetric cryptographic algorithms considered by the enforcement models investigated in this manuscript. For each cryptographic operation, we reported the number of millisecond and the number of Megacycles required for its execution. Since asymmetric encryption is used by privacy enforcement models only to protect symmetric keys of length 256 bit, all tests were done by using 2048-bits RSA keys and by repeating the crypto operations over blocks of random data having a size comparable to such keys. 
Consequently, the output of the asymmetric encryption operation has size equal to that of the RSA key length (i.e., 2048 bits), even when the input data size to encrypt is less than 2048 bits.

The test platform exploited for our experiments is a PC equipped with an Intel Core i7- 2.20 GHz CPU, 8GB or RAM, running Linux Ubuntu. In addition, the cryptographic algorithm based on symmetric schema has been configured to use the CTR configuration block mode. 
%Ubuntu 12.04-LTS 64-bit

\begin{table}
\small
\centering
\caption[Cost of the asymmetric encryption algorithms]{Cost of the asymmetric encryption algorithms provided by the Crypto++ Library \label{tab:encryptionCostAs}}
\begin{tabular}{ccc}
\hline
\textbf{Operation} & \textbf{Milliseconds} & \textbf{Megacycles} \\ \hline
RSA 2048 Encryption &	0.16	&0.29 \\ 
RSA 2048 Decryption &	6.08	& 11.12 \\ 
RSA 2048 Signature &	6.05 &	11.06 \\ 
RSA 2048 Verification &	0.16 &	0.29 \\ \hline
\end{tabular}
\end{table}

\subsection{Content publish}
\label{sec:contentPub}
In this subsection we evaluate the  cost of publishing  contents in a group in the three enforcement models. In general, this cost depends on the size of the group, $n$, and on the number of contents to publish, $p$.
%\st{, regardless of the group type.}

\begin{figure}[bt] 
\centering 
        \includegraphics[width=0.50\textwidth]{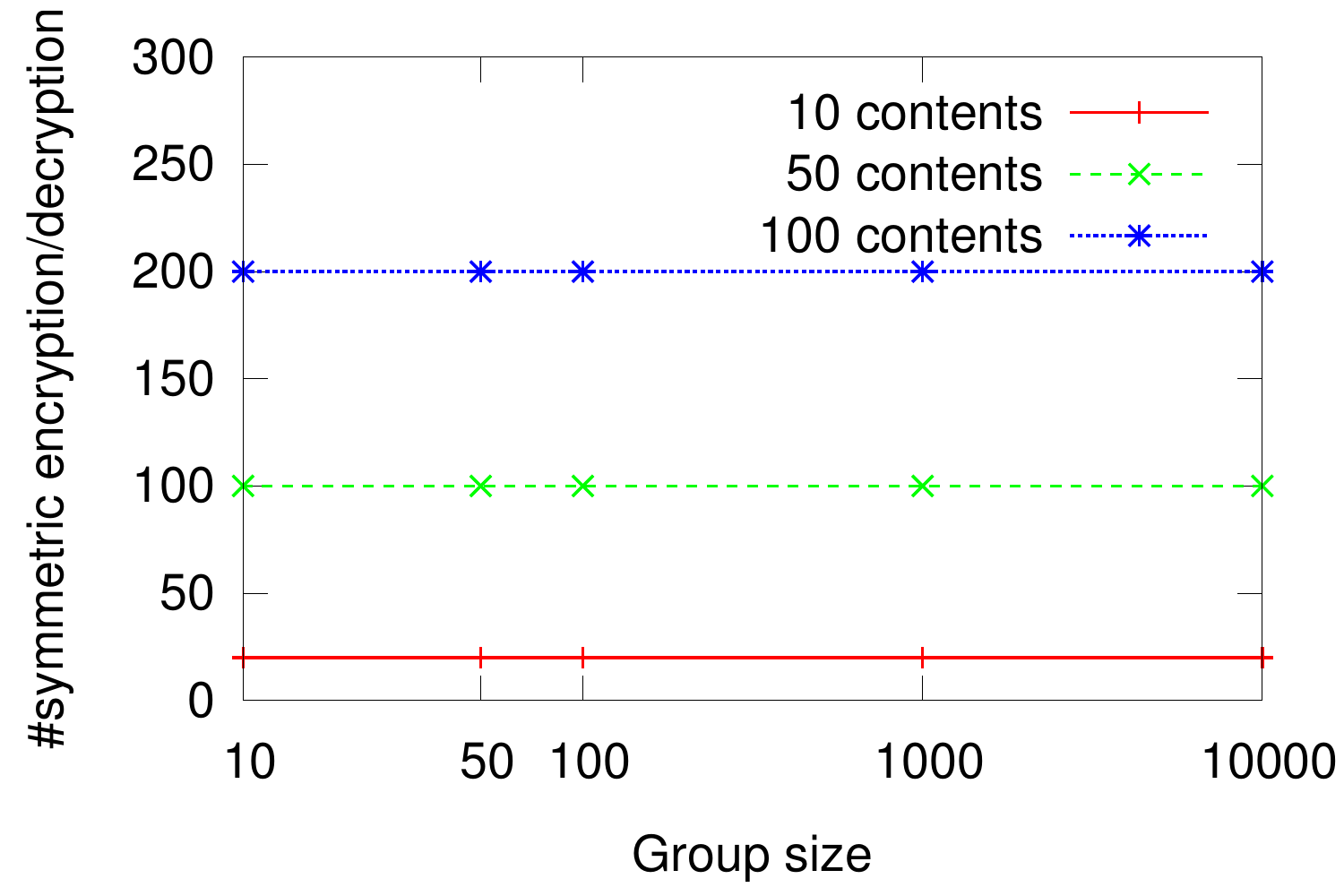}
        %}
\caption[Computation time for the publication of a content]{Evaluation of the number of cryptographic operations taken by \lkhBased~ and \aclBased~ enforcement model for the publication/retrieval of different number of contents.\label{fig:costPublishLKH}}
\end{figure}

For the \lkhBased~ enforcement model, %(see Figure \ref{fig:costPublishLKH}) 
the publication of $p$ contents on a group of size $n$ requires a number of symmetric encryption operations equals to twice the amount of content published, i.e., $2\cdot p$.
%, see Figure \ref{fig:costPublishLKH-based}. 
Indeed, the content producer 
creates a new symmetric content key for each content, performs a symmetric encryption operation to encrypt the content \textcolor{black}{(100KB)} with this content key, and a second symmetric encryption operation to encrypt the content key \textcolor{black}{(256 bits)} with the group key.
As a result, two symmetric encryption operations are executed for each content published
%\st{, i.e., the number of symmetric keys encrypted by the content author is the same as the number of contents encrypted} 
(see Figure \ref{fig:costPublishLKH}).

%\st{To access each content $c$, a group member $o$ has to decrypt the symmetric content keys of $c$ by using the symmetric group key, and then $o$ decrypts the content by using the symmetric content key previously obtained.}

The cost for publication of contents of the \aclBased~ enforcement model is the same as the \lkhBased. Indeed, in the \aclBased~ enforcement model as well, a group is paired with a symmetric group key which is used to securely distribute the symmetric content keys exploited to encrypt the contents published on the group. 
In both the \lkhBased~ and \aclBased~ enforcement models, the encrypted content and the related encrypted content key are then replicated on the peers of the DOSN, for instance by using a DHT.
%\st{However, in contrast to the \lkhBased~ enforcement model which exploits tree data structure, the symmetric group key of the \aclBased~ approach is distributed to the group members by using a list data structure.}

\begin{figure*}[tb] 
\centering 
\subfigure[Time for creating the privacy policy \label{fig:timePolicyCreation}]{
        \includegraphics[width=0.47\textwidth]{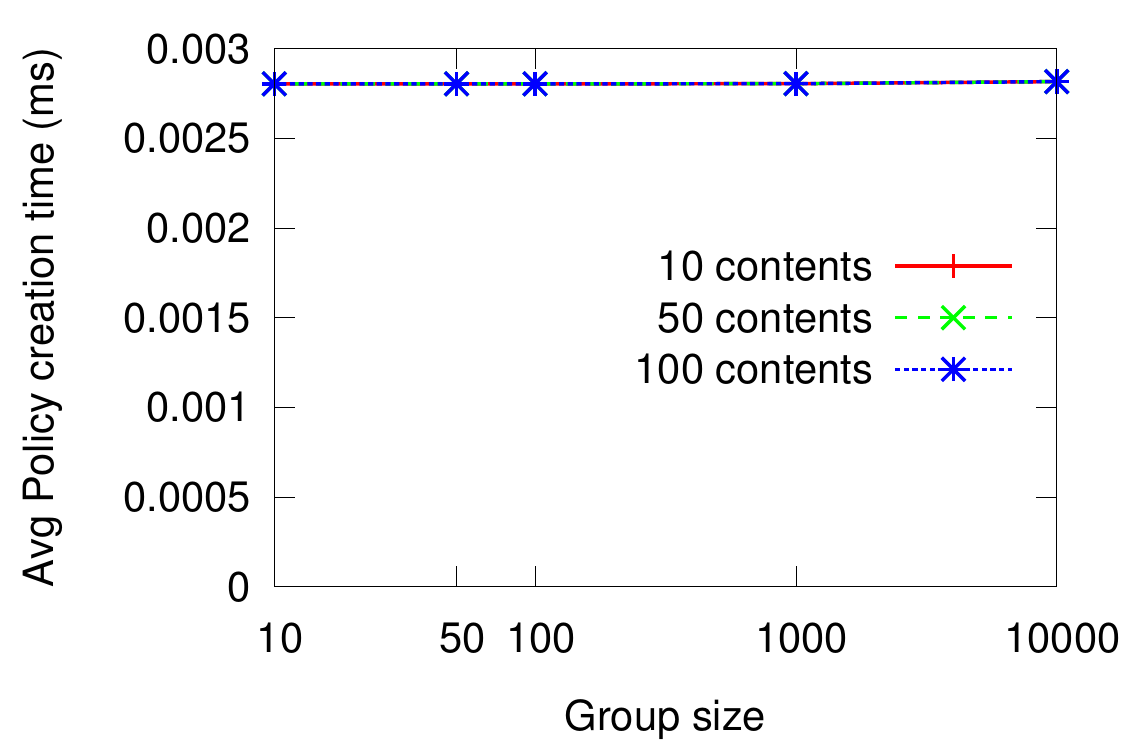}}
\subfigure[Size of the privacy  policy\label{fig:sizePolicy}]{
        \includegraphics[width=0.47\textwidth]{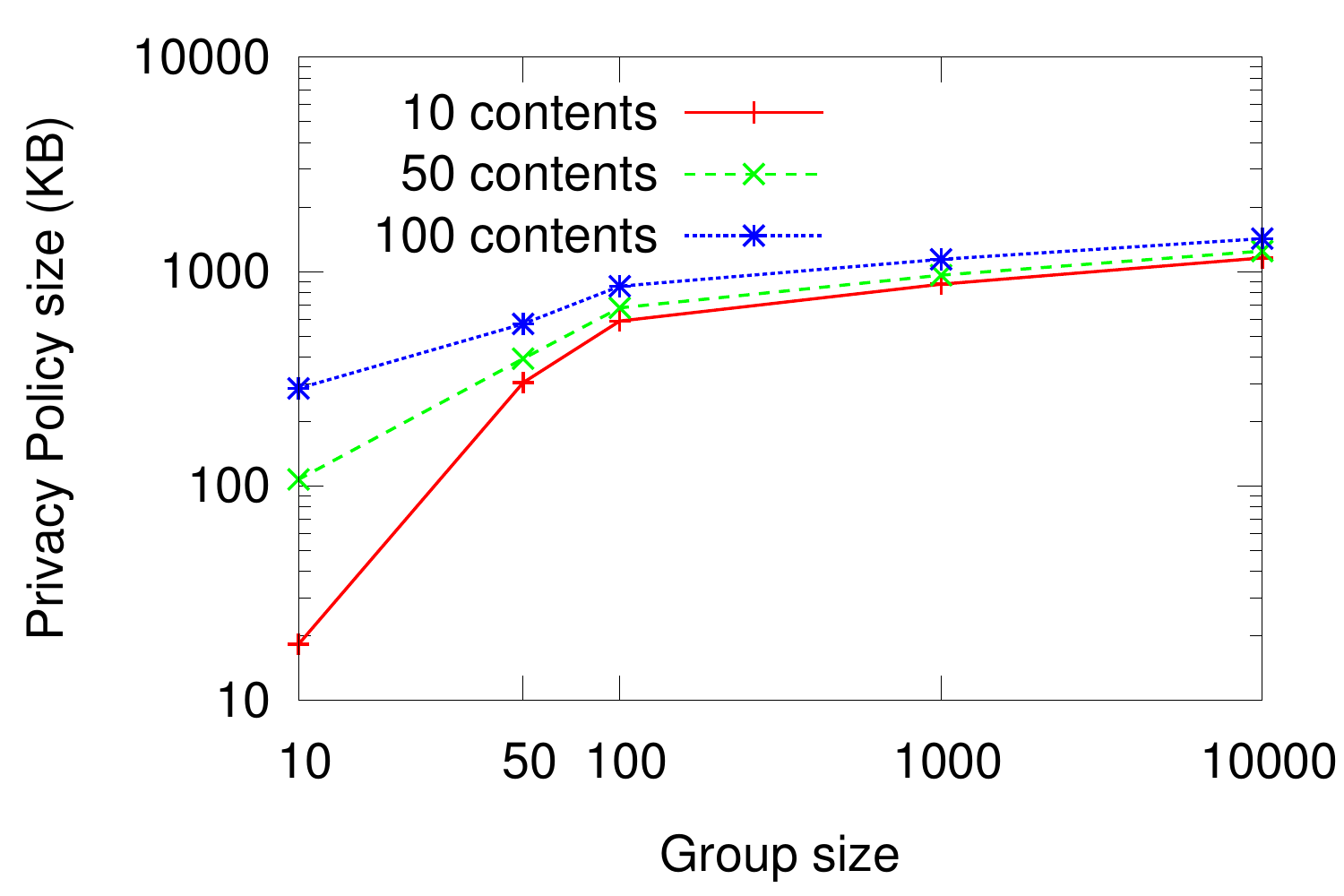}}
\subfigure[Time for evaluating the privacy policy\label{fig:costPublishPolicyBased-based}]{
        \includegraphics[width=0.47\textwidth]{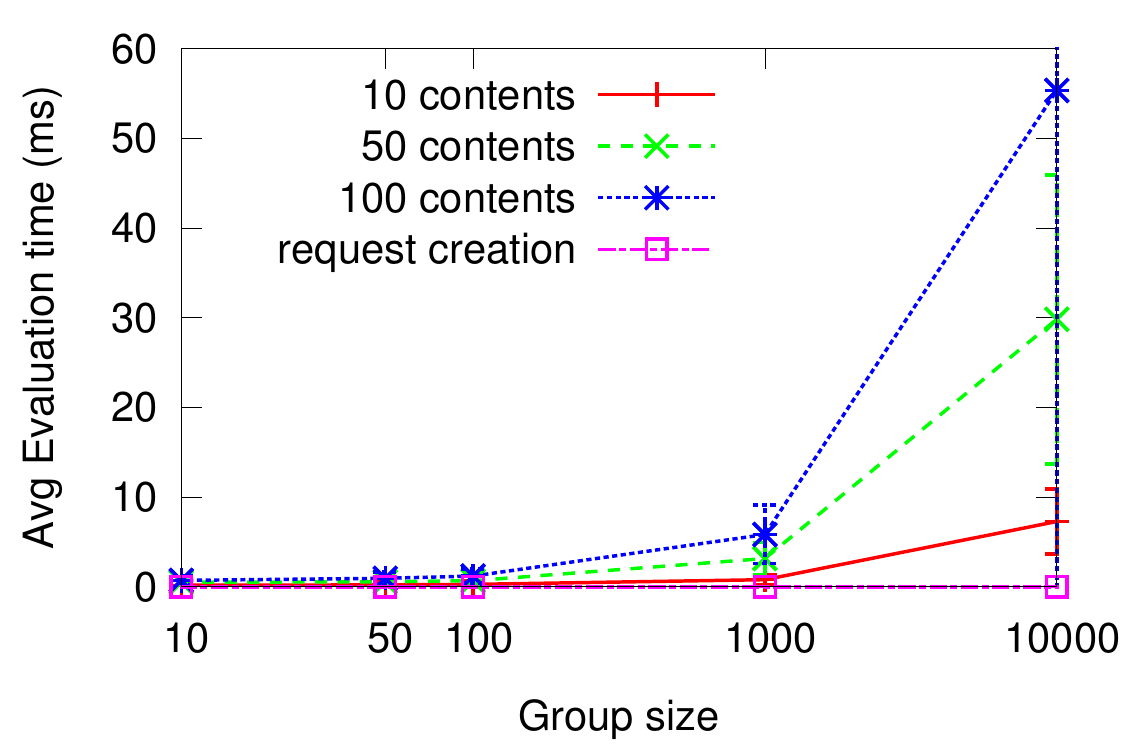}}
\caption[Computation time and message size of the \policyBased~]{\policyBased~ enforcement model: Figure \ref{fig:costPublishPolicyBased-based} shows the time taken by the user in order to evaluate privacy policy on groups with different number of users and contents. Figure \ref{fig:sizePolicy} shows the total size (in KB) taken by the privacy policies paired to the groups, while Figure \ref{fig:timePolicyCreation} shows the average time required for the creation of privacy policies.}
\end{figure*}

%\st{In the \policyBased~ enforcement model, users exploits privacy policy for each content published on a group, in order to specify who can access it. 
%In general, the privacy policy must be created and properly initialized by the  content producer for the content $c$, in order to grant the access to the right group of users.} 
Publishing a content in a group the  \policyBased~  enforcement model requires \textcolor{black}{the creation of the related rule to be embedded in the group privacy policy.} %\st{and a number of evaluations of such policy to decide where the replicas of the content will be allocated.}}
As previously explained, in order to compare the implementation of the \policyBased~enforcement model proposed in \cite{de2015privacy,de2016privacy,de2017privacy} with the implementations of the \aclBased\ and \lkhBased~ones, where the privacy preferences definition is based on groups,
in our evaluation tests we  enforce a XACML policy 
%\st{which is defined by the group owner and} 
which mimics a group by checking whether the ID of the user requesting the access to a content is included in the list of the IDs of the members of the group.
\textcolor{black}{Consequently, the peers used to host the replicas of each content are simply chosen from such list of IDs, thus not requiring additional computational cost w.r.t. the other two enforcement models where the replicas can be stored anywhere.}
%the related privacy policy is evaluated to choose only those peers belonging to users that are actually allowed to access it. In particular, being $k$ the degree of replication required by the DOSN, and given the list of DOSN peers currently online, the \policyBased~ enforcement model evaluates the policy  iteratively on each peer of such list until $k$ allowed peers are found. Hence, in the best case, the policy is evaluated  $k$ times.
A copy of the privacy policy of a group is stored on the PAP of each peer hosting contents of such group, and such  policy is evaluated every time a user requests to access the content, in order to ensure that only the right group members can access it. 
Figure \ref{fig:timePolicyCreation} shows the average time (in ms and 95\% C.I.) required for the creation of a privacy policy intended for groups of different number of members and varying the number of contents. 
The time introduced for the creation of the privacy policy is negligible and it increases with the number of users of the group. 
Figure \ref{fig:sizePolicy} shows the total size (in KB) taken by the privacy policies paired to groups consisting of different number of users and contents. 
In particular, the graph has a log scale and it clearly indicates that the size of a privacy policy paired to a group depends on both the number of users and  contents already published in the group.
\textcolor{black}{Indeed, each content published in the group is paired to a rule of the group privacy policy which grants access to such content to the right members of the group. }
In order to evaluate the overhead introduced by the group privacy policy enforcement when accessing a content, Figure \ref{fig:costPublishPolicyBased-based} shows the average time needed to create the XACML request (denoted as request creation) and to evaluate the group privacy policy varying the number of members of the group and the number of contents published within the group. 
The time required to create the XACML request is constant because it contains mainly the identifier of the applicant and the identifier of the requested content.
Instead, the evaluation time of the group privacy policy depends on both the number of contents published in the group and the number of authorized users listed in the policy.
%\st{(i.e., all the members of the group when the content was created.} 

%\st{because during the evaluation of an access request, privacy policies related to a group must be retrieved using the custom XACML Policy Administration Point module} 
%\st{Finally, conditions defined on the policies are checked and the corresponding authorization decision returned is returned as a result of the evaluation.}
%by using symmetric encryption (y-axis) for groups of different size (x-axis)

\begin{figure}[t] 
\centering 
\subfigure[Time taken for the publication of a content
        \label{fig:timeContentPublish}]{
        \includegraphics[width=0.45\textwidth]{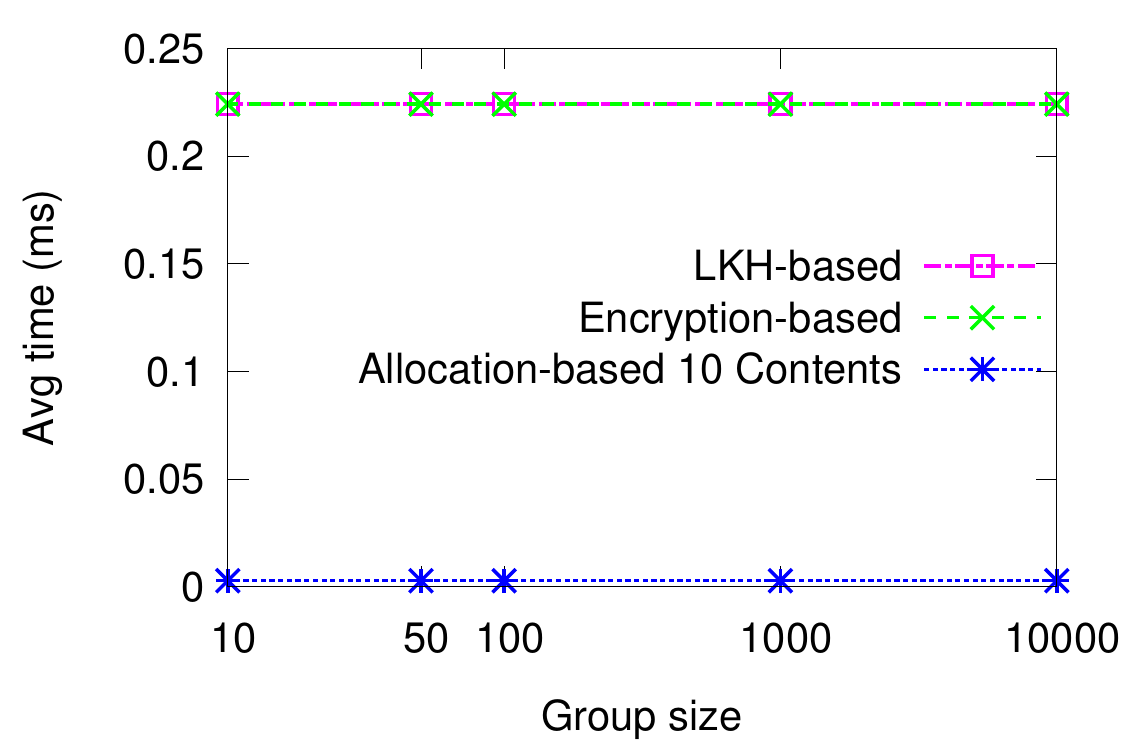}}
\subfigure[Number of KB sent by users for the publication of a content
        \label{fig:sizeContentPublish}]{
        \includegraphics[width=0.45\textwidth]{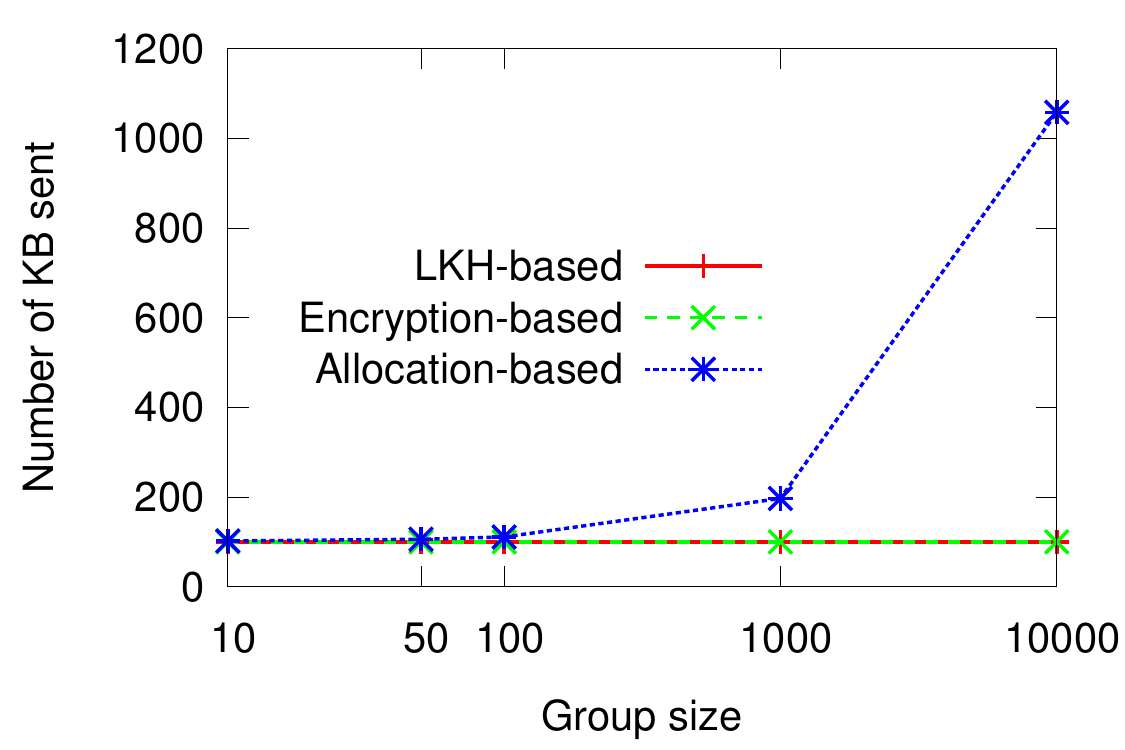}}
\caption[Computation time and messages size of the \lkhBased~ and \aclBased~]{Evaluation of the both time and size of messages needed for the publication of a content of 100KB by for each approach. For the \lkhBased~ and \aclBased~ enforcement models we used AES/CTR (256-bit key) and RSA 2048.\label{fig:comparisonContentPublish}}
\end{figure}

Figure \ref{fig:comparisonContentPublish} summarizes the costs introduced by the three enforcement models for publishing a content on a group G of $n$ members.
The plot does not include the cost for replicating the content on a number of peers of the DOSN.
%\st{The cost of the publication of a content only affects the content publisher and it does not require the use of asymmetric encryption operation.}
As shown in Figure \ref{fig:timeContentPublish}, the \lkhBased~ and the \aclBased~ enforcement models take the same time for the publication of the content, i.e., 0.225 ms,
regardless of both the size of the group and the number of contents already published in the group. As previously explained, the number of symmetric encryption operations involved during the publication of a content is equals to 2: an encryption operation on the symmetric content key and an encryption operation on the corresponding content. 
%The number of symmetric keys initialized by the content publisher or saved by a group member is equals to 1, i.e., the symmetric content key of the new content.\todo{questo paragrafo e' un po' una ripetizione di quento detto prima...} 
The most part of the overhead introduced by \aclBased~ and \lkhBased~ approaches during the publication is spent on content encryption. As shown in Figure \ref{fig:sizeContentPublish}, the \lkhBased~ and the \aclBased~ enforcement models result in the same number of data sent. Indeed, the content publisher sends at most 2 messages: the former to retrieve the last version of the group key 
and the latter to publish \textcolor{black}{the new encrypted content along with the encrypted content key}. As a result, since the size of symmetric keys is fixed to 256 bits in our tests, the amount of data sent depends mainly on the size of the content (which is 100KB in our tests).

Instead, the cost of publishing a content in the \policyBased~ enforcement model  (see Figure  \ref{fig:timeContentPublish}) %\st{introduces an additional overhead for the publication of a content with respect to the \lkhBased and the \aclBased~ approaches. This overhead is required in order} 
is due to
%\st{: \textit{i)}}
\textcolor{black}{the creation of the new rule of the  group privacy policy that grants the access right to the published content to all the current members of the group.}
%\textcolor{red}{In particular, being $k$ the degree of replication required by the system, and given the list of DOSN peers currently online, the \policyBased~ enforcement model the policy is evaluated iteratively on each peer of such list until as necessary at least $k$ times, in the unlikely hypothesis that each of the $k$ times the policy is satisfied.}
Since we already observed that the policy creation time depends mainly on the number of group members, we fixed the number of contents already published in the group to 10.
In Figure
\ref{fig:timeContentPublish} 
we observe that the time taken by \policyBased~ enforcement model is considerably lower than those of the other two approaches and, in general, the overall time required by the approaches still remain negligible.
However,  this plot does not include the time required to replicate the content on a number of peers of the DOSN networks because this time depends on a number of variable factors (e.g., the current availability of online peers). 
As shown in Figure \ref{fig:sizeContentPublish}, the \policyBased~ enforcement model results in a higher number of data sent.
Indeed, the content publisher has to send the updated group privacy policy which includes a new rule for the new content,
whose size depends on the number of users in the group. For instance, in the case of a group consisting of \num{10000} members, the identity information of each member must be considered during the definition of the privacy policy.
Hence, the size of the policy and, consequently, the amount of data sent, grow with the number of group members.
%\st{and the size of the privacy policy in is an order of magnitude lower than the number of group members.}

\subsection{Group Join}
\label{sec:evalJoin}
This section evaluates the cost of the three enforcement models to perform a group join operation, i.e., to add a new user to a group, varying the size of such group. 
In particular, in the following, we evaluate separately the cost of the join operation both in the case when the backward secrecy property is guaranteed, and when it is not ensured. 

\begin{figure*}[!t] 
\centering 
\subfigure[Group owner\label{fig:joinTimeBSgroupOwner}]{
        \includegraphics[width=0.47\textwidth]{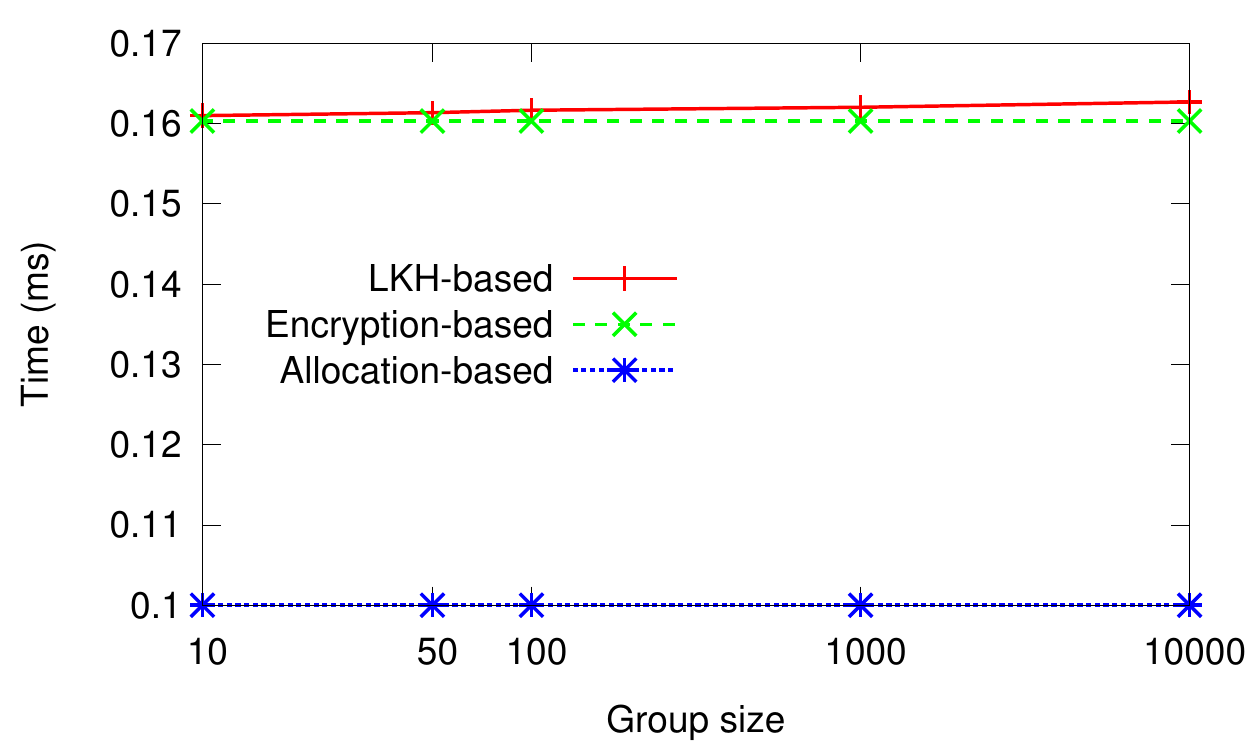}}
\subfigure[Group owner\label{fig:joinSizeBSgroupOwner}]{
        \includegraphics[width=0.47\textwidth]{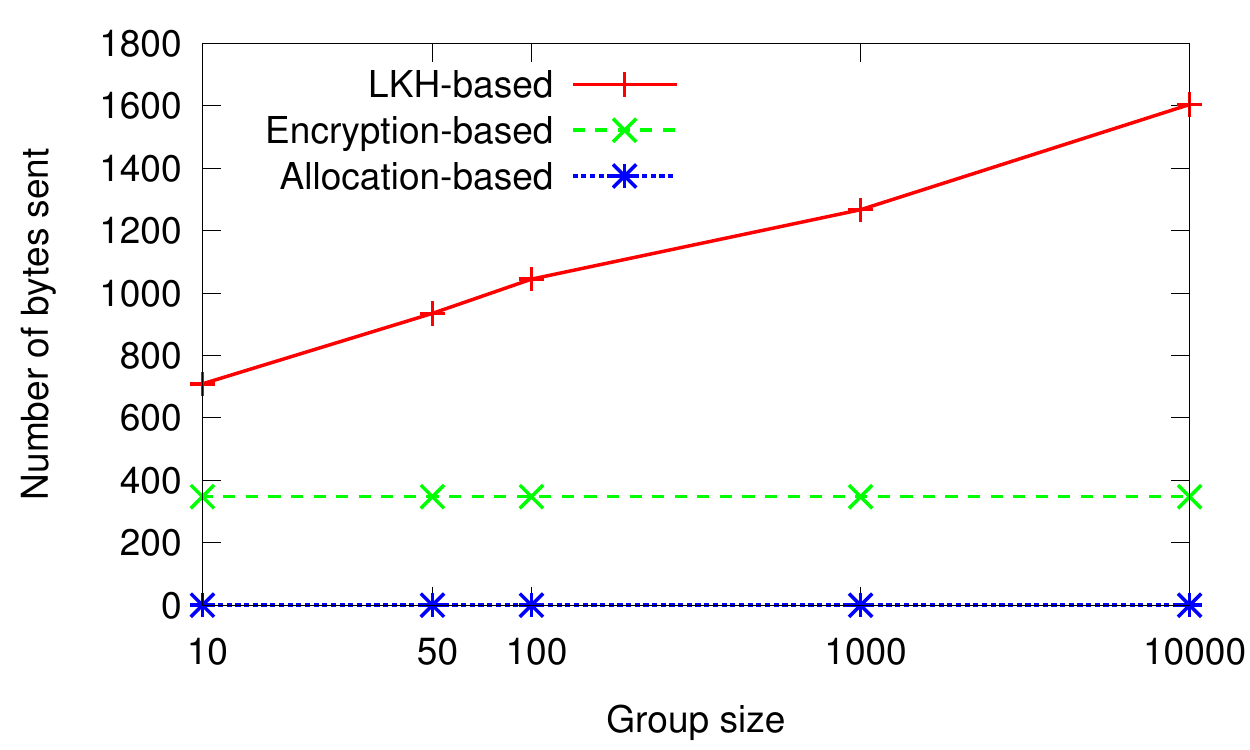}}
\subfigure[Joining user\label{fig:joinTimeBSjoiningUser}]{
        \includegraphics[width=0.47\textwidth]{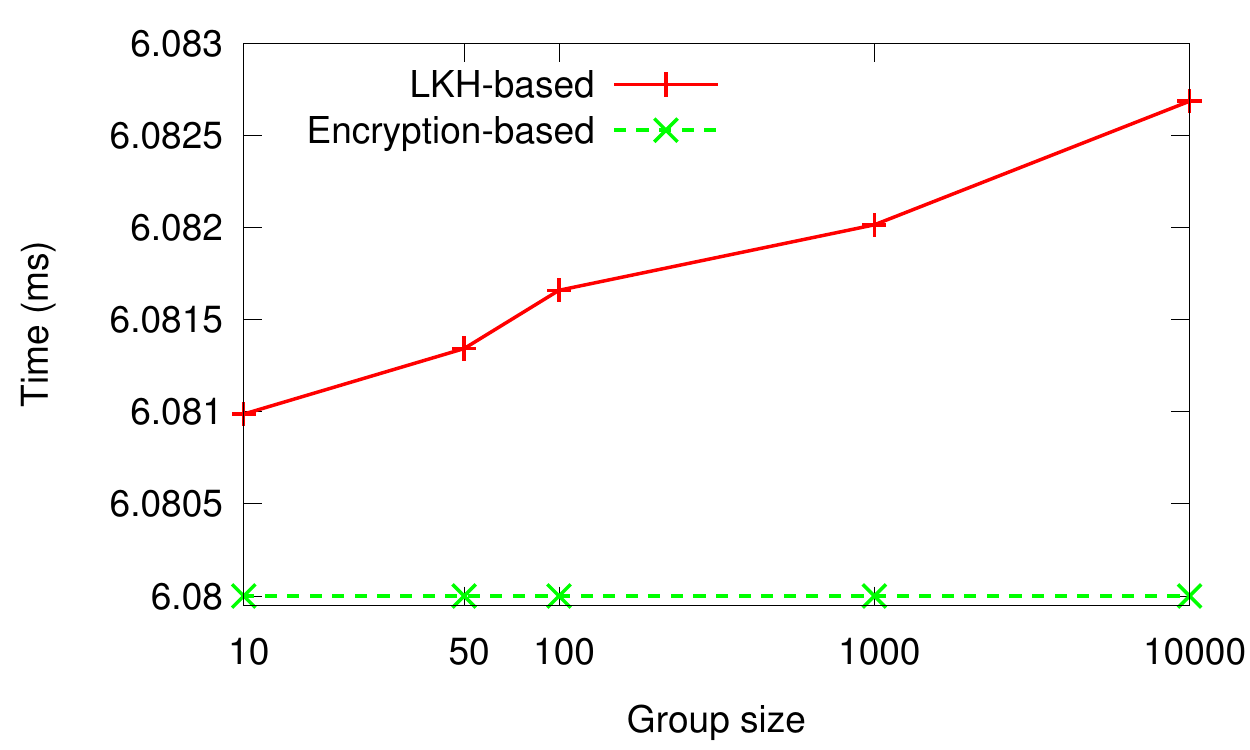}}
\subfigure[Joining user\label{fig:joinSizeBSjoiningUser}]{
        \includegraphics[width=0.47\textwidth]{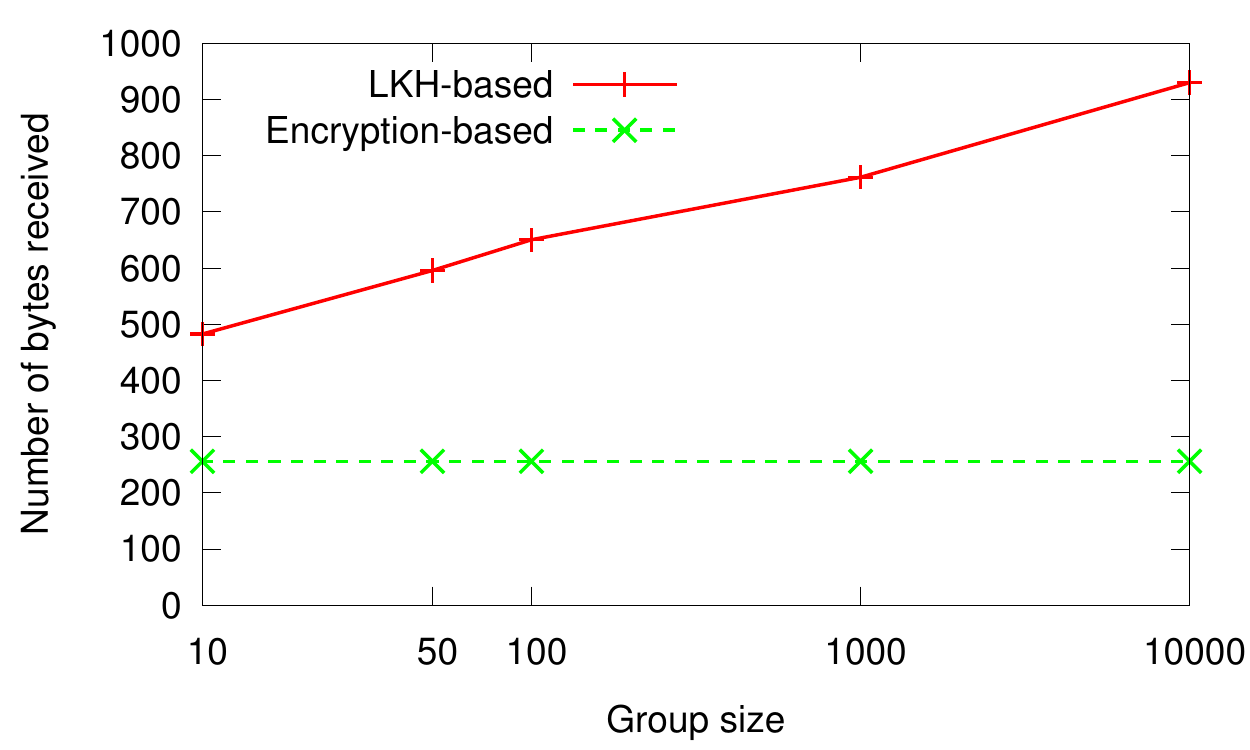}}
\subfigure[Existing members of the group\label{fig:joinTimeBSotherMember}]{
        \includegraphics[width=0.47\textwidth]{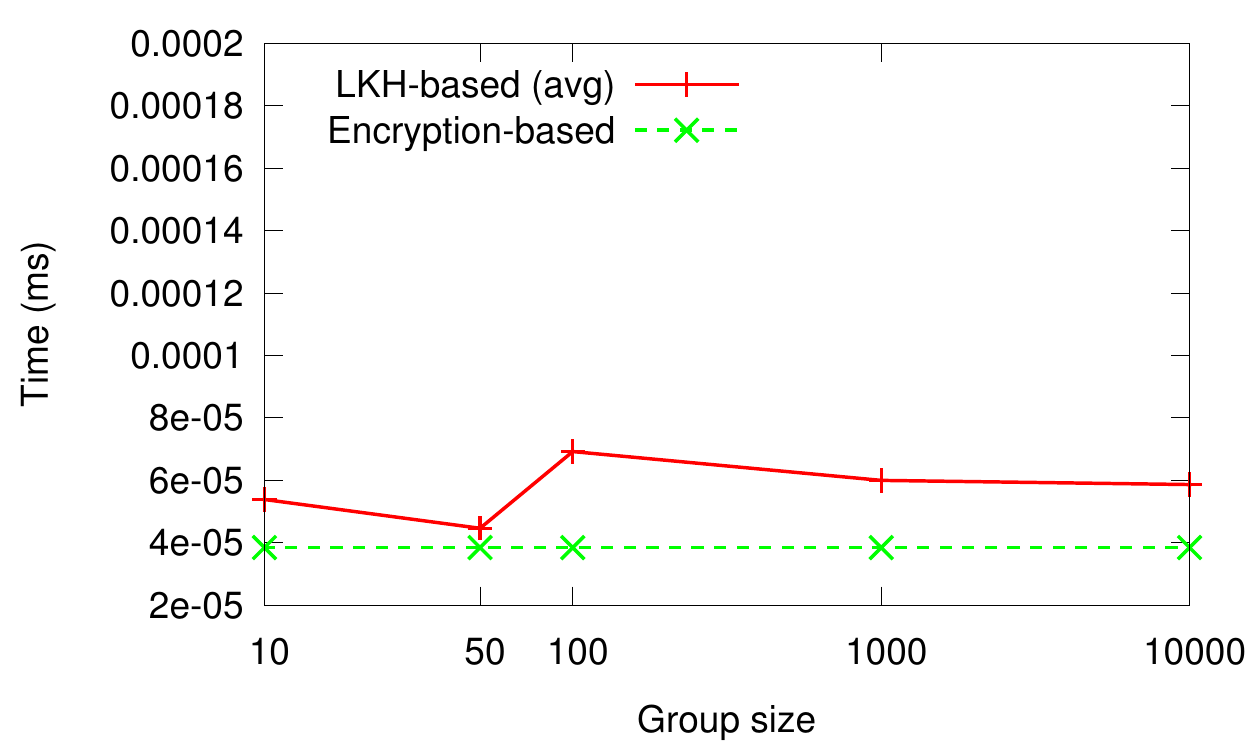}}
\subfigure[Existing members of the group\label{fig:joinSizeBSotherMember}]{
        \includegraphics[width=0.47\textwidth]{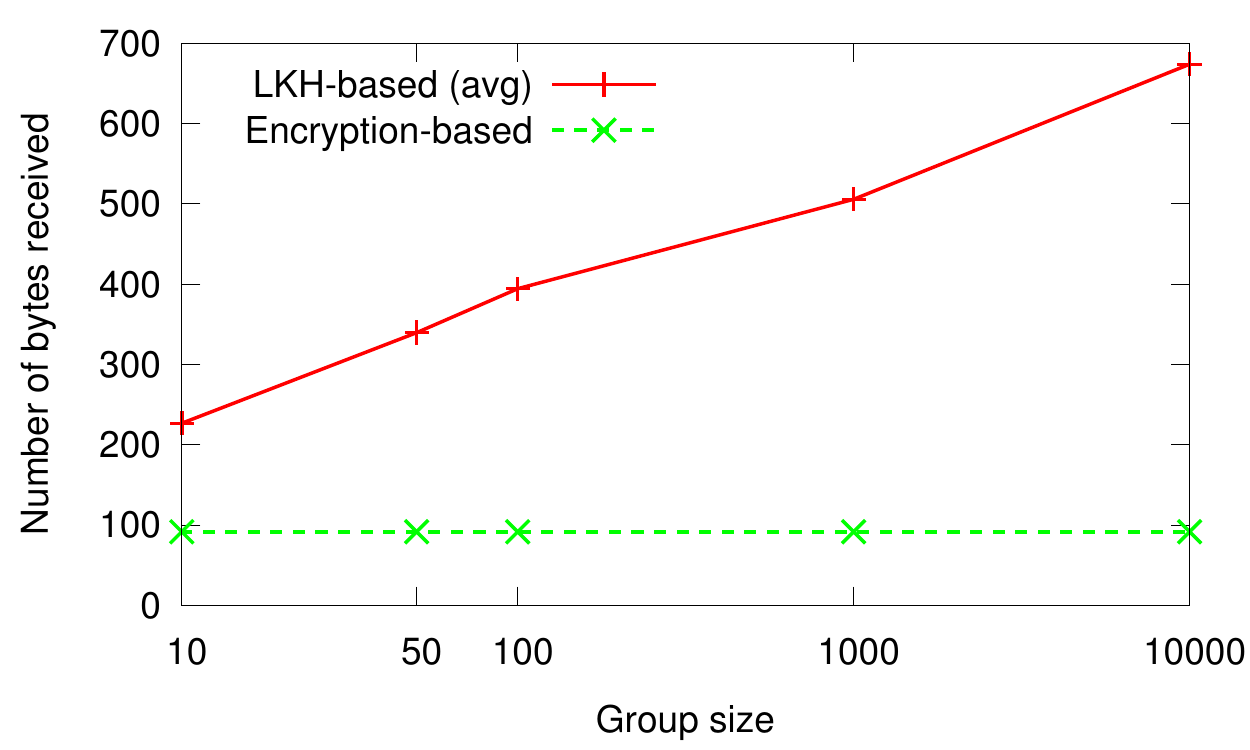}}
\caption[Evaluation of the join with backward secrecy]{Evaluation of the time and the number of bytes spent by the group owner, the joining user, and the existing members for the join with backward secrecy of a user to groups with different numbers of members.\label{fig:joinBWTime}}
\end{figure*}

\paragraph{Join with backward secrecy (G2)} 
The join operation ensures the backward secrecy property when the new member of the group is not allowed to access the contents previously published for the group.
Figure \ref{fig:joinBWTime} shows the average time and the number of bytes spent by the different actors of the DOSN due to the procedure performed to add a user to a group, varying the number of members of such group.

In order to ensure the backward secrecy
in the \aclBased~ and in the \lkhBased~ enforcement models, each time the group owner $o$ adds a new user $a$ to the group $G$,   $o$ also changes the symmetric key of the group $G$ in order to prevent $a$ from accessing the contents previously shared with the existing members of $G$. 
%Figure \ref{fig:joinBWTime} shows the average time and the number of bytes spent by the different actors of the DOSN due to the procedure performed to add a user to a group, varying the number of members of such group.   
%\st{The \aclBased~ enforcement model enables join of a user with backward secrecy. In such a case, the group owner creates a new symmetric group key in order to protect contents already published in the group and sends it to both the joining user and the existing members of the group.} 

If the \aclBased~ enforcement model is adopted, the new group key is securely distributed to the new user by asymmetrically encrypting it using the new user's public key, while the old $n$ members of the group receive the new group key symmetrically encrypted utilizing the old symmetric group key. 
As a result, the group owner performs one symmetric and one asymmetric encryption operations and sends a total of 2 encrypted symmetric keys.
As shown by the Figure \ref{fig:joinTimeBSgroupOwner}, the time needed by the group owner in order to add a user to the group is constant and does not depend on the number of group members. The number of bytes sent by the group owner is also constant (about 350 bytes) and it is shown in Figure \ref{fig:joinSizeBSgroupOwner}. 
In particular, an asymmetrically encrypted packet of 256 bytes is sent from the group owner to the joining user in order to communicate the new symmetric group key, while the remaining bytes are necessary to distribute the new symmetric group key \textcolor{black}{(and other information)} to the old members (through symmetric encryption).
The joining user, instead, has to retrieve the new symmetric group key and to decrypt it by using her/his private key. 
For this reason, the number bytes received by the joining user (see Figure \ref{fig:joinSizeBSjoiningUser}) is equal to the size of the asymmetric public key (2048 bits) and the most of the time is spent for the asymmetric decryption (see Figure \ref{fig:joinTimeBSjoiningUser}).
Similarly, \textcolor{black}{each of } the already existing members of the group have to retrieve the encrypted message containing the new  symmetric group key and to decrypt it by using the old symmetric group key (see Figure \ref{fig:joinTimeBSotherMember} and \ref{fig:joinSizeBSotherMember}).

In the case of \lkhBased~ approach, the group owner $o$ creates a new key for the group, updates the LKH tree as described in Section \ref{sec:privacyControl}, and properly distributes the new keys to the joining user and to the already existing group members.
In particular, the group owner $o$ creates a new leaf node of the LKH tree $KT(d,h,G)$ to represent $a$, and changes 
%\st{In order to ensure backward secrecy,} 
the symmetric keys of the nodes on the path $P_a$ from the root (included) of the key tree to the new leaf representing $a$. 
%\textcolor{red}{\st{Since the maximum height of the key tree is equal to $h$, the sequence of nodes on the path $P_a$ is at most equal to $h+1$. Instead, in the case of a perfectly balanced key tree on the $n$ members of the group $G$, the number of nodes on the path $P_a$ will be proportional to $log_d(n)$.} }
%\st{are updated, i.e., the related symmetric keys are changed.} 
\textcolor{black}{Hence, the number of nodes on $P_a$ is, at most, equal to $h+1$, where $h$ is the height of the LKH tree. Since we assumed the LKH tree balanced, we have that  $h =\lceil	 log_d(n)\rceil	$.}

\textcolor{black}{To communicate the nodes on $P_a$ to $a$, the group owner encrypts the relevant node contents 
%\st{(including the symmetric key paired with the node)}
and uses the Private Mailbox of $a$.
In particular, the groups owner asymmetrically encrypts %the symmetric individual key of $a$ 
the content paired with the leaf representing $a$ (including the symmetric individual key of $a$) with the public key of $a$, and symmetrically encrypts the other $h$ 
%\st{$\approx log_d(n)$}
nodes of $KT (d, h, G)$ on $P_a$ with the symmetric individual key of $a$.} 
As a result, the group owners performs $O(h)$ encryption operations with the symmetric schema and only one encryption operation with the asymmetric schema.
In addition, to communicate to the previously existing group members the $h$ keys refreshed along the path %\st{of the joining user}
\textcolor{black}{$P_a$}, $o$ encrypts each node 
on this path with its old symmetric key, 
%\st{(which are at most equal to $O(h))$} 
\textcolor{black}{thus performing at most $O(h)$ symmetric encryption operations.}
Hence, the total number of %\st{keys}
\textcolor{black}{nodes}
encrypted by the group owner with a symmetric schema is equals $O(2\cdot h)$, while the total number of keys encrypted with an asymmetric schema is equal to 1. Figure \ref{fig:joinTimeBSgroupOwner} shows the amount of time required by group owner in order to set up the new symmetric keys and to perform such encryption operations.
\textcolor{black}{nodes} encrypted using symmetric schema and 1 
%\st{symmetric key} 
\textcolor{black}{node} protected by using asymmetric encryption schema based on 2048 bits key length.

\textcolor{black}{The joining user $a$ receives the $h+1$ encrypted nodes along $P_a$ of the key tree and executes $O(h)$ symmetric decryption operations and one asymmetric decryption operation on them.} As shown by Figure \ref{fig:joinTimeBSjoiningUser}, the majority of time is spent on the asymmetric decryption operation. Instead, the sum of bytes received by the joining user is due to both the asymmetrically encrypted packet of 256 bytes and the symmetrically encrypted $h$ nodes along the path \textcolor{black}{$P_a$} %\st{of the joining user }
%about half of those sent by the group owner 
(see Figure \ref{fig:joinSizeBSjoiningUser}).

The existing members of the group retrieve 
%\st{also the symmetric keys} 
the updated \textcolor{black}{nodes along $P_a$},
%\st{the path of the joining user}, \st{which consist of  the new symmetric keys,} 
each encrypted with the corresponding old symmetric key. Each member of the group has to decrypt only the involved  \textcolor{black}{nodes} along its path. 
Figure \ref{fig:joinTimeBSotherMember} shows the %\st{overall}
\textcolor{black}{average} time required by the existing members of the group in order to decrypt the 
%\st{new symmetric keys} 
\textcolor{black}{updated nodes of their interest}. 
The members of the group have to decrypt only the involved keys along their path (i.e., at most $h$ keys in the case the member is %\st{located on the same sub-tree}
\textcolor{black}{paired with a sibling node} of the one paired to the joining user). As shown by Figure \ref{fig:joinSizeBSotherMember}, the sum of bytes received by a group member is less than those received by the joining user, i.e., equals to $O(h)$ %\st{symmetric keys}
\textcolor{black}{nodes} encrypted by using symmetric schema.

In the \policyBased~ enforcement model the group owner has to update the membership information by considering the new user in the group. As a result, a new content that will be published on the group will be paired to a privacy rule that considers also the identity of the new joining user. As shown by the Figure \ref{fig:joinBWTime}, the join of a user with the \policyBased~ enforcement model involves only the group owner and it does not require expensive operation because the group owner update only local membership information about the group.
\begin{figure*}[t] 
\centering 
\subfigure[Group owner\label{fig:groupOwnerJoinWithoutBSTime}]{
        \includegraphics[width=0.47\textwidth]{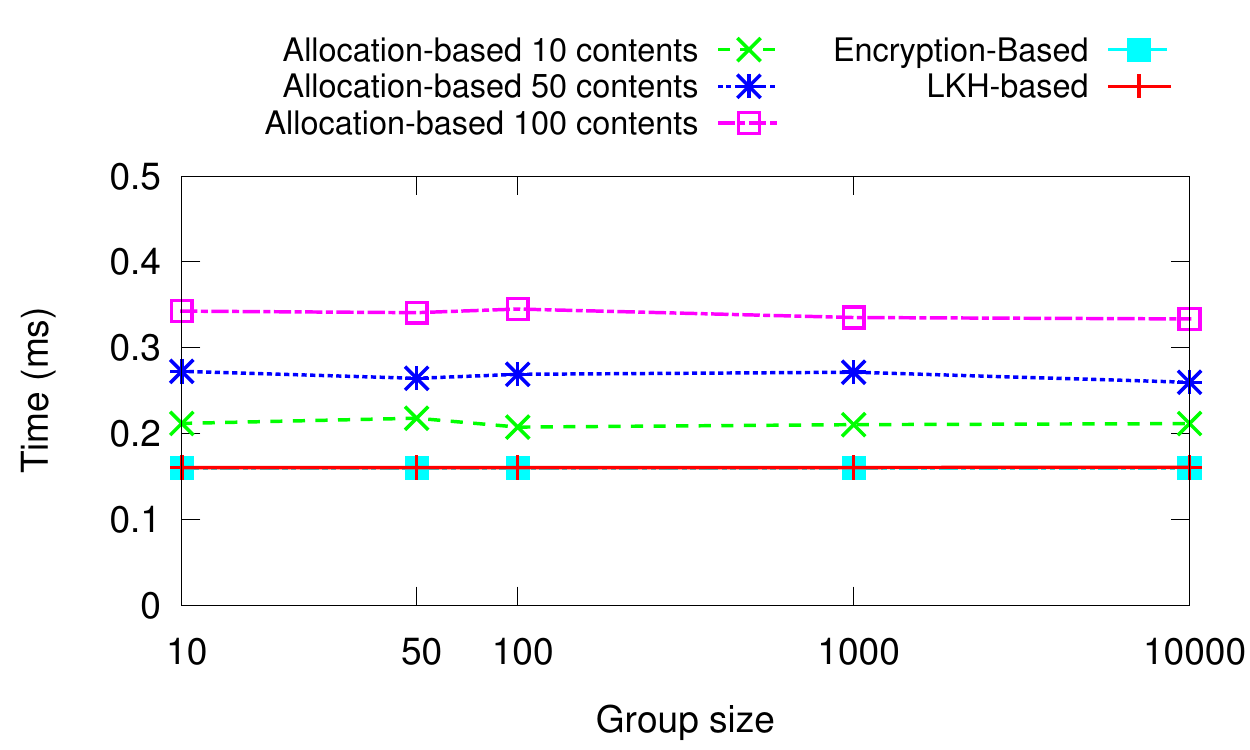}}
\subfigure[Group owner\label{fig:groupOwnerJoinWithoutBSSize}]{
        \includegraphics[width=0.47\textwidth]{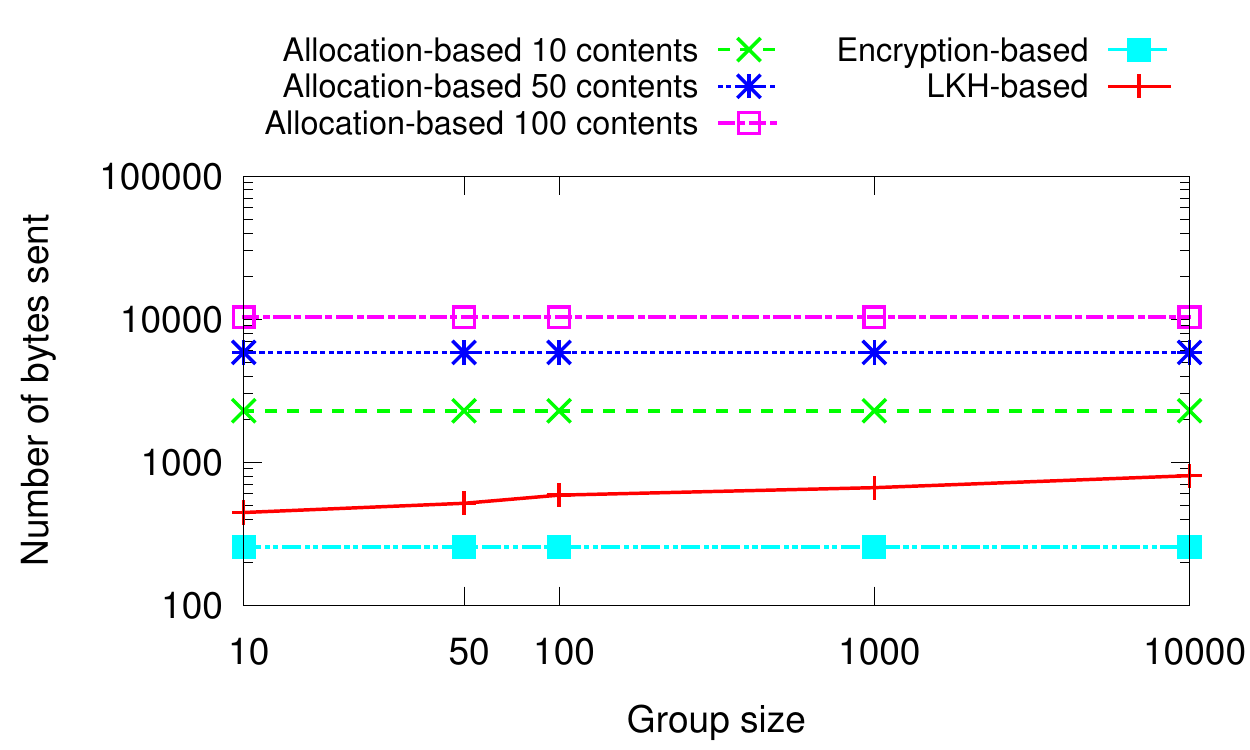}}
\subfigure[Joining user\label{fig:joiningUserJoinWithoutBSTime}]{
        \includegraphics[width=0.47\textwidth]{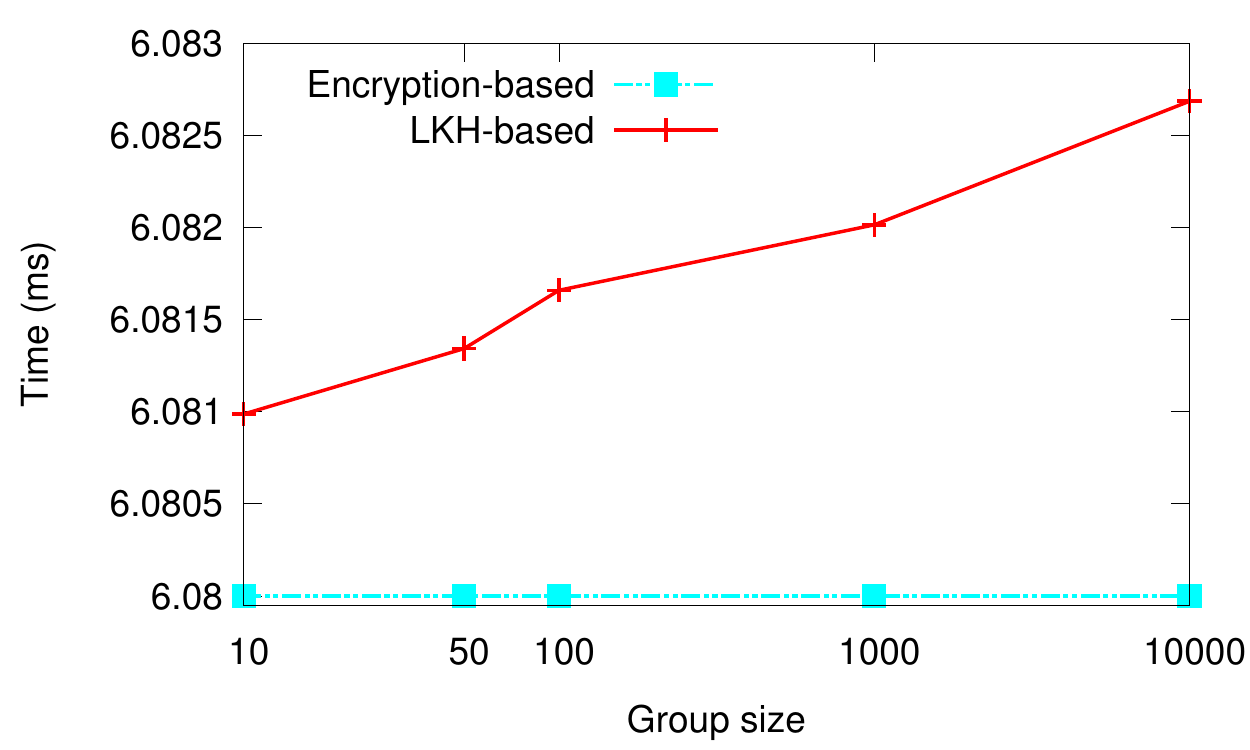}}
\subfigure[Joining user\label{fig:joiningUserJoinWithoutBSSize}]{
        \includegraphics[width=0.47\textwidth]{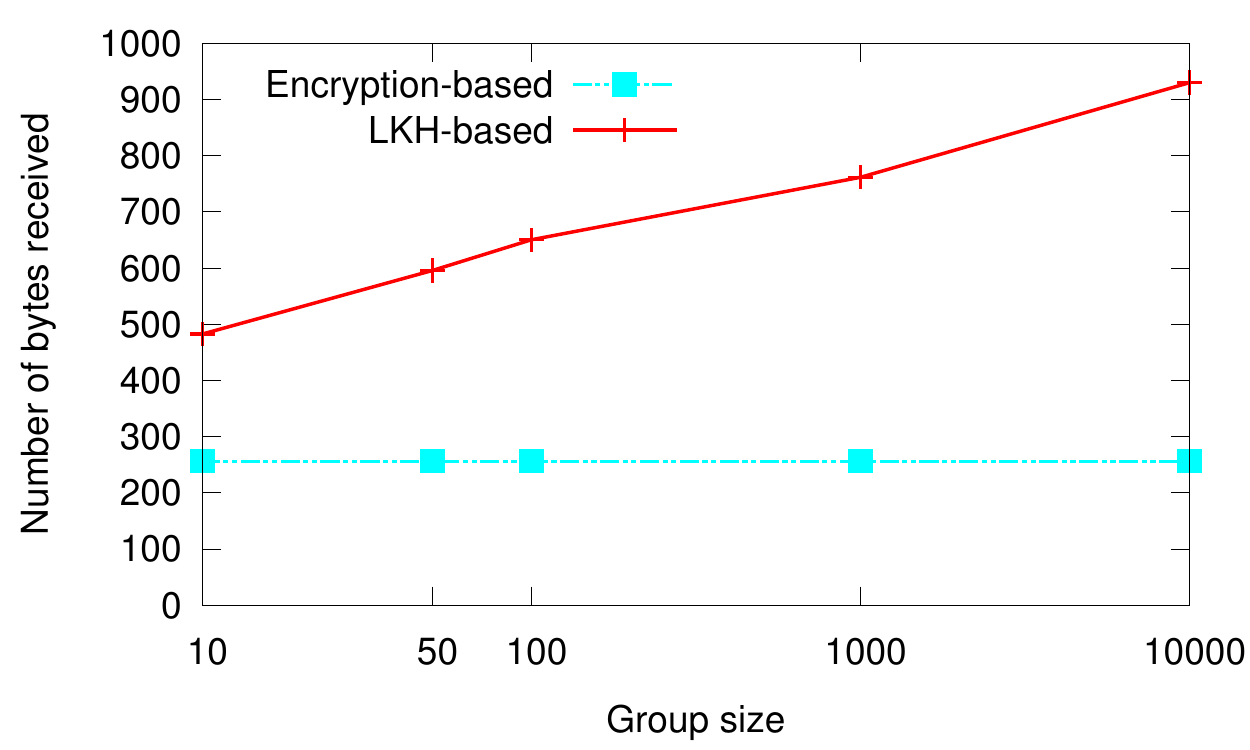}}
%\subfigure[Existing members of the group\label{fig:otherMemberJoinWithoutBSTime}]{
%        \includegraphics[width=0.47\textwidth]{images/otherMemberJoinWithoutBSTime.pdf}}
%\subfigure[Existing members of the group\label{fig:otherMemberJoinWithoutBSSize}]{
%        \includegraphics[width=0.47\textwidth]{images/otherMemberJoinWithoutBSSize.pdf}}
\caption[Evaluation of the join without backward secrecy]{Evaluation of the time and the number of bytes spent by the group owner and the joining user for the join without backward secrecy of a user to groups with different numbers of members.\label{fig:joinNBWTime}}
\end{figure*}

\paragraph{Join without backward secrecy (G3, G4)} When the join operation does not support backward secrecy, a joining user can access all the contents previously published in the group. %Table \ref{tab:joinNBSComparison} summarizes the analytic cost of each approach from the perspective of the group owner, the joining user, and the existing members of the group.
The \lkhBased~ and the \aclBased~ enforcement models can easily provide a join without backward secrecy 
\textcolor{black}{by simply communicating the current symmetric group key to the joining user $a$.}
%\st{, i.e.,  by avoiding to change the group key when a new user joins the group. }
To this aim, with the \lkhBased~ approach, the group owner $o$ creates a new leaf node of the key tree $KT(d,h,G)$ for $a$. %\st{the new user}. 
The symmetric keys of all the nodes of the key tree, including the root, remain the same.
In order to send to the joining user the nodes of the key tree she/he needs, \textcolor{black}{the same approach previously described for the join with backward secrecy is adopted.}
\textcolor{black}{In particular, the leaf corresponding to  the joining user (embedding the individual symmetric key of such user) is asymmetrically encrypted with her/his public key,} while 
%\st{the symmetric keys of}
the nodes on the path \textcolor{black}{from the father of the leaf paired to the joining user to the root} (including the group key) are symmetrically encrypted with such individual symmetric key.
%Eventually, a new intermediate node is created because of the increases of the height of the key tree during the join of a user. The new intermediate node is communicated to the existing members of the group by using the symmetric keys already shared with them on the $d$ children nodes. 
As a result, 
the total number of %\st{keys}
\textcolor{black}{nodes}  sent by the group owner is \textcolor{black}{at most $h+1$}.
%\st{where $h$ $=log_d(n)$ is the maximum height of the key tree.}}
In particular, 
the group owner executes $O(h)$ encryption operations with the symmetric schema and only one encryption operation with the asymmetric schema. Figure \ref{fig:groupOwnerJoinWithoutBSTime} shows the time required by group owner to %\st{in order to set up the new symmetric keys} 
%\st{and to perform such encryption operations.}
add a new user to a group.
The cost is mainly due to the execution of the asymmetric encryption operation.
The message created by the group owner for the joining user contains $O(h)$ nodes protected with symmetric encryption and a leaf node protected by using asymmetric encryption based on 2048 key length. As shown by Figure \ref{fig:groupOwnerJoinWithoutBSSize}, the number of bytes sent by the group  owner 
%\st{does not increase linearly with the number $n$ of group members.}
is logarithmic with respect the number of group members.
%because the message involves the communication of $O(log_d(n))$ symmetric keys.}
%\st{Since the message created by the group owner for the joining user contains $O(log_d(n))$ \st{$O(h)$} symmetric keys,  Figure \ref{fig:groupOwnerJoinWithoutBSSize} shows that the number of bytes sent by the group increases logarithmically with the number $n$ of group members.}
%\st{owner is logarithmic with respect the number of group members because the message involves the communication of $O(log_d(n))$ symmetric keys.} 
The joining user receives the $h+1$ nodes along her/his path, hence executing  $O(h)$  symmetric decryption operations, 
%\st{which is proportional to the height of the tree ($h$)} 
while the number of asymmetric decryption is equals to one. As shown by Figure  \ref{fig:joiningUserJoinWithoutBSTime}, 
the majority of time is spent to perform the asymmetric decryption operation (which costs considerably more that the corresponding encryption operation), while the sum of bytes received by the joining user, shown in Figure \ref{fig:joiningUserJoinWithoutBSSize}, is equal to the sum of bytes sent by the group owner (see Figure \ref{fig:groupOwnerJoinWithoutBSSize}).

%\st{As shown by Figure} \ref{fig:joinNBWTime},
We observe that, \textcolor{black}{supposing that a reorganization of the key tree is not required to insert the new user's leaf,} the execution of the join operation without backward secrecy does not involve the other members of the group because all the symmetric keys of the key tree remain the same.

Instead, the \aclBased~ enforcement model requires only the distribution to the joining user of the symmetric group key, which, to maintain its confidentiality, is asymmetrically encrypted before being transferred by using the joining users's public key. As a result, the group owner perform only one asymmetric encryption operation and sends a single encrypted message of 256 bytes, which consists of the symmetric group key encrypted by using  a public key of 2048 bits (see Figure \ref{fig:groupOwnerJoinWithoutBSTime} and \ref{fig:groupOwnerJoinWithoutBSSize}). 
The joining user retrieves such message containing the encrypted group key (see Figure \ref{fig:joiningUserJoinWithoutBSSize}) and decrypts the symmetric group key by using her/his private key (see Figure \ref{fig:joiningUserJoinWithoutBSTime}). 
%\st{As in the previous case, }
The join of a user does not introduce any cost for the other members of the group. %\st{ if the backward secrecy is not guaranteed.} For this reason, the amount of the processing time required for the join of a user without backward secrecy does not depend on the number of group members or on the number of contents published in the group. 
%\st{involved only asymmetric operations}. 
%The number of bytes exchanged by the involved users is also constant and almost equals to the length of the symmetric key.

Finally, in order to add a user to a group with the \policyBased~ enforcement model, the group owner adds the joining user to the list of the identities of the current group members to enable the user to see the contents that will be published on the group from that moment on. 
In addition, 
%\st{the group owner ensures a join without the backward secrecy property by modifying (or by overwriting)}
in order to allow the joining user to access the contents that have already been published in the group, the group owner properly modifies also
 the privacy rules paired to the contents already published in the group. %\st{This can be done by exploiting the built-in functionalities that the privacy policy language provided for this purpose. A first approach is to change the privacy policies in the privacy policy set by adding the identity of the new user. However, the time required for such modifications can be quite high because the number of privacy policies is typically equals to the number of contents published in the group. A second more efficient approach is to create a new privacy policy for the joining user, that grants access to all the contents published in the group. As a result, we chose to use the latter approach by allowing the group owner to create a new privacy policy for the joining user.} 
Hence, in the \policyBased~ enforcement model, the execution of the join operation without backward secrecy involves the group owner only.
%\st{that adds the member to the group and creates the privacy policy for him.}
Figure \ref{fig:groupOwnerJoinWithoutBSTime} shows the time spent by the group owner in order to add the joining user to the 
%\st{group and to create the privacy policy for him,} 
current group member list and to update the policies of the contents already published in the group,
while Figure \ref{fig:groupOwnerJoinWithoutBSSize} shows the number of bytes (in log scale) sent by the group owner as a result of the join operations.
The plot, confirms that the time and the number of bytes sent by the group owner depends only on the number of contents published in the group, because the 
%\st{privacy policy for the new user}
group owner has to grant to the joining user the access to all the contents already published in the group. 
For the join without backward secrecy, the \policyBased~ enforcement model does not introduce any cost for the joining user and for the existing members of the group.

\subsection{Group Leave}
\label{sec:groupLeave}
In this section, we evaluate the cost for removing a member from a group (leave operation) varying the number of members and contents.
The leave operation always guarantees the forward secrecy property, i.e., the evicted member will not be able to access the contents that will be published in the group after the execution of the leave operation.

\begin{figure*}[tbp] 
\centering 
\subfigure[Group owner\label{fig:groupOwnerLeaveWithoutBSTime}]{
        \includegraphics[width=0.47\textwidth]{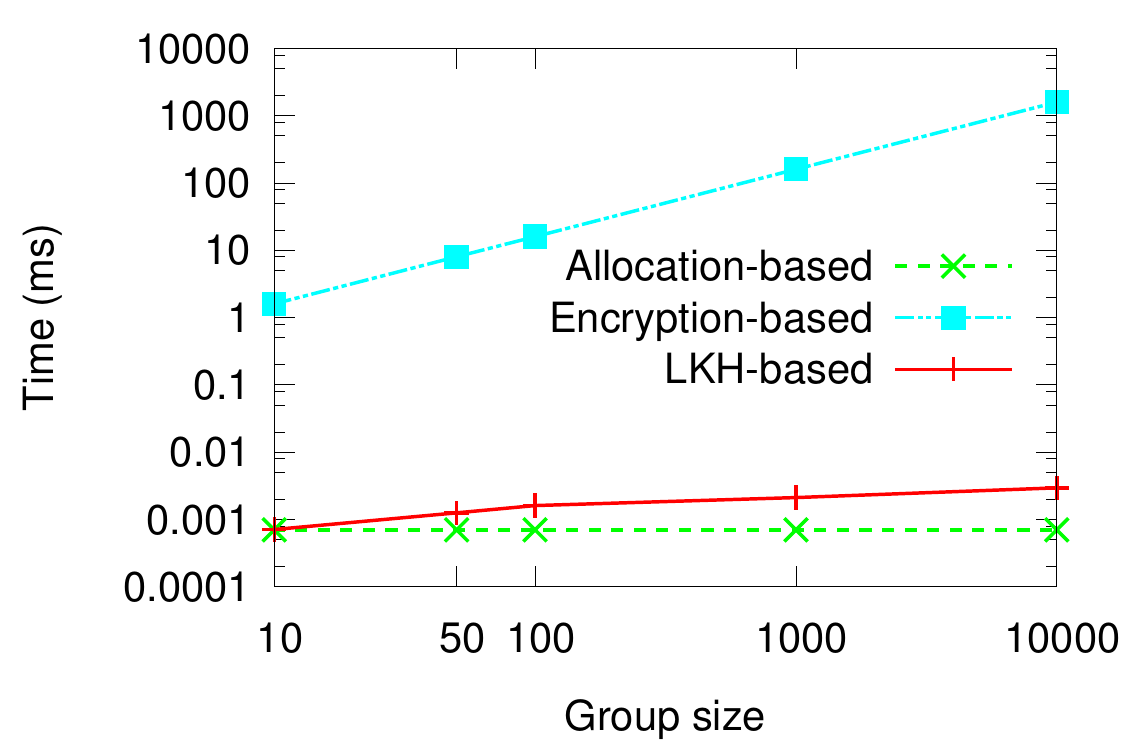}
        }
\subfigure[Existing members of the group\label{fig:otherMemberLeaveWithoutBSTime}]{
        \includegraphics[width=0.47\textwidth]{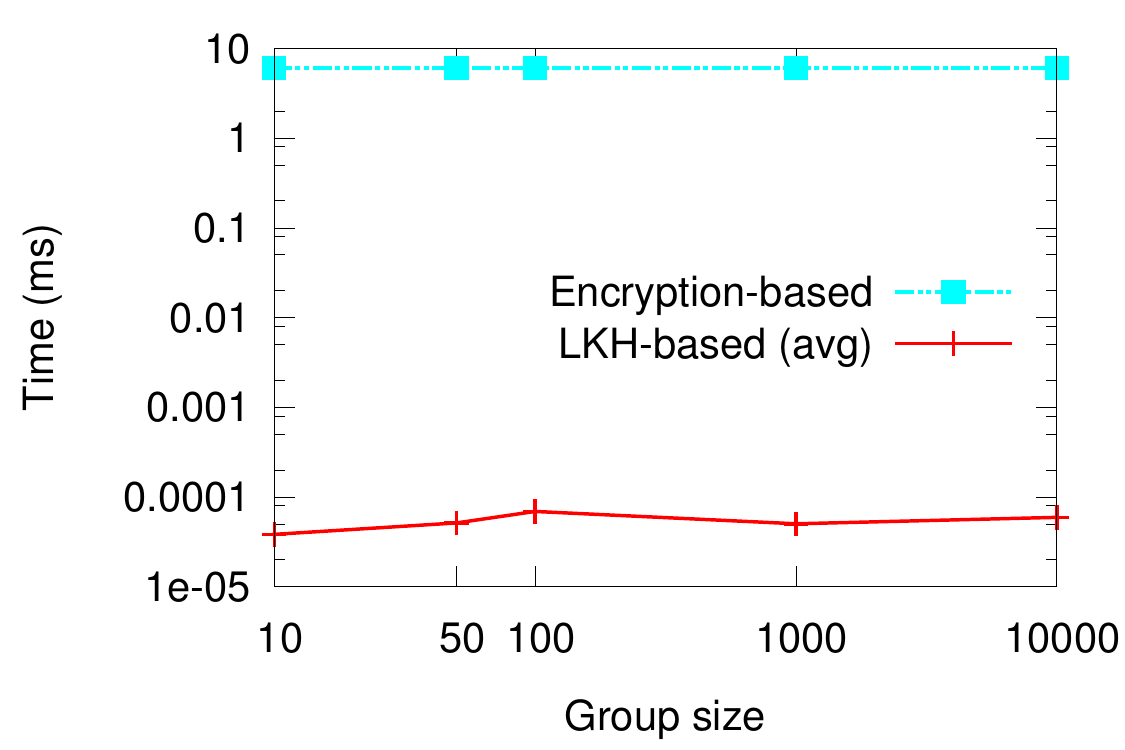}
        }
\subfigure[ Group owner\label{fig:groupOwnerLeaveWithoutBSSize}]{
        \includegraphics[width=0.47\textwidth]{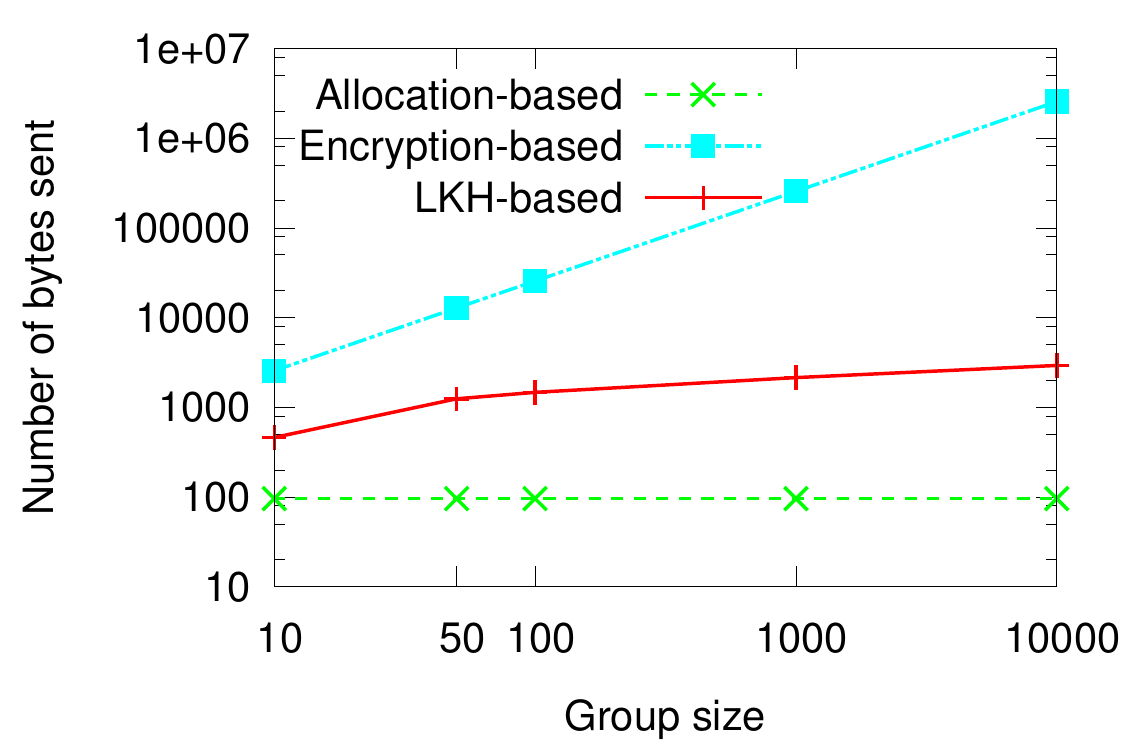}
       }
\subfigure[Existing members of the group\label{fig:otherMemberLeaveWithoutBSSize}]{
        \includegraphics[width=0.47\textwidth]{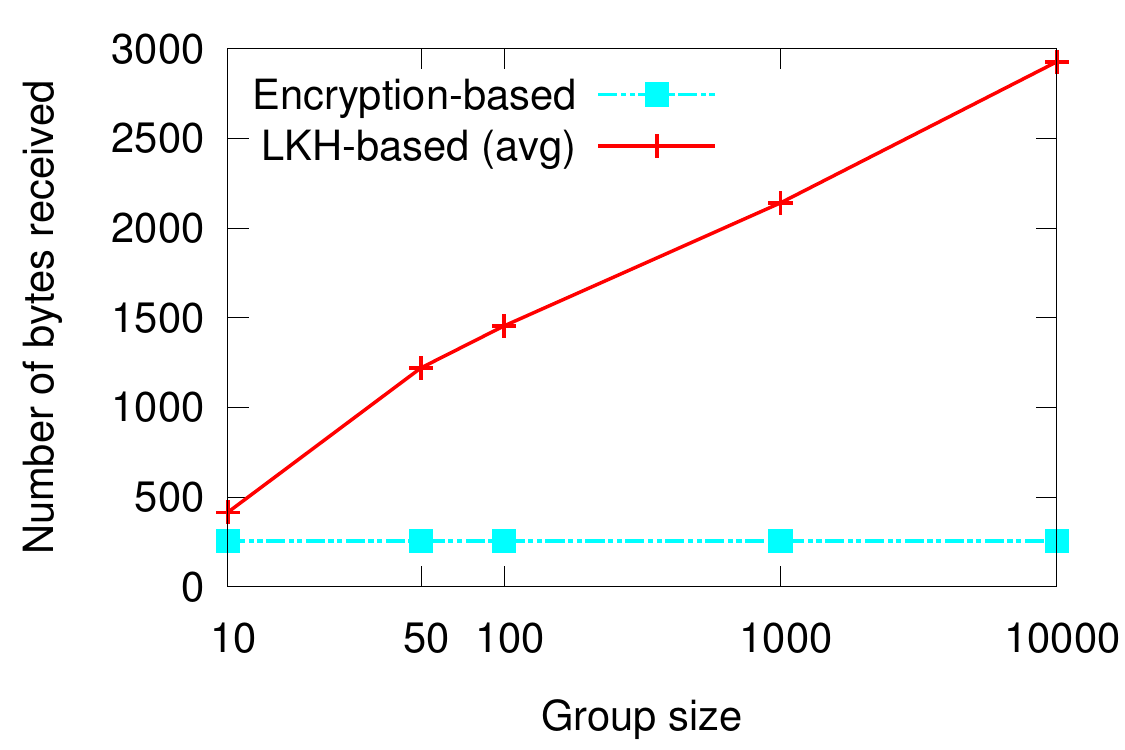}
        }
\caption[Evaluation of the leave without backward secrecy]{Evaluation of the time and the number of bytes spent by the group owner and the existing members for the leave without backward secrecy of a user from groups with different numbers of members.\label{fig:leaveNBWTime}}
\end{figure*}

\paragraph{Leave without backward secrecy (G2,G4)}\label{sec:leaveWithoutBS}
When the leave operation does not guarantee backward secrecy, the member removed from the group is still able to access the contents published in the group before her/his leave. 

In order to remove a member from the group $G$ with the \lkhBased~ enforcement model, the group owner $o$ deletes from the key tree $KT (d, h, G)$ stored on her/his local peer the leaf node corresponding to the removed user, $r$. To guarantee the forward secrecy property, the group owner refreshes the symmetric keys on the path $P_r$ from the father of the removed leaf to the root (including the symmetric group key). Since the height of the key tree is equal to $h$, the number of nodes on the path $P_r$ is, at most, equal to $h$.
The 
%\st{new symmetric key paired with the}
\textcolor{black}{updated} father node of the removed leaf is symmetrically encrypted exploiting the (at most) $d-1$ keys paired with its remaining children nodes, i.e., the siblings of the removed leaf, in order to securely communicate it to the group members paired to such nodes.
Recursively following this approach, i.e., exploiting the symmetric keys paired with the $d$ children nodes, the new %\st{symmetric key}
\textcolor{black}{version} of each node 
on the path $P_r$ is symmetrically encrypted and securely communicated to the related users. %left in the group 
%by exploiting the symmetric keys paired with the children nodes.
%, which are already shared with the current members. Indeed, each new symmetric key on the path is encrypted with the symmetric key of each child node. 
As a result, the number of symmetric keys created by the group owner is equals to $O(h)$ while the number of symmetric encryption operations is at most $O(d\cdot h)$. 
Hence, the group owner creates one message, the leave notification message, which consists of at most $O(d\cdot h)$ encrypted %\st{symmetric keys}
\textcolor{black}{nodes} 
and embeds them on the Group Message List in order to send it to the remaining group members.
Indeed, Figure \ref{fig:groupOwnerLeaveWithoutBSTime} shows that the time (in log scale) required by the group owner in order to \textcolor{black}{refresh the symmetric keys and encrypt the related nodes} does not increase significantly with the size of the group.
Similarly, Figure \ref{fig:groupOwnerLeaveWithoutBSSize} 
clearly indicates that the amount of bytes sent by the group owner is also logarithmic with respect the size of the group and it is at most equals to $O(d\cdot h)$.
Each of the existing members of the group has to retrieve the %\st{encrypted symmetric keys} 
\textcolor{black}{leave notification message}
and decrypt only the new symmetric keys on her/his path, i.e., at most $O(h)$ symmetric keys. As shown by Figure \ref{fig:otherMemberLeaveWithoutBSTime}, the time spent by a member of the group in order to decrypt the symmetric keys along her/his path is negligible and also depends on the position of the removed user in the key tree. 
As regards the average number of bytes received by the existing members of the group (see Figure \ref{fig:otherMemberLeaveWithoutBSSize}), \textcolor{black}{each member reads the full leave notification message}.
%Indeed, all the symmetric keys encrypted during the leave of a user are securely stored in a message, which must be retrieved by all group's members from the Group Message List.}
Finally, the 
%\st{removal} 
leave operation 
%\st{of a user only}
does not introduce any cost for the removed user 
%\st{ affects the group owner and the member of the group}
and it does not require the use of asymmetric encryption operation.

When the \aclBased~ enforcement model is used, in order to implement the leave operation without backward secrecy, the group owner updates the symmetric group key and communicates the new group key to the members left in the group to protect future contents published in the group. However, the contents already published in the group remain encrypted with the old symmetric group key. As a result, the removed user can still access them by using the old symmetric group keys stored on her/his devices. 
The new symmetric group key is asymmetrically encrypted with the individual public key of each member remaining in the group. Hence, the group owner performs a number of asymmetric encryption operations which is linear with respect to the number of group members (i.e., $O(n)$). Finally, the group owner sends an encrypted copy of the new symmetric group key to each member of the group. As shown by Figure \ref{fig:groupOwnerLeaveWithoutBSTime}, the time required by the group owner for the encryption of the new symmetric group key is quite big because it involves only asymmetric encryption operations. In addition, the bytes (see Figure \ref{fig:groupOwnerLeaveWithoutBSSize}) spent by the group owner in order to distribute the new symmetric group key linearly
increase with the number of group members.\\
Each member of the group retrieves only the encrypted message which contains the symmetric group key encrypted with his/her public key (equal to 256 bytes). The member decrypts the new symmetric group key by using her/his private key and stores it on her/his device. As a result, each member left in the group performs only an asymmetric decryption operation and store only one symmetric key. 
Indeed, the amount of the processing time required by the remaining group members for the leave of a user without backward secrecy is constant and does not depend on the number of group members or on the number of contents published in the group (see Figure \ref{fig:otherMemberLeaveWithoutBSTime}). 
The number of bytes retrieved by a group member is also constant and equals to the length of the public key (see Figure \ref{fig:otherMemberLeaveWithoutBSSize}). Instead, the user removed from the group does not perform other any operations to exit from the group.

The \policyBased~ enforcement model can be easily adapted to provide the leave operation without backward secrecy. In order to remove a user from the group, the group owner deletes the affected user from the list of the identities of the group members. 
%\st{stored on her/his device}. 
As a result, the new contents which will be published in the group will be paired to a privacy rule that does not consider the identity of the removed member. In contrast, the privacy rules paired to the contents already published in the group remain the same and they still provide access to the removed user. 
As shown by the experimental results in Figure \ref{fig:groupOwnerLeaveWithoutBSTime} the time required by the group owner in order to request the removal of a user from the list of the identities of the members of a group is negligible. 
%\st{Indeed, the leave operation takes a very short amount of time because the identity of a user has a constant size.}
In fact, Figure \ref{fig:groupOwnerLeaveWithoutBSSize} shows that the number of bytes sent 
by the group owner is constant and it does not depend on the number of users.
%\st{the message or contents in the group because the removed user is not cancelled from the policies paired with old data since it must be still able to access them. }
Finally, adopting the \policyBased~ enforcement model,  the leave without backward secrecy does not introduce any cost for the %\st{joining user and for the}
existing members of a group.

\begin{figure*}[tbp] 
\centering 
\subfigure[Group owner\label{fig:groupOwnerLeaveWithBSTime}]{
        \includegraphics[width=0.48\textwidth]{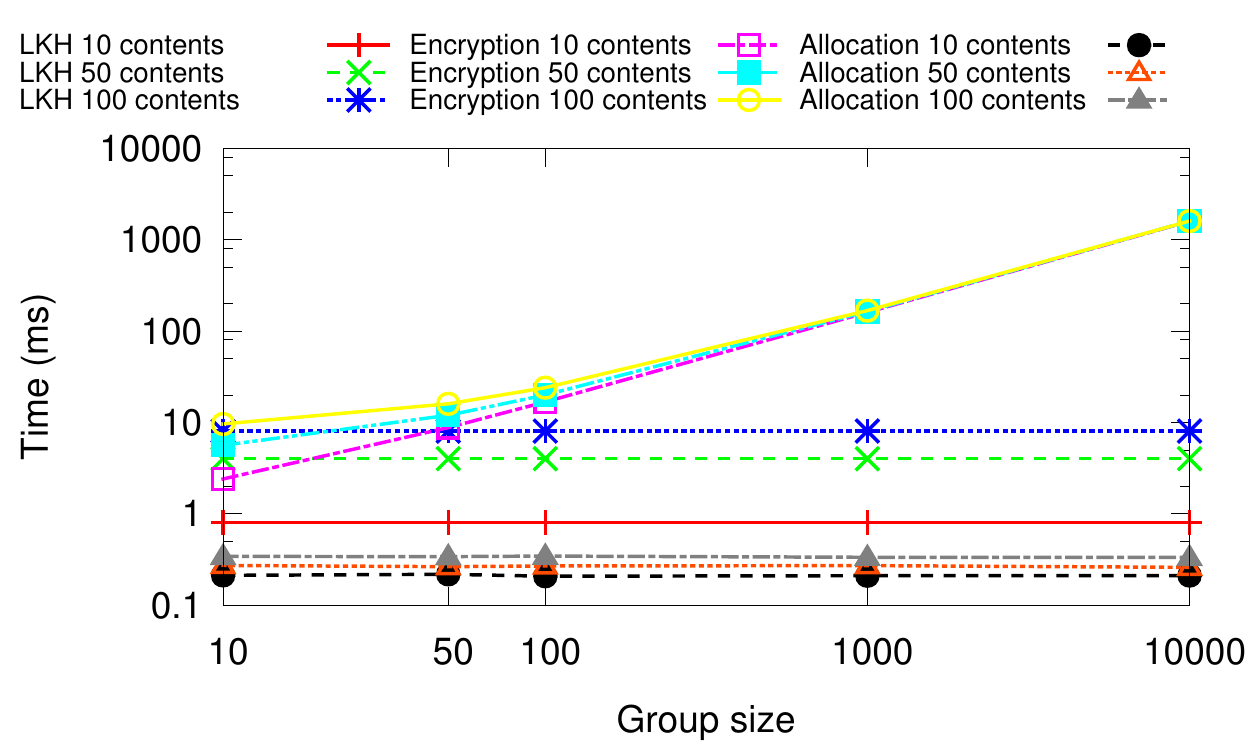}}      
\subfigure[Existing members of the group\label{fig:otherMemberLeaveWithBSTime}]{
        \includegraphics[width=0.48\textwidth]{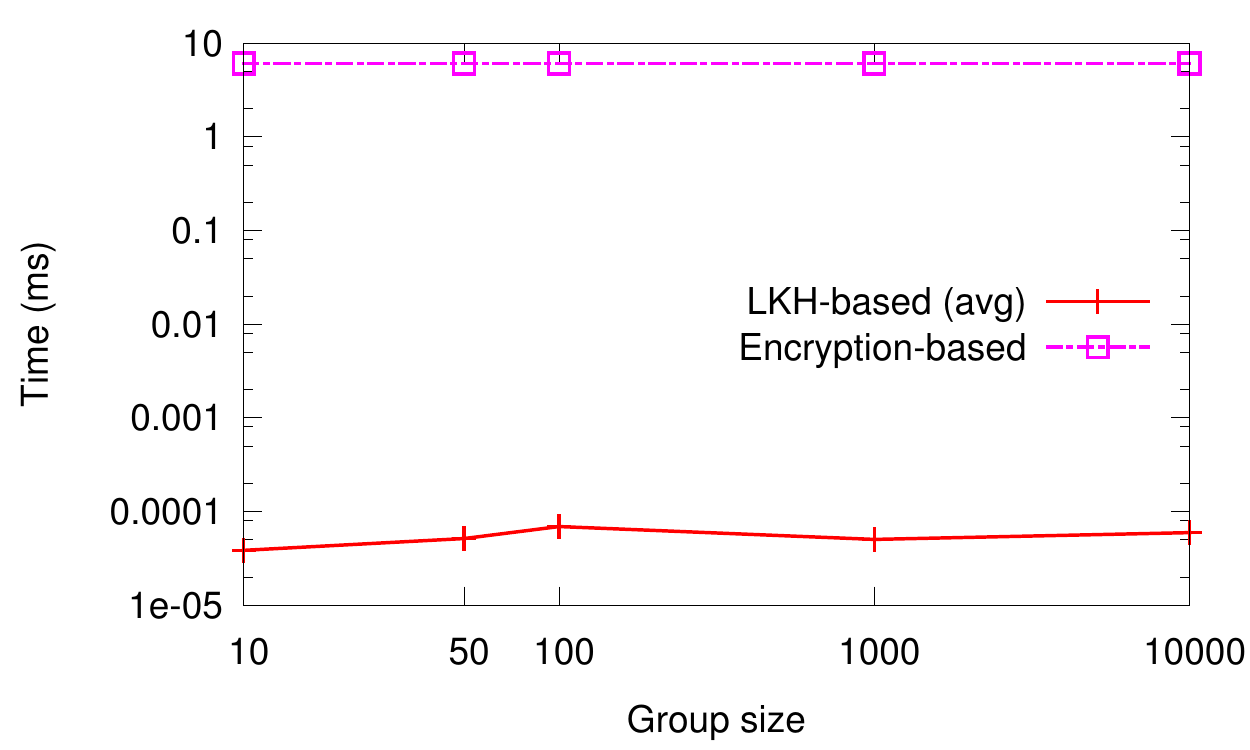}}
\subfigure[Group owner\label{fig:groupOwnerLeaveWithBSSize}]{
        \includegraphics[width=0.47\textwidth]{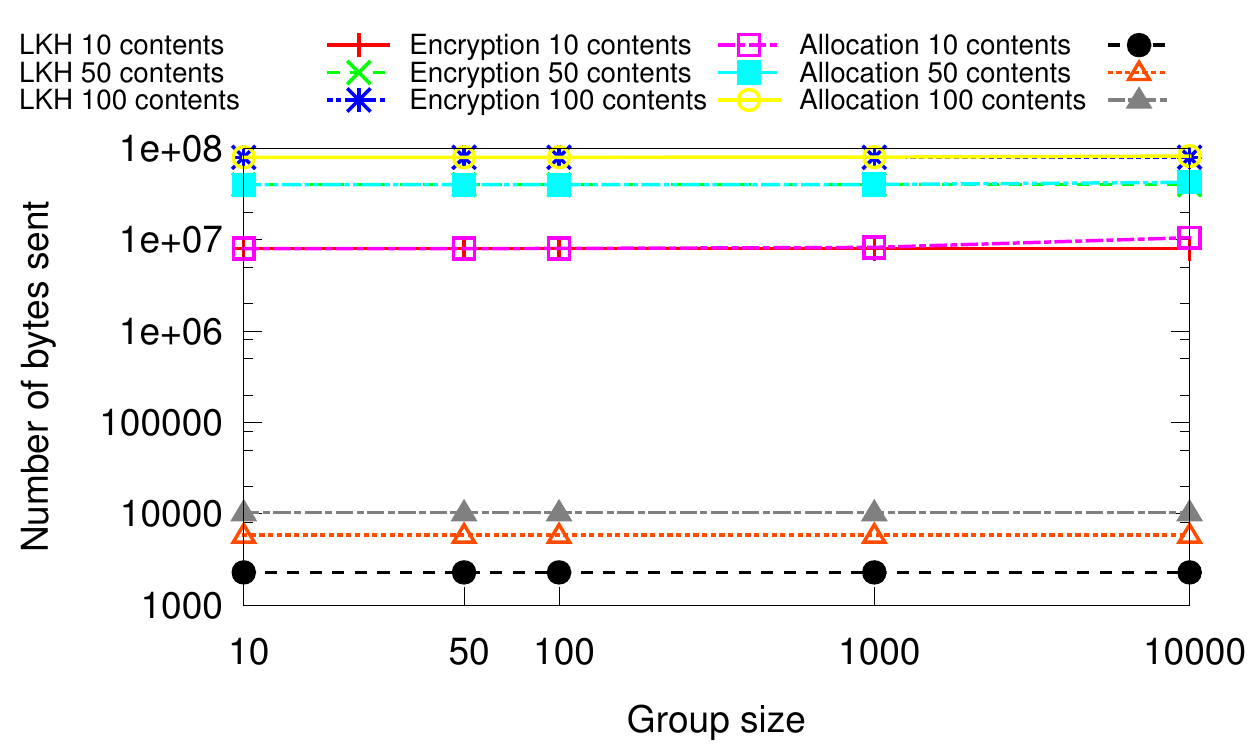}
        }
\subfigure[Existing members of the group\label{fig:otherMemberLeaveWithBSSize}]{
        \includegraphics[width=0.48\textwidth]{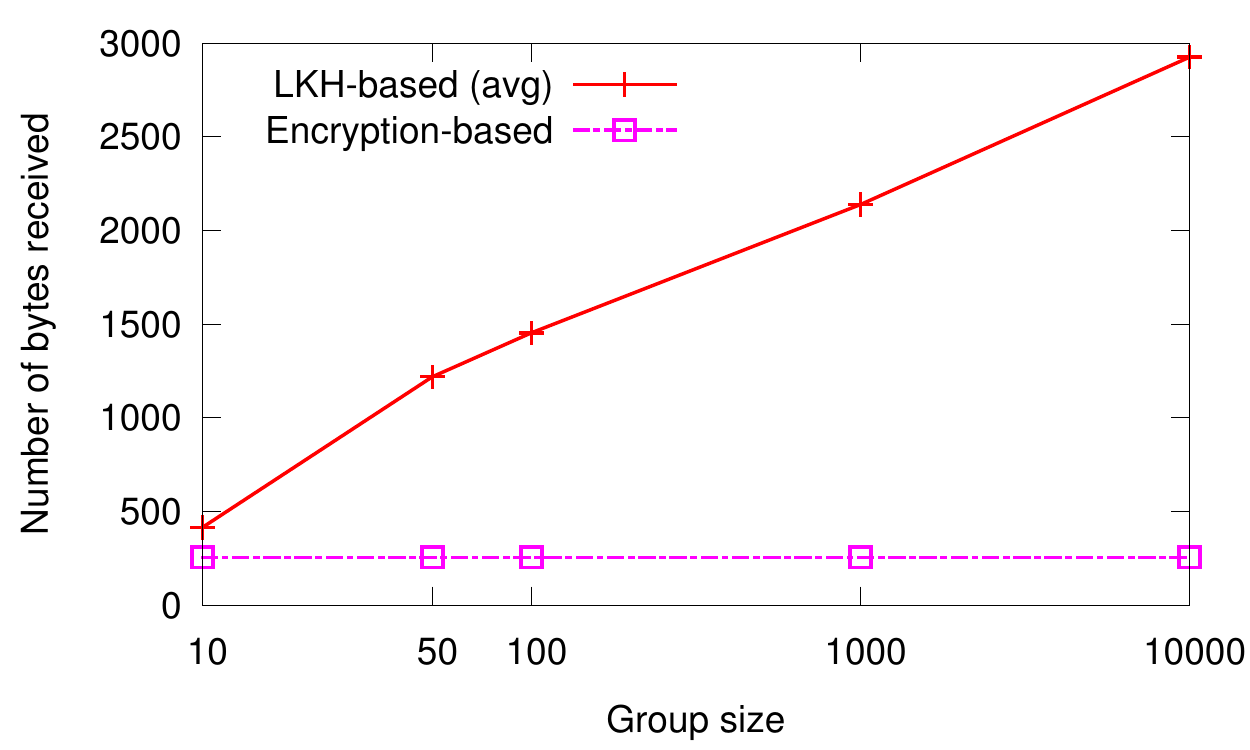}       }
\caption[Evaluation of the leave with backward secrecy]{Evaluation of the processing time and the number of bytes spent by the group owner and the existing members for the leave with backward secrecy of a user from groups with different numbers of members and contents.\label{fig:leaveBWTime}}
\end{figure*}

\paragraph{Leave with backward secrecy (G3)}
%\st{\textcolor{blue}{As we explained in Sections \ref{sec:scenario} and \ref{sec:privacyControl}, it is not possible to define a solution for guaranteeing the backward secrecy property for the leave operation, because the removed users could have made local copies of the contents of the group before being removed from the group itself. Hence, although the group contents would not be accessible anymore for removed users through the DOSN interface, they would be available because of the local copies. However, since to the best of our knowledge  there are two DOSNs which thus implementing groups of type Gx, in the following we give an evaluation of the leave with backward secrecy in the specific case of groups of type G3 (i.e., where the group join operation does not guarantee the backward secrecy property).}}

In order to ensure the backward secrecy property, the leave operation must guarantee that the removed member cannot access anymore the contents already published in the group. 
When the \lkhBased~ enforcement model is adopted, the forward secrecy is guaranteed by properly updating the key tree as previously described for the leave without backward secrecy.

However, this is not sufficient to guarantee the backward secrecy property as well, because 
the symmetric contents keys of the $p$ contents published in the group before executing the leave operation are \textcolor{black}{potentially} known to the removed user, as well as the old group key with which they are encrypted. 
As a result, the removed user can still access them. Hence, in order to enforce the backward secrecy on the removed user, the group owner must retrieve the $p$ contents published in the group and re-encrypt them by creating and using $p$ new symmetric content keys. Such $p$ new symmetric content keys must then be symmetrically encrypted using the new group key.
%\st{(for a total of $O(2p)$ symmetric encryption operations)}.
As a result, the removed user cannot access the new versions of the $p$ contents because they are encrypted with a new content key and a new group key which have not been revealed to her/him.
%\st{Obviously, this solution works only if all copies of the previous versions of such contents have been permanently deleted.}
\textcolor{black}{Again, we recall that this solution does not really protect content privacy, as explained in Section \ref{sec:scenario}}.
%\st{In addition, the group owner creates only one message for the existing members of the group and $2p$ messages to update the contents of the  group.} 
Hence, the total number of symmetric encryption operations performed by the group owner is at most equal to $O(2\cdot p+d\cdot h)$, where $O(2\cdot p)$ operations are necessary to ensure backward secrecy \textcolor{black}{($p$ operations are executed on contents and $p$ operations are executed on keys)}, while  $O(d\cdot h)$ operations \textcolor{black}{(executed on key tree nodes)} are needed to ensure forward secrecy.
Figure \ref{fig:groupOwnerLeaveWithBSTime} shows the processing time required by the group owner in order to remove a user while guaranteeing backward secrecy from groups having different number of users (10, 50, 100, \num{1000}, and \num{10000}) and contents 
(10, 50, and 100).
In our experimental setup, the processing time spent by the group owner depends mainly on the number of contents in the group. Indeed, \textcolor{black}{if we compare the time required for updating and encrypting the nodes of the key tree only (shown in Figure \ref{fig:groupOwnerLeaveWithoutBSTime} for the leave without backward secrecy) with the total time shown in Figure \ref{fig:groupOwnerLeaveWithBSTime}, we observe that} the re-encryption of published contents takes more time than those required to update and distribute 
%\st{the symmetric keys} 
\textcolor{black}{the updated nodes} of the key tree. 
In addition, Figure \ref{fig:groupOwnerLeaveWithBSSize} shows that the number of bytes sent
by the group owner in order to remove a user with backward secrecy mainly depends on the number $p$ of the contents published in the group that the group owner has to re-encrypt. 
Indeed, \textcolor{black}{for all number of contents $p$ and for all the number of group members $n$, more than }  
%\st{about} 
99\% of the total bytes sent by the group owner are necessary to re-encrypted contents already published on the group.

The existing members of the group retrieve the updated nodes of the key tree from the leave notification message stored in the Group Message List. As shown by Figure \ref{fig:otherMemberLeaveWithBSSize}, the number of bytes received in the Group Message List and read by each member of the group is logarithmic with respect the number of users $n$ and correspond to, at most, $O(d\cdot h)$  \textcolor{black}{encrypted nodes of the key tree}.
%In addition, the number of bytes received \st{from a} \textcolor{red}{by a} member \st{of} \textcolor{red}{remaining in} the group, as a result of the leave operation with backward secrecy, does not depend on the number of the contents $p$ in the group. 
Each member $m$ decrypts from the leave notification message at most $O(h)$ nodes, 
%\st{i.e., all those nodes that are on the path towards the root of the key tree.}
i.e., all the nodes that are in common on the paths from the root towards the two leaves representing $m$ and the leaving member.
In addition, the number of bytes read by a member remaining in the group, as a result of the leave operation with backward secrecy, does not depend on the number of the contents $p$ in the group.
Indeed, Figure \ref{fig:otherMemberLeaveWithBSTime} shows that the processing time required by a group member in order to decrypt the updated nodes is logarithmic with respect the size $n$ of the group.

In the \aclBased~ enforcement model, in order to implement the leave operation with backward secrecy, the group owner removes the identity of the user from the group members list and updates the symmetric group key with a new one. For each of the $n$ members left in the group, the group owner encrypts the new symmetric group key with the individual public key of a member and sends the resulting packet having size 2048 bits to the group member. As a result, the group owner executes $O(n)$ asymmetric encryption operations and sends $O(n)$ encrypted messages having a constant size that depends on the length of the public key.
%In order to communicate the encrypted symmetric group key copies to the corresponding users, the group owner \textcolor{red}{stores them in the Group Message List,} creating a message of size $O(n)$.
Finally, the group owner retrieves the $p$ contents already published in the group and re-encrypts each content of the group with a new symmetric content key which is, in turn, encrypted with the latest symmetric group key. As a result, the group owner creates $2p$ new symmetric keys for the contents and performs $2p$ symmetric encryption operations to secure them. 
Figure \ref{fig:groupOwnerLeaveWithBSTime} shows the total processing time required by the group owner for the leave operation with backward secrecy on groups of different number of members $n$ and contents $p$. The results clearly indicate for the groups having $n<<p$, i.e., the number of group contents is considerably higher than the number of group members, the processing time mainly depends on the symmetric encryption operations on the contents. However, when $n \geq p$ 
the processing time mainly depends on the asymmetric encryption operations required to communicate the new symmetric group key to the members of the group. The number of bytes sent by the group owner is mainly affected by the number of contents $p$ of the groups (see Figure \ref{fig:groupOwnerLeaveWithBSSize}).  
%\st{and it is on average 3\% higher than those sent by using the \lkhBased~ enforcement model.} 
Each existing member of the group retrieves the encrypted message of 256 bytes, containing the new symmetric group key encrypted by using public key of the member.
Indeed, the Figure \ref{fig:otherMemberLeaveWithBSSize} shows how the number of bytes received by each group member is constant and it does not depend on both the number of members and the number of contents in the group. %\st{because it involves only the group members.} 
As shown by the Figure \ref{fig:otherMemberLeaveWithBSTime}, the processing time required by each group member is constant and it is equals to the time of an asymmetric decryption operation.

When the \policyBased~ enforcement model is used, in order to implement the leave operation, the group owner removes the intended user from the group so as to ensure that she/he will not be considered during the publication of future contents (forward secrecy). 
%However, the removed user can still access old contents published in group. 
%Hence, 
Moreover, to ensure the backward secrecy property as well, the group owner updates the privacy rules paired to all the $p$ contents already published in the group, by deleting the identity of the removed users from the list of authorized members.
%\st{creates a new privacy policy for the removed user, that denies access to all the $p$ contents published in the group.}
The amount of processing time required to update the privacy policy, as well as its size, depends on the number of contents $p$ published in the group. Indeed, Figure \ref{fig:groupOwnerLeaveWithBSTime} shows that the processing time spent by the group owner is very low since it does not require expensive cryptographic operations. In addition, it depends only on the number of contents in the group. 
The number of the bytes sent by the group owner as a result of the leave operation is the same of the total size of the privacy policy. In particular, Figure \ref{fig:groupOwnerLeaveWithBSSize} clearly shows that the number of bytes only depends on the number of contents in the group. Finally, the leave operation with backward secrecy does not involve the existing members of the group but only the group owner.

\section{Discussion}
\label{sec:discussion}
The experimental results shown in the previous section provide important \mbox{information} regarding the overhead introduced by the 3 privacy enforcement models discussed in this manuscript.
In this section, we highlight the differences among these privacy enforcement models in term of costs %and benefits 
introduced for each operation. Table \ref{tab:generalComparison} compares the performance of the 3 models by
%exploited by current DOSNs 
showing the analytical costs 
%required by the 3 enforcement models in order to implement
of the group join and leave operations, and of the content publication on each group type from the perspective of the group's owner. 
%Since the execution cost of symmetric encryption depends has a very different execution cost w.r.t. the asymmetric one, 
In the cost expressions reported in Table \ref{tab:generalComparison} we denote by $E_{S_k}$ the cost of encrypting a symmetric key using symmetric encryption, by $E_{S_c}$ the cost of encrypting a content using symmetric encryption, and by $E_{AS}$ the cost of encrypting a symmetric key using asymmetric encryption.
Instead, $n$ is the number of member of the group, $p$ is the number of contents published in the group, $d$ represents the maximum number of children nodes of a node in the LKH-based enforcement model \textcolor{black}{ and $h$ is the height of the key tree.}
Moreover, $z$ and $w$ indicate the complexity to add/remove a user to/from the member lists, respectively.

\begin{table*}[tb]
\footnotesize
\centering
\caption{Comparison of the analytic cost of each privacy enforcement model \label{tab:generalComparison}}
\begin{tabular}{clclll}
\hline
\multirow{2}{*}{\textbf{Cost}} & \textbf{Ope-} & \textbf{Group} & \multicolumn{3}{c}{\textbf{Enforcement Models}} \\ \cline{4-6}    
& \textbf{ration} &\textbf{Type}& \multicolumn{1}{c}{\textbf{Encryption}} & \multicolumn{1}{c}{\textbf{LKH}} & \multicolumn{1}{c}{\textbf{Allocation}} \\ 
%\multirow{2}{*}{\textbf{Cost}} & \multirow{2}{*}{\textbf{Opera}} & \textbf{Group} & \multicolumn{3}{c}{\textbf{Enforcement Models}} \\ \cline{4-6}    
%& &\textbf{Type}& \multicolumn{1}{c}{\textbf{Encryption}} & \multicolumn{1}{c}{\textbf{LKH}} & \multicolumn{1}{c}{\textbf{Allocation}} \\ 
\hline
\multirow{4}{*}{}&join&  G2 & $O(E_{AS}+E_{S_k})$ & $O(2\cdot h \cdot E_{S_k}+E_{AS})$& $O(z)$ \\ \cline{2-6} 
 &join & G3, G4 &$O(E_{AS})$&$O(h \cdot E_{S_k} +E_{AS})$& \textcolor{black}{$O(p \cdot z+z)$}\\ \cline{2-6} 
Time & leave &  G3 &$O(p \cdot E_{S_c}+$& $O(p\cdot E_{S_c} +p\cdot E_{S_k}+ $& \textcolor{black}{$O(p\cdot w+w)$}\\ %\cline{2-6} %SPEZZATA DA PAOLO
& & & $p \cdot E_{S_k}+n\cdot E_{AS})$ & $d\cdot h\cdot E_{S_k})$ & \\ \cline{2-6} %SPEZZATA DA PAOLO
 &leave& G2, G4 & $O(n\cdot E_{AS})$ & $O(d\cdot h\cdot E_{S_k})$ & $O(w)$ \\ \cline{2-6} 
 %Time & leave &  G3 &$O(p \cdot E_{S_c}+p \cdot E_{S_k}+n\cdot E_{AS})$& $O(p\cdot E_{S_c} +p\cdot E_{S_k}+ d\cdot h\cdot E_{S_k})$& \textcolor{black}{$O(p\cdot w+w)$}\\ \cline{2-6} 
 %&leave& G2, G4 & $O(n\cdot E_{AS})$ & $O(d\cdot h\cdot E_{S_k})$ & $O(w)$ \\ \cline{2-6} 
\multicolumn{1}{l}{} & publ. & - & $O(E_{S_c}+E_{S_k})$ & $O(E_{S_c}+E_{S_k})$ & $O(n)$ \\ \hline
%& join  & G1, G2 & $O(1)$ & $O(2\cdot log_d(n)+1)$ & $O(1)$ \\ \cline{2-6} 
%& join & G3, G4 & $O(1)$ & $O(log_d(n)+1)$ & $O(p)$\\ \cline{2-6} 
%Transmission & leave & G1, G3 & $O(p+n)$ & $O(p + d\cdot log_d(n))$ &$O(p)$ \\ \cline{2-6} 
%\textbf{} & leave & G2, G4 & $O(n)$ & $O(d\cdot log_d(n))$ & $O(1)$ \\ \cline{2-6} 
%\multicolumn{1}{l}{} & publication & - & $O(1)$ & $O(1)$ & $O(c)$ \\ \hline
\end{tabular}\\
{\scriptsize \textit{n} = number of group members; \textit{p} = number of contents in the group; $d$ = the arity of the key tree; $h$ = the height of the key tree; $E_{AS}$ = cost of encrypting a symmetric key using asymmetric encryption; $E_{S_k}$ = cost of encrypting a symmetric key using symmetric encryption; $E_{S_c}$ = cost of encrypting a content using symmetric encryption; $z$ = cost of adding a user to the member list of the group; $w$ = cost of removing a user from the member list of the group}
\end{table*}

From Table \ref{tab:generalComparison}, we observe that the time complexity introduced by the Encryption-based enforcement model for the group join operation on groups of type G2, G3, and G4 is constant, i.e., it does not depend neither on the number of members belonging to the group nor on the number of content published it the group. 
As a matter of fact, in case of group of type G3 or G4, the cost of the join operation is due to one asymmetric encryption operation, $E_{AS}$, involving the existing group key, and its transmission to the new member. 
In case of groups of type G2 a new group key is created, and one symmetric encryption operation (which costs $E_{S_k}$) 
%using the previous group key 
is performed to securely transmit such new group key to the existing users, while one asymmetric operation (which costs $E_{AS}$) is executed to transmit the same new group key to the new member. 
For what concerns the group leave operation, its execution cost when the group does not support backward secrecy (i.e., groups of type G2 or G4) depends on  $n$, the number of members of the group, because $n$ asymmetric encryption operations (which cost $n \cdot E_{AS}$), are executed to distribute the new group key to the remaining users.
In case of group of type G3, where backward secrecy is supported, the cost is higher and also depends on the number of contents $p$ already published in the group because, besides redistributing the new group key ( which costs $n \cdot E_{AS}$), each content is also re-encrypted using a new symmetric content key (the additional cost is $p \cdot E_{S_c}$) and each new symmetric content key is encrypted by using the new symmetric group key (the additional cost is $p \cdot E_{S_k}$). 
Finally, the cost for the publication of a content is also constant and it does not depends on the number of members in the group but it depends only on the size of the published content. Indeed, the content is encrypted  with a new symmetric content key (which costs $E_{S_c}$). Instead, the new symmetric content key is encrypted by using the symmetric group key (which costs $E_{S_k}$).
%\begin{table}[tb]
%\centering
%\caption{Comparison of the analytic cost of each privacy enforcement model  where: \textit{n}=number of group members; \textit{p}=number of contents in the group; \textit{c}=complexity of the privacy policy; $d$=the maximum number of children nodes of the LKH\label{tab:generalComparison}}
%\begin{tabular}{clclll}
%\hline
%\multirow{2}{*}{\textbf{Cost}} & \multirow{2}{*}{\textbf{Operation}} & \textbf{Group} & \multicolumn{3}{c}{\textbf{Enforcement Models}} \\ \cline{4-6} 
%& &\textbf{Type}& \multicolumn{1}{c}{\textbf{Encryption}} & \multicolumn{1}{c}{\textbf{LKH}} & \multicolumn{1}{c}{\textbf{Allocation}} \\ 
%\hline
%\multirow{4}{*}{}&join& G1, G2 & $O(1)$ & $O(2\cdot log_d(n)+1)$& %$O(1)$\\ \cline{2-6} 
% &join & G3, G4 &$O(1)$&$O(log_d(n)+1)$&$O(p)$\\ \cline{2-6} 
%Time & leave & G1, G3 &$O(2p+n)$& $O(2p + d\cdot log_d(n))$& $O(p)$ \\ \cline{2-6} 
% &leave& G2, G4 & $O(n)$ & $O(d\cdot log_d(n))$ & $O(1)$  \\ \cline{2-6} 
%\multicolumn{1}{l}{} & publication & - & $O(1)$ & $O(1)$ & $O(c)$ \\ \hline
%& join  & G1, G2 & $O(1)$ & $O(2\cdot log_d(n)+1)$ & $O(1)$ \\ \cline{2-6} 
%& join & G3, G4 & $O(1)$ & $O(log_d(n)+1)$ & $O(p)$\\ \cline{2-6} 
%Transmission & leave & G1, G3 & $O(p+n)$ & $O(p + d\cdot log_d(n))$ &$O(p)$ \\ \cline{2-6} 
%\textbf{} & leave & G2, G4 & $O(n)$ & $O(d\cdot log_d(n))$ & $O(1)$ \\ \cline{2-6} 
%\multicolumn{1}{l}{} & publication & - & $O(1)$ & $O(1)$ & $O(c)$ \\ \hline
%\end{tabular}
%\end{table}

The cost of the LKH-based enforcement model for the group join operation
depends on the height $h$ of the key tree, i.e., which is logarithmic in the number $n$ of the group's members because we assumed the LKH tree balanced.
%in the case of join on group G1, G2, G3, and G4
%for the 4 groups types.
As a matter of fact, in case of group of type G3 or G4, it involves at most  $O(h)$ symmetric encryption operation (which cost $h \cdot E_{S_k}$), for the keys along the path from the root of the key tree to the leaf representing the joining user, plus one asymmetric encryption operation (which costs $E_{AS}$) concerning the individual symmetric key to be sent to the new user. 
For group type G2, since the keys on the path from the root of the key tree to the leaf of the joining user are updated, they must be distributed to both the old members of the groups and the joining user, by encrypting them with different keys. As a result, at most $O(h)$ keys on the path must be encrypted twice with respect the join on group G3, or G4.
Instead, the cost introduced by the leave operation on group G2 and G4 depends on both the height $h$ of the key tree and the maximum number of children nodes $d$ (which is constant), because each new key on the path from the root of the key tree to the removed leaf must be symmetrically encrypted with the keys on the $d$ children nodes. Such cost, which is at most equal to $O(d \cdot h \cdot E_{S_k})$, is also present in the case of leave on groups of type G3, where the contents already published in the group must be re-encrypted, thus introducing an additional cost of $O(p \cdot E_{S_c})$ for the encryption of the contents and $O(p \cdot E_{S_k})$ for the encryption of the new contents keys. 
Finally, the cost for the publication of a content is the same as for the \aclBased~model.

The cost of the Allocation-based enforcement model in the case of join on group G2 is equal to the cost of adding the user to the member list of the group, i.e., $O(z)$, while in the case of leave on group G2, and G4, the cost of removing the user from the member list is equal to $O(w)$.
%because the group owner has only to add the information about the new/removed user in the group. 
Instead, in the case of join on group G3 and G4, as well as in the case of the leave on group G3, \textcolor{black}{the current members list must be updated by adding or removing the involved user. In addition, the privacy policies on the contents $p$ already published in the group must properly modified in such a way that \textcolor{black}{the user added to the group can access the contents already published in the group, and the user removed from the group cannot access such contents anymore.}}
%the cost depends on both the number of contents $p$ already published in the group because the policy must be properly 
Finally, in the case of publication of a contents, the cost depends on the complexity of the privacy policy of the group which, in turn, depends on the number of current members of the group, $n$. 
%boolean conditions on attributes that included in the policy 
%privacy rules included in the policy and that require to be evaluated.

\begin{figure*}[tbp] 
\centering 
%\subfigure[Radar plot of the cost of each privacy enforcement model\label{fig:radarPlot}]{
%        \includegraphics[width=0.45\textwidth]{images/radarPlotTime.png}
%        }
%\subfigure[Radar plot of the cost of each privacy enforcement model\label{fig:radarPlot}]{
%        \includegraphics[width=0.45\textwidth]{images/radarPlotTimeLogScale.png}
%        }
\subfigure[n=4000 p=2000\label{fig:radarPlotPassive}]{
        \includegraphics[width=0.47\textwidth]{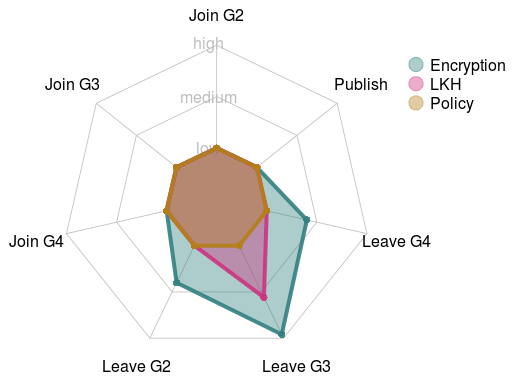}
        } 
\subfigure[n=4000 p=4000\label{fig:radarPlotNormal}]{
        \includegraphics[width=0.47\textwidth]{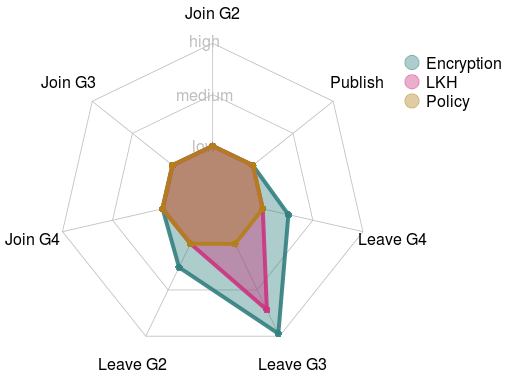}
        }         
\subfigure[n=4000 p=8000\label{fig:radarPlotActive}]{
        \includegraphics[width=0.47\textwidth]{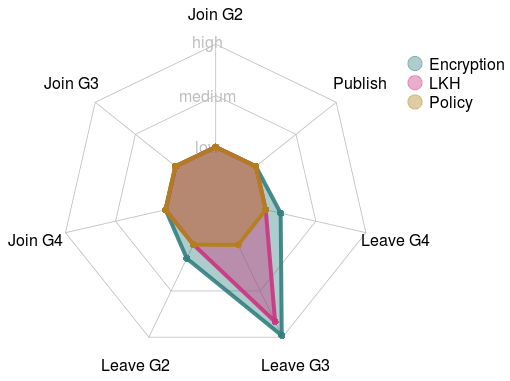}
        }
%\subfigure[n=4000 p=8000\label{fig:radarPlotActive}]{
        %\includegraphics[width=0.48\textwidth]{images/sampleplot.png}
 %       }   
\caption{Comparison of the execution time of the join, leave and publish operations in the three enforcement models.\label{fig:comparisonDiscussion}}
\end{figure*}

In order to numerically compare the 3 enforcement models in a realistic scenario, a graphical and more intuitive representation of such costs is given by the plot in Figure \ref{fig:comparisonDiscussion}, where we measured the time taken by the enforcement models to perform the operations on different types of groups.
%how users behave when they connect to OSNs is crucial for designing user-centered systems. For these reasons, we characterize the temporal properties of users and social groups of users because they allow to evaluate and lead to a better site design of the performance of our approaches.
In particular, we identified three different categories of groups based on both the number of group's members $n$ and the number of contents $p$ published on the groups: passive, normal, and active groups. 
Passive groups category consists of groups with low level of activity where the number of contents $p$ published by groups members is less than the number of members in the group (i.e., $p<<n$) while in normal groups category the number of contents $p$ published by group's members is more or less the same as the number of members $n$ (i.e., $p\simeq n$) and it represent groups with a balance level of activity among members. Finally, active groups category consists of groups with a high level of activity where the number of contents $p$ published on the group is higher than the number of group's members (i.e., $p>>n$).
We derive the average number of groups' members from \cite{de2017logical}, where the size of 18 real Facebook groups of different types has been monitored and analyzed over time. Based on the results of \cite{de2017logical}, we assume that each group consists of about $n=4000$ members.
Consequently, for what concerns the number of posts published in the group, we suppose: \textit{i)} $p=2000$ contents in the case of passive groups, \textit{ii)} $p=4000$ contents in the case of normal groups, and \textit{$p=8000$} contents in the case of active groups. Furthermore, we set the maximum number of children nodes $d$ of the key tree to $4$.
%\st{and the maximum height of the tree to $8$}
%\st{We assume that the size of each content is equals to 100KB\footnote{\url{https://www.facebook.com/help/266520536764594}}.}
%GIA' DETTO IN PRECEDENZA
%In order to numerically compare the costs introduced by the three enforcement models in a realistic scenario, 
We performed the experiments using the same test platform we exploited for the experiments  %\st{and the same symmetric and asymmetric scheme of the Crypto++ library already} 
we described in Section \ref{sec:evaluation}.
To ease the plot reading, the performance results obtained by each group are normalized by using the min-max normalization, and the cost axes report three values: \textit{low} (close to the heptagon center), \textit{medium}, and \textit{high} (in the heptagon borders), which correspond to increasing amounts of time.

%These costs are derived from the analytical cost model of Table \ref{tab:generalComparison}, they reflect the different overhead introduced by the enforcement models. 

%Since privacy enforcement models must be adapted on a case-by-case basis, depending on the characteristics of each group, we provided useful guidance for planning the most appropriate and efficient privacy enforcement model.

%OR

%Since privacy enforcement models expose different characteristics, they must be evaluated on a case-by-case basis, depending on the type of group. For this reason, it is important to provide useful guidance for planning the most appropriate and efficient privacy enforcement model. 

%G1
%a group where each member u can access only the content published since u was joined to the group while, in the case of removal, such contents are no longer accessible to u.
%G2 a member u of the group can access only the contents published after the addition of u to the group. When u is removed from the group, such contents can still be accessed by u.
%G3,G4  a member u of the group can access both the contents published after the addition of u and the contents already in the group before the addition of u.
% In the case of group G3, when u is removed from the group, such contents are no longer accessible while in the case of G4 they are can still be accessed by the removed member.
%Each type of groups ensure that a removed member cannot access future contents that will be published in the group.
%Groups of type G1 are suitable to be implemented through the \policyBased~ and

%FORSE SI POTREBBE DIRE CHE TIPO DI ENCRYPTION PRENDIAMO IN CONSIDERAZIONE PER PRENDERE I TEMPI DEI PLOT

%The plot clearly indicates that the three enforcement models expose different performance which must be evaluated on a case-by-case basis, depending on the type of group. 
%In particular, for the join operation, the three enforcement models expose a similar cost. For instance, in Figure \ref{fig:joinTimeBSgroupOwner} we can see that, in case of groups of type G2, the cost of the join operation in the \aclBased\ and in the \policyBased~models are constant and they take, respectively, about 0.16ms and 0.10ms. The cost of the same operation in the \lkhBased~model, instead, depend on the number of group members, and in our experiments takes about 0.16ms.

The three plots clearly indicate that, for the group join operation, the three enforcement models expose a similar cost, while for the group leave operation the three models expose different performance which must be evaluated on a case-by-case basis, depending on the type of group. 
For instance,  in case of groups of type G2, the cost of the join operation in the \aclBased\ and in the \lkhBased~models is about 0.16 ms.
As a matter of fact, the additional cost introduced by the \lkhBased~enforcement model with respect to the \aclBased\ one due to the management of the key tree is negligible in our experiments (where $n$ = \num{4000}) with respect to the cost of Asymmetric Encryption.
The \policyBased~model, instead, takes 0.10 ms (see Figure \ref{fig:joinTimeBSgroupOwner}).
Also in the case of groups of type G3 and G4, the cost of the join operation in the \aclBased\ and \lkhBased\ models is mainly due to the Asymmetric Encryption, thus resulting to be 0.16 ms in both cases, while in the \policyBased~model the cost is 3.2 ms for groups of $p=2000$ contents and 10.8 ms for groups of $p=8000$.

The execution of the group leave operation on groups of type G2 and G4 when the \aclBased\  approach is adopted takes about 640 ms, because of the number $n=4000$ of Asymmetric Encryption operations, while the same operation executed on groups of type G3 requires also the re-encryption of the contents already published in the group, i.e., about 4221 ms in the case of group with $p=8000$ contents.\\
In the case of the \lkhBased\  approach, the leave operation on groups of type G2 and G4 takes only 0.006 ms, because it requires a limited number of symmetric encryption operations, while the leave operation on group G3 requires also the re-encryption of the contents already published in the group, i.e., about 3581 ms in the case of group with $p=8000$ contents.
The execution of the group leave operation adopting the \policyBased\ approach on group G2 and G4 takes about 0.1 ms, because it only needs to remove a member from member list, while the leave operation on groups of type G3 requires also the modification of the privacy rule paired to each published content. As a result, the cost %of the leave operation on group G3 
depends on the number of contents on the group and it ranges between 3.29 ms for group with $p=2000$ contents and 10.8 ms for group with $p=8000$ contents.
Summarizing, in the three plots shown in Figure \ref{fig:comparisonDiscussion}, the normalized cost of the group join operation results to be \emph{low} for any type of group, for any enforcement model, and for any category of group.

In the case of leave operation, regardless of the group type, the plots clearly show that, in our experiments, the \aclBased~enforcement model introduces the highest cost. %\st{which linearly depends on the number of group's members}.
As a matter of fact, in case of groups of type G3 the normalized cost results to be \emph{high} for the three group categories \textcolor{black}{and it is dominated by the time taken for re-encrypting all the contents of the group.}
Instead, in case of groups of type G2 and G4, the normalized cost results to be close to \emph{medium} for passive groups and in between \emph{medium} and \emph{low} in case of normal and active groups. 
Instead, the \lkhBased~ and the \policyBased~ enforcement models are alternative solutions providing lightweight implementations of the leave operation.
%on group's types G1, G2, G3, and G4. 
In particular, under the assumptions on the values of $n$ and $p$ we made, the normalized cost of the leave operation in the \policyBased~enforcement model always results to be \emph{low}, while the normalized cost of the \lkhBased~model results to be \emph{low} for the three group categories only for groups of type G2 and G4. 
Instead, for groups of type G3, the normalized cost of the leave operation in the \lkhBased~model results to be \emph{medium} in case of passive groups, and between \emph{medium} and \emph{high} in case of normal and active groups.
\textcolor{black}{Such costs reflect the number of contents in each group because it is necessary to re-encrypt again all of them.} 
%\st{because it requires only the definition of a privacy policy which denies access to the old contents published in the group while the \lkhBased~ enforcement model takes a medium overhead for the modification of the key tree.}
%Furthermore, in the case of the leave on the group's types G1 and G3, the proposed enforcement models expose a similar cost which depends mainly on the number of contents previously published in the group. 
%\st{Indeed, the \policyBased~ enforcement model requires only the modification of the privacy policy of the group in order to deny the access to the user who leaved the group, while the \lkhBased~ and \aclBased~ enforcement models create a new group key which is distributed to the members of the groups by exploiting either asymmetric encryption (in the case \aclBased~)  or symmetric encryption (in the case \lkhBased~). For this reason, the overhead introduced by the \aclBased~ is in general higher than the overhead of the \lkhBased~ approach, for both passive, normal, and active groups.}
As a result, the \aclBased~enforcement model is well suited to manage group where either the leave operation is not permitted or it does not often occur, because the leave operation is very  expensive. 

In the case of the publication of a content, the \policyBased, the \lkhBased, and the \aclBased~enforcement model introduce a similar overhead. However,  one of the main differences between the \policyBased~and the \lkhBased~ or the \aclBased~models is that in the former the contents of users are stored and maintained according to the users' privacy preferences while in the last ones models the contents of users can be stored on any peers of the DOSN because they are encrypted only for authorized members. 

%QUI FORSE SI POTREBBE FARE UN DISCORSO SUL FATTO CHE NEL POLICY BASED OGNI TANTO BISOGNA CREARE NUOVE REPLICHE DEI CONTENUTI, E QUESTO E' UN COSTO ADDIZIONALE ANCHE SE ASINCRONO.
Finally, another important aspect of the proposed models is the availability of protected contents. The contents protected using both the \aclBased~and the \lkhBased~enforcement models can be stored and replicated on any peers of the system because they are encrypted with a shared asymmetric key known only to authorized members. Instead, the \policyBased~enforcement model can store the contents only on some trusted peers of the system and the availability of such contents depends on the peers selected as replica. As a result, \policyBased~enforcement could not ensure the availability of the contents and it introduces a further hidden cost for managing replica's peers.

%The \aclBased~ enforcement model is well suited to manage group where the leave operation is not permitted or does not often occur because it is very  expensive. In contrast, the \lkhBased~ enforcement model is able to manage groups with a high number of users, even in the presence of a high number of addition and removal of users.

\section{Future directions}
\label{sec:futureDirection}
In the following, we draw future research directions related to privacy enforcement models in DOSNs and, based on the previous comparison, we also identify some open issues related to the privacy enforcement models which deserve further investigation.

\subsection{\textcolor{black}{Open issues}}
\noindent
\textcolor{black}{Current researches focus their attention on the following major challenges:}\\

%\vspace{0.2cm}\\
\textcolor{black}{\emph{How to reduce the  computational cost when a leave occurs? }%in the case of a leave of a member from a group?} 
%\todo[inline,color=yellow]{a causa del formato ci veniva sia il "?" che il ".", che e' brutto. Ho messo emph e un vspace invece di paragraph e ora mi pare meglio. }
The reasons for a user to leave (or to be removed from) a group can be different (e.g., the member is not interested in the topic discussed in the group anymore, 
%\st{the private key of the member gets compromised}
%\todo[inline,color=yellow]{questa della chiave privata vale solo per un modello, quindi la leverei. Ho aggiunto che l'utente potrebbe non essere piu' interessato al gruppo}
or the member behaves improperly) and the group membership revocation must be securely and efficiently implemented by any privacy enforcement model. 
However, achieving a low computational cost for group membership revocation remains a relevant challenge because of the high number of members typically involved in groups. 
A recent approach used to mitigate this problem is to design privacy enforcement models that define some order relationships among either the members of a group or the members of different subgroups, making constant the computing overhead required to remove a member from a group in the best case scenario. However, such approach exposes an high computational overhead in the worst case scenarios, and rely on the fact that such scenarios are unlikely to happen \cite{tiloca2020group}.}\\

%\vspace{0.2cm}

\textcolor{black}{\emph{How to enhance the security level of privacy enforcement models?} Since some privacy enforcement models rely on encryption algorithms, cryptographic key management, digital signatures, and hashing functions, then the reliability and security of such models heavily depend on such algorithms and their configuration parameters. For the introduced  reasons, privacy enforcement models used in current DOSNs must follow security recommendations provided by IT advisors, government organizations, or cybersecurity agencies about the cryptographic keys size and the algorithms to be used. For instance, the National Institute of Standards and Technology (NIST) develops cybersecurity standards and encourages the adoption of guidelines, best practices, and other recommendations that help system administrators and users to secure their data and applications.\\ %\st{In addition to security recommendations and best practices, designers of privacy enforcement models must also plan to start software bug tracking and monitoring in order to notify users with the associated security issues.}\\
%\todo[inline,color=yellow]{a me la frase sbarrata sembra troppo generale, non è una cosa tipica dei privacy enforcement mechanisms, quindi non la metterei}
However, our analysis reveals that most of the existing DOSNs provide groups with  public membership information, i.e., they do not hide the identity of the group members to the other users of the DOSN. In particular, the most part of current DOSNs do not support groups having protected or private membership because the group membership information is disclosed through key management and interactions between members. 
%Instead, a group having protected membership must ensure that only current group members are aware of the group, while a group with private membership guarantees that only the administrators of the group can view current group members. 
%\todo[inline,color=yellow]{il paragrafo precedente non e' chiarissimo. Forse si puo' eliminare la spiegazione dei gruppi protetti e privati, e focalizzarsi sul dire semplicemente che il sistema usato per dare le chiavi del gruppo alle persone rivela i membri dei gruppi a tutti. }
%\todo[inline]{ok}
To address the above issue, privacy enforcement models proposed by current DOSNs could adopt onion routing  \cite{graffi2020libresocial} and anonymization techniques to provide mitigation against identity disclosure \cite{masinde2020peer}.
%In contrast to DOSNs, Table \ref{tab:mapping} indicates that centralized OSNs are able to easily implement groups with protected and private membership because all the information of the group are stored and managed by the service providers.
}\\

%\vspace{0.2cm}

\textcolor{black}{\emph{How to \textcolor{black}{dynamically} adapt the characteristics of privacy enforcement models \textcolor{black}{to the user's needs}?} 
%\todo[inline,color=yellow]{adattarli a cosa? Lo scriverei nel titolo}
%\todo[inline]{aggiunto}
Users connect to OSNs by using several devices with different characteristics, some of them having limited computational and storage capacity. Consequently, adaptive privacy enforcement models are necessary to deal with heterogeneity of users devices, demands, and usage behaviour.\\ For instance, the pervasive penetration of mobile devices (such as smartphones and tablet) has changed the ways in which people connect to the DOSNs \cite{stuedi2011contrail} and introduced several challenges, such as finding privacy enforcement models with limited computational and memory resources requirements. Indeed, due to technical constraints, some mobile devices cannot support the use of specific cryptography operations and/or are not able to store the whole cryptographic material needed by the privacy enforcement models \cite{schwittmann2013privacy}. 
Additionally, users can configure the DOSN to be executed on different devices having either pay-as-you-go resource management model (e.g., virtual machine provided by cloud storage provide) or pre-paid resource management model (e.g., mobile devices with a limited plan for data download). Consequently, users have to pay for resources used by the DOSNs. 
%and the users' devices have only a limited amount of them.
%\todo[inline,color=yellow]{il secondo pezzo della frase "and the users' devices have only a limited amount of them" non e' relativo al pagamento, quindi lo leverei.}
In such a case, in particular in the \policyBased~enforcement model where the allocation strategy is driven by the privacy preferences expressed by the content publisher,
a malicious publisher can exhaust the resources of other users by sharing a massive number of contents with them. 
To prevent this from happening, the privacy enforcement models 
%INIZIO ULTIMO TAGLIO PAOLO
%(and in particular the \policyBased~enforcement model which leverages users' personal resources) 
%FINE ULTIMO TAGLIO PAOLO
should prevent denial of service attacks by introducing context-aware adaptation strategies that monitoring consumed resources \cite{schaub2018context}.}\\

%\vspace{0.2cm}

\emph{How to support the collaborative administration of groups?}
Large groups can have a very complex internal structure consisting of several members who have different roles and collaborate with each other for managing groups' activities.
Our analysis reveals that privacy enforcement models having collaborative administration policy are not supported by current systems (see Table \ref{tab:mappingDOSNs}), hence they need to be integrated within current DOSNs. Indeed, the administration of a group is an important characteristic 
%which affect the performance 
of the privacy enforcement model because it regulates who can authorize or really initiate the execution of the user join or leave operation.\\
In addition, in the case of a group with collaborative administration, the sequence of operations generated by the administrators could raise some conflicts and new recovery mechanisms must be designed to keep the group information updated and to resolve inconsistency \cite{such2016resolving}.
For instance, multimedia contents produced in OSNs, such as photos or videos, can have multiple owners (e.g., several users may appear in the same picture \cite{such2017photo}) and each of them may want to define a different access policies for the shared item. 
Collaborative access control models \cite{paci2018survey} were proposed to enable security policy definition \cite{carminati2011collaborative}, negotiations \cite{such2016privacy}, and conflicts resolution \cite{such2016resolving} in a collaborative way. Unfortunately, existing collaborative access control models have been designed to support few application scenarios (e.g., healthcare) and they do not consider the requirements of DOSNs' users.

\subsection{Blockchain-based privacy enforcement model}
%non in linea con il nostro obbiettivo ma interessante per gli spunti di ricerca futuri su privacy dei contenuti
Recently, Distributed Ledger Technology (DLT) \cite{natarajan2017distributed} has gained considerable attention from industry and academia due to its applicability in several relevant scenarios. 
The blockchain is one of the most popular type of DLT, allowing to establish trust in an open and decentralized environment while providing a rich set of tools that cover various tasks (e.g., smart contract and token). Several works have been proposed in the literature exploiting blockchain to enhance the privacy of contents published by DOSNs' users. 
However, some characteristics of the blockchain (e.g., immutability) could conflict with some privacy regulations (such as the General Data Protection Regulation which requires that users can delete all personal information \cite{ahmed2020towards}) and further studies are needed to understand which privacy requirements can be accommodated by the privacy enforcement models implemented in DOSNs \cite{ahmed2020gdpr}.

%Current solutions exploit blockchain technology to manage the reward-penalty schemes on contents published by users, verify or protect access to users' contents. 
%\todo[inline,color=yellow]{i lavori elencati sotto non mi pare che siano molto rilevanti per la protezione della privacy, a parte quello che distribuisce le chiavi con la blockchain. Quello che mette l'hash sulla blockchain riguarda l'integrity dei contenuti.}
%\todo[inline]{qui ho cercato di dare un quadro generale su come viene utilizzata la blockchain nella gestione dei contenuti delle OSNs.}
With regard to content privacy,  most of the existing DOSNs use the blockchain as an infrastructure to verify the correctness of the  published contents. 
%\todo[inline,color=yellow]{la frase precedente non torna con la successiva. LA precedente sembra che dica che i veri contenuti sono memorizzati altrove, e tipo solo l'hash sulla blockchain. LA frase successiva dice che i contenuti stessi sono sulla blockchain. Abbiamo mica cancellato qualcosa per errore?}
For instance, the DOSNs proposed in \cite{jiang2019bcosn,chen2021blockchain} leverage blockchain to ensure the integrity of the contents stored on a distributed storage service based on DHT \cite{rhea2005opendht}. Indeed, the blockchain is used to store an hash of each content, while users can use the hash values stored in the blockchain to query the corresponding data on the DHT. A similar approach is recommended in \cite{xu2018building}, where InterPlanetary File System (IPFS) \footnote{Available at: \url{https://ipfs.io}} is used as distributed data storage service while the blockchain is used to verify the integrity of the contents.

In contrast to the previous approaches which store on the blockchain only an hash pointer, some DOSNs prefer to store the original contents on the blockchain. For instance, Steemit\footnote{Steemit: \url{https://steemit.com/}}, Hive \footnote{Hive: \url{https://hive.io/}}, Sapien\footnote{Sapien: \url{https://www.sapien.network/}}, and  SocialX\footnote{SocialX: \url{https://socialx.network/}}, utilize blockchain to store textual contents and to regulate compensation of users for their actions (such as, creating a content or reviewing an item). Since contents are stored on the blockchain, each user can retrieve and verify them \cite{zeng2019decentralized,jiang2019bcosn,chen2021blockchain}. However, privacy enforcement models are not considered in these platforms because anyone can see all contents without restrictions . %Indeed, the blockchain is mainly used to regulate compensation of users, while large data (such as images) are stored in a separate layer, implemented by a distributed storage service provided by a third-party storage providers \cite{zeng2019decentralized,jiang2019bcosn,chen2021blockchain}.

%\todo[inline]{Use blockchain to verify users' contents}

%\todo[inline,color=yellow]{qui bisogna specificare subito che i contenuti sono da un'altra parte e sono modificabili da chiunque}
%In contrast to the previous approaches which store the original contents on the blockchain, several works prefer to alleviate memory consumption of the blockchain by storing on it only an hash of the content while the original content is hosted by a data storage service. For instance, the DOSN proposed in \cite{jiang2019bcosn,chen2021blockchain} exploits blockchain to ensure the integrity of the contents stored on a distributed storage service based on DHT \cite{rhea2005opendht}.
%The blockchain is used to store an hash of the content while users can use the same hash value stored in the blockchain to query the corresponding data on the DHT. A similar approach is proposed in \cite{xu2018building}, where InterPlanetary File System (IPFS) \footnote{Available at: \url{https://ipfs.io}} is used as distributed data storage service while the blockchain is used to verify the integrity of the contents.\\
%\todo[inline,color=yellow]{Questo sopra e' l'esempio che cercavo prima. Allora questo esempio va messo dove io ho scritto il commento precedente. Vi torna?}
%\todo[inline]{aggiunto}

Authors of \cite{zeng2019decentralized} proposed a solution to ensure the correctness of contents published in the DOSN by focusing on the scalability issues introduced by blockchain \cite{zamani2018rapidchain}. The original content is stored unencrypted on several peers of the DOSNs while the hash of each content is recorded on a shard, i.e., a parallel blockchain which records transactions relevant to the content owner. Since many shards exist in the DOSNs, the state of each shard is hashed and recorded in the blocks of the main chain by using Verifiable Random Functions \cite{MicaliRV99}.
%\todo[inline,color=yellow]{cosa e' questa subchain?}
%\todo[inline]{spiegato meglio}

%\todo[inline,color=yellow]{non capisco come il meccanismo della reward sia integrato con la protezione della privacy dei contenuti}
%\todo[inline]{Nulla, ma sono le uniche OSN  deployate che integrano la blockchain.}

%\todo[inline]{Use blockchain to manage the  reward-penalty schemes on contents published by users}

%\todo[inline]{Use blockchain to protect access to users' contents}
Blockchain technology can be used to protect access to users' contents while ensuring the quality of the services. In particular, DOSNs can exploit Proof-of-Work (PoW) to prevent the storage service from Denial-of-Service (DoS) attacks resulting from large amounts of requests sent, preventing overload of requests. 
For example, in \cite{biedermann2014proofbook} the user  requesting a content has to solve a PoW, whose complexity depends on both the load of the network and the users' request rate, and it acts as a stamp used to pay for the delivery of requests.
%\todo[inline,color=yellow]{fatta leggera modifica alla frase, rimane corretta, vero?}
%\todo[inline]{si}

Another interesting approach to protect access to users' contents is to employ the blockchain as a medium to distribute cryptographic keys, tokens, or policies between friends. 
In particular, the cryptographic keys or tokens must be securely stored on the blockchain, and they can be retrieved by authorized users in order to decrypt contents stored on the storage layer. For instance, in \cite{jiang2019bcosn,dang2021sharing,he2021efficient,javed2021petchain} the contents are encrypted by using secret key and stored on a third-party storage providers while the blockchain contains the secret key encrypted with the asymmetric public keys of the users authorized to access such contents. Instead, some approaches \cite{chakravorty2017ushare,arquam2021blockchain} store an identifier of the encrypted content on the blockchain along with a token which is used to monitor the propagation of information and to limit the number of re-shares that can be performed with that content. Whenever a user re-shares the content, she/he makes another transaction which is registered on the blockchain only if the token value is greater than 0.\\
As per privacy policies, they are stored on the blockchain with the aim of being enforced by an external access control system. For instance, the blockchain can provide users with the ability to define ACL that can be evaluated by the storage system \cite{rahman2020blockchain}, while in \cite{lax2021blockchain} the blockchain is in charge of storing both the privacy preferences of a user in order to determine whether the privacy settings assigned by the social network are compliant with those declared by the user.

\section{Conclusion}
\label{sec:conclusion}
In this paper we analysed the enforcement models used in current Decentralized Online Social Networks (DOSNs) to assess the provided level of protection of the contents that are published in social groups. 
%In particular, we analyzed the existing enforcement models for content privacy protection in DOSNs and we performed an extensive evaluation of the approaches adopted by DOSNs to protect the privacy of contents shared to group of users. In order to evaluate the proposed models, we provided to DOSNs' users the capability to create groups of different types where content delivery typically occurs from one user (sender) to one of these (possibly large) groups of friends (receivers). We measured the performance achieved by enforcement models in providing these types of groups.
In order to analyse such models, we evaluated for each of them the cost of executing the three typical operations of a DOSN (i.e., group join, group leave and content publish), on different types of groups.

The obtained results reveal that the \aclBased~enforcement model is affected by a serious drawback: it is not scalable because  the number of asymmetric encryption operations to be executed to remove a user from a group is linear with the number of users belonging to that group, and this could cause a relevant overhead in case of large groups.
We showed that the \lkhBased~enforcement model can solve the above issue by exploiting the  Logical Key Hierarchy model in order to optimize the costs of the operations (i.e., dramatically reducing the cost of the group leave operation while slightly increasing the one of the group join operation) thus enabling dynamic groups management. 
%In particular, this enforcement model reduces the costs required by current DOSNs for guaranteeing privacy of contents by affecting: the number of encryption operations required each time a user is removed or added to a group of size, and the number of messages.
However, the \lkhBased~enforcement model behaves like the \aclBased~model when the groups are quite static %in the case of quite static groups 
because they both require to encrypt the contents with a symmetric schema. %Furthermore, an addition cost for the management of the LKH is introduced during the join operation.

The \policyBased~ enforcement model is an alternative
approach that can be used to solve the cited problem because it is not based on content encryption. This approach allows users to define privacy policies on their contents by using a privacy policy language and the allocation of users' contents is performed on the basis of the privacy policy defined by the content owner. However, the \policyBased~enforcement model introduces some overhead required for privacy policy definition and evaluation (which depends on the complexity of the policies). 
In addition, in order to ensure high availability of the published contents, this approach also needs to consider the availability patterns (online/offline) of the users, in order to allocate the contents on the peers that will probably be  online for more time.\\ 

The above findings are supported by a thorough  analysis and  extensive experimental campaigns. Finally, research directions on how to tackle the highlighted problems and further open issues in DOSNs are also provided.

%\begin{acknowledgements}
%If you'd like to thank anyone.
%\end{acknowledgements}
\section*{Acknowledgements}
This publication was partially supported by awards NPRP-S-11-0109-180242
from the QNRF-Qatar National Research Fund, a member of
The Qatar Foundation. The findings  are solely responsibility
of the authors.
\balance
\bibliography{IEEEabrv,sample2}
%\bibliographystyle{ACM-Reference-Format}
%\bibliography{sample2}
\end{document}